\documentclass[fleqn,usenatbib]{mnras}

\usepackage{hyperref}   
\hypersetup{colorlinks=true,linkcolor=blue,citecolor=blue,filecolor=blue,urlcolor=blue}

\usepackage{pdflscape}  
\usepackage{afterpage}
\usepackage{newtxtext,newtxmath}
\usepackage[compatibility=false]{caption}
\usepackage{subcaption}
\usepackage{lscape}
\usepackage{enumitem}
\usepackage{amssymb}
\usepackage{pifont}
\usepackage{enumitem}
\usepackage{multirow}
\usepackage{booktabs}
\usepackage{upgreek}

\defcitealias{Peng2010_galfit}{Peng et al. 2010}
\defcitealias{Peng2010}{Peng et al. 2010}
\defcitealias{Darvish2016}{Darvish et al. 2016}
\defcitealias{Tacchella2015}{Tacchella et al. 2015}

\usepackage[T1]{fontenc}
\usepackage{ae,aecompl}
\usepackage{xcolor,colortbl}

\usepackage{cleveref}
\crefname{section}{$\S$}{$\S\S$}
\Crefname{section}{$\S$}{$\S\S$}

\usepackage{graphicx}   
\usepackage{amsmath}    
\usepackage{amssymb}    
\usepackage[flushleft]{threeparttable}

\newcommand{\Sers}{S\'{e}rsic }

\title[KDS II: Tully-Fisher evolution since ${\it z} \simeq 4$]{The KMOS Deep Survey (KDS) II: The evolution of the stellar-mass Tully-Fisher relation since ${\it z} \simeq 4$}

\author[O.J. Turner et al.]{
O. J. Turner,$^{1\thanks{E-mail: turner@roe.ac.uk (OJT)}}$
C. M. Harrison,$^{2}$
M. Cirasuolo,$^{2,1}$
R. J. McLure,$^{1}$
J. Dunlop,$^{1}$\newauthor
A. M. Swinbank,$^{3,4}$
A. L. Tiley$^{3}$
\\
$^{1}$SUPA\thanks{Scottish Universities Physics Alliance}, Institute for Astronomy, University of Edinburgh, Royal Observatory, Edinburgh EH9 3HJ\\
$^{2}$European Southern Observatory, Karl-Schwarzschild-Str. 2, 85748 Garching b. M{\"u}nchen, Germany\\
$^{3}$Centre for Extragalactic Astronomy, Durham University, South Road, Durham, DH1 3LE, U.K.\\
$^{4}$Institute for Computational Cosmology, Durham University, South Road, Durham, DH1 3LE, U.K.
}

\date{Accepted XXX. Received YYY; in original form ZZZ}

\pubyear{2017}

\begin{document}
\label{firstpage}
\pagerange{\pageref{firstpage}--\pageref{lastpage}}
\maketitle

\begin{abstract}
We use KMOS Deep Survey (KDS) galaxies, combined with results from a range of spectroscopic studies in the literature, to investigate the evolution of the stellar-mass Tully-Fisher relation since $z \simeq 4$.
We first establish the slope and normalisation of the local rotation-velocity -- stellar-mass ($V_{\textrm{C}} - M_{\star}$) relationship using a reference sample of nearby spiral galaxies; thereafter we fix the slope, and focus on the evolution of the velocity normalisation with redshift. 
The rotation-dominated KDS galaxies at $z \simeq 3.5$ have rotation velocities $\simeq -0.1$\,dex lower than local reference galaxies at fixed stellar mass.
By fitting 16 distant comparison samples spanning $0 < {\it z} < 3$ (containing $\simeq 1200$ galaxies), we show that the size and sign of the inferred $V_{\textrm{C}}$ offset depends sensitively on the fraction of the parent samples used in the Tully-Fisher analysis, and how strictly the criterion of `rotation dominated' is enforced (i.e. the median $V_{\textrm{C}}/\sigma_{\textrm{int}}$ of the samples, where $\sigma_{\textrm{int}}$ is the intrinsic velocity dispersion).
Confining attention to subsamples of galaxies that are especially `disky' results in a consistent positive offset of $\Delta V_{\textrm{C}} \simeq +0.1$ dex, however these galaxies are not representative of the evolving-disk population at $z \geq 1$.
Using the KDS galaxies we investigate the addition of pressure support, traced by velocity dispersion, to the dynamical mass budget by adopting a `total' effective velocity of form $V_{\textrm{tot}} = (V_{\textrm{C}}^{2} + 4.0\sigma_{\textrm{int}}^{2})^{0.5}$.
The rotation-dominated and dispersion-dominated KDS galaxies fall on the same locus in the total-velocity versus stellar-mass plane, removing the need for debate over the precise selection threshold for rotation-dominated galaxies.
Applying this approach to the comparison samples, we find total-velocity versus stellar-mass relation offsets in the range $+0.08$ to $+0.15$ dex in total-velocity zero-point ($-0.30$ to $-0.55$ dex in stellar-mass zero-point) from the local Tully-Fisher relation at $z \geq 1$, consistent with steady evolution of $M_{\star}$/$M_{\textrm{dyn}}$ with cosmic time.
\end{abstract}




\section{INTRODUCTION}

Typical star-forming galaxies are usually defined as those that lie on the relatively tight relationship between star-formation rate (SFR) and stellar mass $(M_{\star})$ that is observed over a wide redshift range \citep[e.g.][]{Daddi2007,Noeske2007,Elbaz2007}.
As well as the arrival and departure of galaxies from this sequence, due to the combination of processes which replenish and quench star formation  \citepalias[e.g.][]{Peng2010,Tacchella2015}, the mean physical properties of typical star-forming galaxies evolve over time.
This is manifest in the evolution of: the main-sequence normalisation \citep[e.g.][]{Whitaker2012,Whitaker2014}; the mass-metallicity relationship normalisation \citep[e.g.][]{Erb_2006,Maiolino2008,Cullen2014}; disk sizes \citep[e.g.][]{Trujillo2007,VanderWel2014a} and dynamical properties \citep[e.g.][]{Cresci2009,Wisnioski2015,Harrison2017,Swinbank2017}.  \\

Much of the observational progress over the last decade can be attributed to the advent of integral-field spectroscopy, a technique in which an array of spectra can be collected across a given spatial region, allowing for spatially-resolved measurements of the dynamical and chemical properties of both star-forming and quiescent galaxies to be made \citep[e.g.][]{Epinat2008a,ForsterSchreiber2009,Gnerucci2011,Epinat2012,Troncoso_2014,Wisnioski2015,Stott2016,DiTeodoro2016,Swinbank2017,Turner2017}.
In tandem with this, the internal gas properties (e.g. kinematics, metallicity gradients) can now be predicted by high-resolution cosmological simulations for 100s-1000s of galaxies \citep[e.g.][]{Schaye2015,Genel2015,Lagos2017,Swinbank2017}.   
However, these models require several `sub-grid' assumptions to describe the many processes which govern how galaxies evolve, such as nonlinear feedback from star formation and active galactic nuclei \citep[e.g.][]{Schaye2015}.
These assumptions must be refined or refuted by comparison to observations, ideally using large samples of galaxies across cosmic time with integral-field spectroscopic data.   \\

Galaxy samples have now been observed using integral-field spectroscopy, spanning the redshift range $0 < {\it z} < 4$, thanks in particular to the Spectrograph for INtegral-Field Observations in the Near Infrared (SINFONI; \citealt{Eisenhauer2003a}) and the multiplexing capabilities of the {\it K}-band Multi-Object Spectrograph (KMOS; \citealt{Sharples2013}).
The number of observations is continually growing, and as a result we are better placed than ever to study the evolving star-forming population over this redshift interval, corresponding to 12 Gyrs of cosmic time (i.e. most of the history of the Universe).
A picture of the dynamical evolution of star-forming galaxies has emerged, in which initially turbulent systems with large intrinsic velocity dispersions ($\sigma_{\textrm{int}}$) `settle' over time \citep[e.g.][]{Law2009,Simons2016}, leading to lower observed $\sigma_{\textrm{int}}$ values and therefore higher ratios of rotation velocity to velocity dispersion ($V_{\textrm{C}}/\sigma_{\textrm{int}}$) with decreasing redshift (e.g. \citealt{Wisnioski2015,Simons2017,Turner2017,Johnson2017}, although see \citealt{DiTeodoro2016} for a discussion of how $\sigma_{\textrm{int}}$ could be overstimated at intermediate redshifts). \\

The evolution of dynamical scaling relations in star-forming galaxies provides information about the partition of the total mass between dark matter, stars and gas, as well as shedding light on the stability of the galaxy disks. 
One important example is the stellar mass `Tully-Fisher' relation \citep{Tully1977,Bell2001} which connects the stellar mass within a galaxy to the rotation velocity, a tracer of the total dynamical mass.
A change in the slope of the relationship with increasing redshift indicates a stellar-mass dependent change to the connection between velocity and total galaxy mass.
A change in the normalisation indicates a redistribution of the total mass in the galaxy between visible and dark components on the scales traced by the observations.
It is mostly accepted that the number of galaxies observed and the data quality are too low to accurately constrain the slope of the relationship at ${\it z}>1$, and so evolution is assessed by fixing the slope to a reference value measured in the local Universe and monitoring shifts in the normalisation \citep[e.g.][]{Puech2008,Cresci2009,Miller2011,Miller2012,Tiley2016,Harrison2017,Straatman2017,Pelliccia2017,Ubler2017}.
However, there is no consensus throughout these studies on how the normalisation of the relationship changes over cosmic time.
There are several systematic effects throughout the analysis which can lead to diverging conclusions, such as the choice of local reference relationship and the sample-selection criteria \citep[e.g.][]{Tiley2016,Harrison2017}.
In this paper we investigate the evolution of the stellar mass Tully-Fisher relation from ${\it z}\sim4$ to ${\it z}=0$, and reconcile discrepant literature results over this range.

Furthermore, it has recently emerged that the rotation velocity may be an inadequate tracer of the total dynamical mass, especially at high redshift, due to the contribution of pressure support to the dynamical mass budget.
For example \cite{Kassin2007} show that the $S_{0.5} = \sqrt{0.5V_{\textrm{C}}^{2} + \sigma_{\textrm{g}}^{2}}$ parameter correlates more tightly with mass than the rotation velocities alone for galaxies over the redshift range $0 < {\it z} < 1.2$, where $\sigma_{\textrm{g}}$ is the integrated velocity dispersion of the galaxies.
\cite{Burkert2010} also show that the addition of a pressure term to the equation of hydrostatic equilibrium can reduce the observed rotation velocities of star-forming galaxies, prompting others to adopt a corrected rotation velocity which accounts for the contribution from pressure \citep[e.g.][]{Newman2013,Ubler2017}.
In \cite{Ubler2017}, a pressure-corrected velocity is explored in the context of the evolution of the stellar-mass Tully-Fisher relation out to $z\sim2.3$, concluding that it is necessary to account for pressure support to truly trace the evolution of dynamical mass with redshift.
These works have sparked a new debate on how best to explore the connection between dynamical and stellar mass over cosmic time.
In this study we expand such investigations out to $z\sim4$. \\

The KMOS-Deep Survey (KDS) \citep{Turner2017}, is a programme which aims to study the dynamical and chemical properties of $\simeq 80$ star-forming galaxies at $z\simeq3.5$.
A detailed study of the dynamical properties of a sample of morphologically isolated KDS galaxies revealed that only one third are dominated by ordered rotation \citep{Turner2017}, as a result of low rotation velocities and high velocity dispersions.
This is significantly lower than the $\sim80-100$ per cent of star-forming galaxies at ${\it z}\leq1$ (e.g. Fig.\,7 of \citealt{Turner2017}) observed to be rotation-dominated.
We concluded that, for the majority of ${\it z} = 3.5$ star-forming galaxies, random motions are prevalent throughout the galaxy disk and, as suggested above, it may be important to account for these when attempting to trace the total dynamical mass of the galaxies. 
In this paper we discuss the stellar-mass Tully-Fisher relation for the KDS galaxies and use a compilation of comparison samples in the literature spanning $0 < {\it z} < 4$ to assess the evolution of the relation.
We also consider the impact of pressure support, traced by velocity dispersions, to the dynamical properties of the KDS and comparison sample galaxies.

The structure of the paper is as follows.
In \cref{sec:sample_and_dr} we give a brief overview of the selection criteria, data reduction and extraction of physical properties for the KDS sample.
In \cref{sec:Tully-Fisher-relation} we present our study of the stellar-mass Tully-Fisher relation for the KDS galaxies, and using our compilation of comparison samples we discuss the dependence of the observed evolution of the relationship on sample-selection criteria. 
In \cref{sec:sigma_contribution} we explore the impact of accounting for pressure support, traced by the velocity dispersions, in the dynamical mass budget of the galaxies, and formulate an effective `total velocity'.
Using the comparison samples, we then study the evolutionary trends of the total-velocity versus stellar-mass relation.    
In \cref{sec:conclusion} we present our conclusions.
Throughout this work we assume a flat $\Lambda$CDM cosmology with (h, $\Omega_{m}$, $\Omega_{\Lambda}$) = (0.7, 0.3, 0.7). 

\section{SURVEY DESCRIPTION, SAMPLE SELECTION AND OBSERVATIONS}\label{sec:sample_and_dr}
The KDS is a survey of the gas kinematics and chemical compositions of 77 star-forming, ${\it z}\simeq3.5$ galaxies observed with KMOS.
Full details of the survey, sample selection, data reduction, dynamical modelling and kinematic parameter extraction can be found in \cite{Turner2017}.
However, below we present a brief overview of the survey, and of the KDS sample used throughout \cref{sec:Tully-Fisher-relation} and \cref{sec:sigma_contribution}.

\subsection{The KDS \& sample selection}\label{subsec:sample_selection}
The KMOS Deep Survey (KDS) is a guaranteed time programme focusing on the spatially-resolved properties of typical star-forming galaxies at ${\it z}\simeq3.5$.
These data have been collected to guide our understanding of early disk formation in terms of both the observed kinematics and the chemistry, as inferred through observations of the ionised gas emission lines.

The 77 KDS target galaxies have previous spectroscopic redshift measurement and were selected to fall in the redshift range $3.0 < {\it z} < 3.8$.
We include regions of both low and high galaxy density, spanning GOODS-S \citep[e.g.][]{Koekemoer2011,Grogin2011,Guo2013} and SSA22 \citep[e.g.][]{Steidel1998}.
These fields are covered by rest-frame ultraviolet to far-infrared high-resolution ancillary photometry, allowing us to infer galactic physical properties with SED fitting and to recover the morphological properties of the galaxies.
The measurement of stellar masses is described in detail in \cite{Turner2017}.
Briefly, we make use of the available multi-wavelength photometry to fit the galaxy SEDs following the procedure described in \cite{McLure2011}.
We use solar metallicity BC03 \citep{Bruzual2003} templates with a \cite{Chabrier2003} Initial Mass Function (IMF), account for dust attenuation using the Calzetti reddening law \citep{Calzetti2000} with dust attenuation allowed to vary freely over the range $0.0<A_{V}<4.0$, and include the effects of strong nebular emission lines according to the line ratios determined by \cite{Cullen2014}.
We focus on the stellar mass range $9.0 < \textrm{log} (M_{\star}/M_{\odot}) < 10.5$, (i.e. the range covered by local reference data, see Fig.\,\ref{fig:rom_tf_relation}), which leads to 4 KDS galaxies with $\textrm{log}  (M_{\star}/M_{\odot}) < 9.0$ being omitted from the subsequent analysis.
The median stellar mass of the galaxies in this range is log($M_{\star}/M_{\odot}) = 9.8$.
We have also verified using both the rest-frame UVJ colour space diagnostic and the star-formation rate versus stellar mass `main-sequence' plot that the KDS target galaxies fall in the loci defined by typical star-forming galaxies at these redshifts \citep{Turner2017}.

\subsection{KMOS observations and data reduction}\label{subsec:kmos_observations}
KMOS is a second-generation Integral-Field Spectrograph (IFS) mounted at the Nasmyth focal plane of Unit Telescope 1 (UT1) at the European Southern Observatory's Very Large Telescope (ESO/VLT).
The instrument has 24 moveable pickoff arms, each with an integrated IFU, which patrol a region 7.2$^{\prime}$ in diameter on the sky, providing considerable flexibility when selecting sources for a single pointing.
The light from each set of 8 IFUs is dispersed by a single spectrograph and recorded on a 2k$\times$2k Hawaii-2RG HgCdTe near-infrared detector, so that the instrument is comprised of three effectively independent modules.

The target galaxies were observed with KMOS in the {\it H} and {\it K}-bands during ESO observing periods P92-P96 using Guaranteed Time Observations (Programme IDs: 092.A-0399(A), 093.A-0122(A,B), 094.A-0214(A,B), 095.A-0680(A,B), 096.A-0315(A,B,C)) with excellent {\it K}-band seeing conditions ranging between $0.5-0.7^{\prime\prime}$.
At these redshifts, the H$\upbeta$,  [\ion{O}{iii}]$\lambda$4959 and  [\ion{O}{iii}]$\lambda$5007 emission lines are shifted into the {\it K}-band and the [\ion{O}{ii}]$\lambda$3727,3729 doublet is shifted into the {\it H}-band.
When used together, these features are rich in both kinematic and chemical information.
The on-source exposure times were in the range 7-10 hours, accumulated using a standard object-sky-object (OSO) nod-to-sky observation pattern, with 300s exposures and alternating $0.2^{\prime\prime}$/$0.1^{\prime\prime}$ dither pattern for increased spatial sampling around each of the target galaxies.
We also observed standard stars to allow for flux calibration of the data products, and the spatial locations of control stars were monitored throughout the observations to determine the shifts which must be applied to each exposure when creating the final object stacks \citep{Turner2017}. \\ 

Data reduction was carried out using the Software Package for Astronomical Reduction with KMOS, (SPARK; \citealt{Davies2013}), implemented using the ESO Recipe Execution Tool (ESOREX) \citep{Freudling2013}, with additional python scripts for non-standard methods (\citealt{Turner2017}).
Following the reduction of each object-sky pair, all exposures were stacked together to create a final datacube for each object, which was flux calibrated using the standard star observations. 
We attempted to make an integrated measurement of the [\ion{O}{iii}]$\lambda$5007 emission line in each cube (the highest S/N line), detecting emission in 62 (81 per cent) of the galaxies in the sample.

\subsection{Morphological and kinematic measurements}\label{subsec:measurements}

To accurately constrain the kinematics of the KDS sample, it was important to measure the galaxy sizes and inclination angles, which allowed us to extract rotation velocities at a known fiducial radius and correct the line-of-sight velocity component.  
We used {\scriptsize GALFIT} \citepalias{Peng2010_galfit} to constrain the half-light radii ($R_{1/2}$) and axis-ratios of the KDS galaxies by fitting exponential light profiles to the {\it Hubble Space Telescope} ({\it HST}) {\it WFC3 F160W} imaging.
In each of the stacked datacubes where an integrated [\ion{O}{iii}]$\lambda$5007 measurement was made, we extracted two-dimensional flux, velocity and velocity dispersion maps by fitting gaussian profiles to the individual spatial-pixels (spaxels) of the cube (see \citealt{Turner2017}).
We classified 47/62 galaxies with an integrated [\ion{O}{iii}]$\lambda$5007 measurement as spatially-resolved, with this being defined as those galaxies where the extent of the [\ion{O}{iii}]$\lambda$5007 flux map is more extended than the Point Spread Function (PSF) of the observations.

For the dynamical modelling, a series of mock datacubes were populated with [\ion{O}{iii}]$\lambda$5007 emission lines in an exponential spatial flux distribution and with a two-dimensional velocity field which followed the arctangent function.
These intrinsic models were convolved with the atmospheric PSF, measured from the observations, and fitted to the observed velocity field, which generated both observed and intrinsic model velocity fields and a map of the beam-smearing corrections to the observed velocity dispersions.
The rotation velocities were extracted from the best-fit intrinsic model at a fiducial radius of $2R_{1/2}$ (corresponding to $\simeq3.4R_{\textrm{d}}$, where $R_{\textrm{d}}$ is the exponential disk scale radius).
The velocity dispersion value for each galaxy was taken as the median of the intrinsic velocity dispersion maps, i.e. the observed map with the beam-smearing correction map subtracted linearly and the intrumental resolution map subtracted in quadrature \citep{Turner2017}.
In summary, the outcome of the dynamical modelling procedure was a measure of the intrinsic rotation velocity, $V_{\textrm{C}}$, and the intrinsic velocity dispersion, $\sigma_{\textrm{int}}$, for each galaxy in the spatially-resolved sample. \\

The high-resolution {\it HST} imaging allowed us to identify galaxies involved in probable late-stage merger events, for which the interpretation of velocity and velocity dispersion fields is complicated.
After identifying KDS targets which are spatially-resolved and removing the merger candidates, we are left with 29 galaxies in the specified mass range (see \cref{subsec:sample_selection}), which we refer to as the `isolated-field sample' throughout the remainder of this paper, and which constitute the KDS sample for the analysis described in the following sections.
The isolated-field sample is further subdivided into `rotation-dominated' with $V_{\textrm{C}}/\sigma_{\textrm{int}} > 1$ (13/29) and `dispersion-dominated' with $V_{\textrm{C}}/\sigma_{\textrm{int}} < 1$ (16/29) galaxies, as a simple method to distinguish between whether ordered rotation or random motions dominate the gas dynamics of the system.
The morphological and kinematic properties of the isolated-field sample are listed in \cite{Turner2017}.

\subsection{Comparison samples}\label{subsec:intro_comparison_samples}
Throughout this work we make use of 5 `local' (${\it z}=0$) and 16 `distant' (${\it z}\sim0.1-3$) comparison samples to establish the evolution of the stellar-mass Tully-Fisher relation.
A detailed description of each of these samples is provided in Appendix \ref{app:comparison_samples}, to which we refer the reader for more information.
In \cref{subsec:vc_m_evolution} and \cref{subsubsec:discussion_v_over_sigma_cuts} we use the comparison samples to explore how sample selection and the choice of ${\it z}=0$ reference sample impacts the conclusions surrounding the evolution of the stellar-mass Tully-Fisher relation.
All of these samples are analysed using the same methodology throughout \cref{sec:Tully-Fisher-relation}, to provide consistent results across a wide redshift range.
Briefly, we obtained the published rotation velocities, velocity dispersions and stellar masses of these star-forming galaxy samples spanning $0 < {\it z} < 3$, correcting the stellar masses to a \cite{Chabrier2003} IMF where required, and monitoring the methods used to measure rotation velocities and intrinsic velocity dispersions.
The samples were carefully-selected to contain typical star-forming galaxies with dynamics measured using similar approaches which seek to correct for the effects of beam-smearing.
In some cases, we do not consider the measured kinematic properties to be directly comparable to the other samples; we highlight these results using grey hollow symbols in the plots throughout \cref{sec:Tully-Fisher-relation} \& \cref{sec:sigma_contribution}.
We list the fit results in Table \ref{tab:fit_results}, and plot the data used for these fits in Figs \ref{fig:velocity_fits} and \ref{fig:v_tot_fits}.  

\section{The evolution of the stellar-mass Tully-Fisher relation}\label{sec:Tully-Fisher-relation}

The stellar-mass Tully-Fisher relation \citep[e.g.][]{Bell2001} connects the stellar mass within a galaxy to the rotation velocity, and hence, in the case of a disk galaxy, the total dynamical mass.
It is thus a powerful tracer of the stellar assembly history of galaxies. 
There is currently a debate in the literature surrounding both the extent of the evolution of this scaling relation and the interpretation of an observed evolution. 
In comparison with the stellar-mass Tully-Fisher relation zero-point derived using local spiral galaxies, some surveys report a varying degree of evolution over the range $0 < {\it z} < 2$ \citep[e.g.][]{Puech2008,Cresci2009,Straatman2017,Ubler2017} for massive, disky galaxies.
Others report very little evolution using samples of star-forming galaxies covering a wide range in stellar mass and showing significant kinematic and morphological diversity \citep[e.g.][]{Miller2011,Miller2012,Harrison2017}.
In the following sections we fit the stellar-mass Tully-Fisher relation to KDS isolated-field sample galaxies and study the evolution of the relationship out to $z \sim 4$.
We do this using the compilation of comparison samples described in \cref{subsec:intro_comparison_samples}, carefully exploring the impact of differing sample-selection criteria on the inferred evolution.
All of the galaxy samples are fitted in the rotation-velocity versus stellar-mass plane and so offsets from the local stellar-mass Tully-Fisher relation are quoted in terms of the velocity zero-point, which differs from the mass zero-points often quoted throughout the literature by a factor of $-1/\upalpha$, where $\upalpha$ is the slope of the relation defined in the following section\footnote{We have confirmed that we recover equivalent offsets when fitting the samples in the stellar-mass versus velocity plane.}.

\subsection{The stellar-mass Tully-Fisher relation for the KDS galaxies}\label{subsec:vc_m_relation}

\begin{figure*}
\centering \hspace{-1.13cm}
\includegraphics[width=\textwidth]{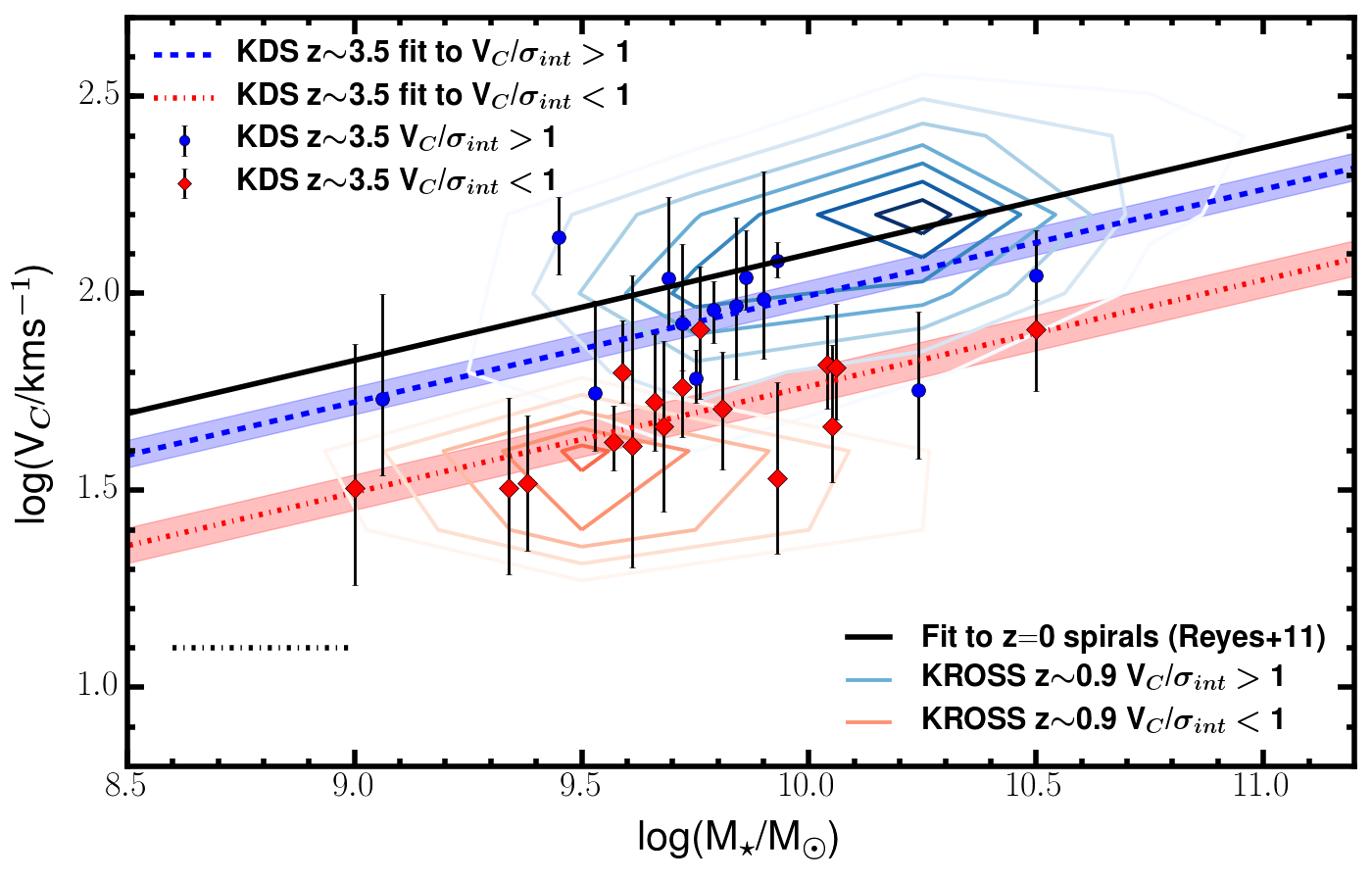}
\caption{Rotation velocity versus stellar-mass for the rotation-dominated (blue-circular) and dispersion-dominated (red-diamond) KDS isolated-field sample galaxies.
The black dot-dashed line in the lower left indicates the $\pm0.2$ dex error on the KDS stellar masses. 
    The solid black line shows the fit to local spiral galaxies from \protect\cite{Reyes2011} using the fitting the relation $\textrm{log}({\it V_{\textrm{C}}}) = \upbeta + \upalpha[\textrm{log}({\it M_{\star}}) - 10.1]$, which constitutes our local reference relation (see Appendix \ref{subsubsec:Reyes_2011}).
    The blue-dashed and red-dot-dashed lines show the fits to the rotation-dominated and dispersion-dominated KDS galaxies respectively, using the fixed slope $\upalpha=0.270$ recovered from the fit to the \protect\cite{Reyes2011} spirals.
    The shaded regions around these lines represent the $1-\sigma$ uncertainty on the fit.
    The blue and red contours show the density of rotation-dominated and dispersion-dominated galaxies from KROSS \protect\citep{Harrison2017}, both starting at 10 and increasing in increments of 10 and 3 respectively.
    The velocity zero-point offset for the rotation-dominated KDS galaxies is $\simeq-0.10$ dex below the local relation, with the dispersion-dominated galaxies $\simeq-0.35$ dex below.}
\label{fig:tf_relation}
\end{figure*}

In Fig.\,\ref{fig:tf_relation} we plot intrinsic rotation velocity against stellar mass for the rotation-dominated and dispersion-dominated KDS subsamples.
For comparison, we plot the density of $z\simeq0.9$ rotation-dominated and dispersion-dominated KROSS galaxies (see Appendix \ref{subsubsec:kross_comp}) in this plane using the blue and red contours respectively.
In order to place the KDS galaxies in context in the rotation-velocity versus stellar-mass plane, it is important to choose a reliable ${\it z}\simeq0$ comparison sample.
We define the local stellar-mass Tully-Fisher relation by fitting the spiral galaxy sample from \cite{Reyes2011}, which contains 189 galaxies covering a wide range in stellar mass.

To fit this sample we use the python package {\scriptsize LMFIT} \citep{Newville2014} which makes use of the Levenberg-Marquardt algorithm.
The relation $\textrm{log}({\it V_{\textrm{C}}}) = \upbeta + \upalpha[\textrm{log}({\it M_{\star}}) - 10.1]$ (as per e.g. \citealt{Reyes2011,Harrison2017}) is fitted to the \cite{Reyes2011} galaxies, with both the slope and intercept allowed to vary.
We carry out the fit 1000 times, perturbing each velocity value by a random number drawn from a gaussian distribution centred on the datapoint and with standard deviation given by the velocity error.
The stellar mass is perturbed in the same way, using a fixed error of 0.2 dex\footnote{0.2 dex is the typical uncertainty on the stellar masses recovered from SED fitting in the KDS isolated-field sample} on each of the stellar mass values.
We take the median of the resultant parameter distributions and recover the $1-\sigma$ error from the 16th and 84th percentiles to find $\upbeta_{{\it z}=0} = 2.127 \pm 0.010$ and $\upalpha_{{\it z}=0} = 0.270 \pm 0.016$, which we use as our reference local relation for all subsequent comparisons.
The normalisation and slope quoted in \cite{Reyes2011} are $\upbeta_{\textrm{Reyes}} = 2.142 \pm 0.004$ and $\upalpha_{\textrm{Reyes}} = 0.278 \pm 0.010$ respectively.
The slope we measure when re-fitting the data is consistent within the errors and the small difference between normalisations is the result of converting between the Kroupa IMF \citep{Kroupa2002} adopted in \cite{Reyes2011} and the Chabrier IMF used throughout this work (more details in Appendix \ref{subsubsec:Reyes_2011}). \\

Other surveys (\citealt{Tiley2016,Straatman2017,Harrison2017,Ubler2017}) have also made use of the \cite{Reyes2011} sample as a local reference, using the fit parameters quoted in \cite{Reyes2011} in order to study the evolution of the stellar-mass Tully-Fisher relation.
Several local samples exist, leading to several possible reference relations which differ from one another in terms of slope and normalisation.
As discussed in Appendix \ref{subsec:local_comparison}, the commonly adopted reference samples span $\simeq0.05$ dex in velocity normalisation, and so the choice of which to use has an impact on the conclusions drawn for the evolution of the stellar-mass Tully-Fisher relation \citep[e.g.][]{Straatman2017}.
For example, when defining the local relation using a fit to spiral galaxies from \cite{Romanowsky2012}, the reference velocity zero-point is $\simeq0.05$ dex higher.
For some samples, this uncertainty could account for a significant fraction of the observed evolution of the stellar-mass Tully-Fisher relation to high redshift (see Figs \ref{fig:rom_tf_relation}, \ref{fig:rom_slope_vel_evolution} and \ref{fig:rom_slope_vel_evolution}), which we discuss further in \cref{subsubsec:vtot_m_evolution}   .  \\

The rotation-dominated and dispersion-dominated galaxies in the KDS isolated-field sample are fitted following the same procedure as above, but with the slope held fixed to the local value of $\upalpha_{{\it z}=0} = 0.270$.
Fixing the slope allows us to focus on the evolution of the normalisation of the relation by applying a consistent functional form at each redshift.
We find $\upbeta_{\textrm{rot},{\it z}=3.5} = 2.026 \pm 0.036$ and $\upbeta_{\textrm{disp},{\it z}=3.5} = 1.790^{+0.042}_{-0.045}$ which are offset from the local relation by $\simeq-0.10$ dex and $\simeq-0.35$ dex in velocity zero-point respectively.
As mentioned above, it is common to quote stellar-mass zero-point offsets in units of log($M_{\star}/M_{\odot}$), which differ from the velocity zero-point offsets by a factor of $-1/\upalpha$.
The KDS velocity zero-point offsets correspond to $+0.33$ dex and $+1.22$ dex offsets in stellar mass for the rotation-dominated and dispersion-dominated galaxies respectively.

We also fit the rotation-dominated and dispersion-dominated KROSS \citep{Harrison2017} galaxies to find $\upbeta_{\textrm{rot},{\it z}=0.9} = 2.112 \pm 0.009$, $\upbeta_{\textrm{disp},{\it z}=0.9} = 1.549 \pm 0.013$.
In \cite{Harrison2017}, the rotation-dominated KROSS galaxies, defined as those with $V_{\textrm{C}}/\sigma_{\textrm{int}} > 1$, are fitted in the velocity versus stellar-mass plane with the slope allowed to vary, reporting fit parameter values $\upalpha_{\textrm{Harrison}} = 0.33 \pm 0.11$ and $\upbeta_{\textrm{Harrison}} = 2.12 \pm 0.04$.
The fixed slope we use and the normalisation we recover are both consistent within the errors with the results from \cite{Harrison2017}. \\

Fig.\,\ref{fig:tf_relation} shows that the rotation-dominated KDS galaxies have lower rotation velocities at fixed stellar mass than the local reference and intermediate redshift KROSS star-forming galaxies.
The KDS rotation-dominated velocity zero-point is offset in the opposite sense to other intermediate redshift studies of the stellar-mass Tully-Fisher relation, which either show no evolution \citep[e.g.][]{Miller2011,Miller2012,Epinat2012,Pelliccia2017,Harrison2017} or evolution of up to $+0.12$ dex in velocity zero-point \citep[e.g.][]{Cresci2009,Straatman2017,Ubler2017}.
The shifts towards higher velocities are usually interpreted as an increase in the ratio of dynamical to stellar mass with increasing redshift, as galaxies have yet to convert their gas mass into stars.
We show in \cref{sec:sigma_contribution} that the KDS offset is likely a consequence of the increasing significance of pressure support at high redshift \citep[e.g.][]{Kassin2012,Simons2017,Ubler2017}, leading observationally to an increase in velocity dispersions at the expense of rotation velocity \citep[e.g.][]{Burkert2010,Simons2017}. \\

To study the inferred evolution of the stellar-mass Tully-Fisher relation, we apply the same fitting analysis to our compilation of 16 distant star-forming galaxy comparison samples with median redshifts in the range $0 < {\it z} < 4$.
We also explore the link between sample-selection criteria and the inferred evolution of the stellar-mass Tully-Fisher relation in the following subsections.

\subsection{Evolution of the stellar-mass Tully-Fisher relation out to $\bmath{z\sim4}$}\label{subsec:vc_m_evolution}

\begin{figure*}
\centering 
\includegraphics[width=\textwidth]{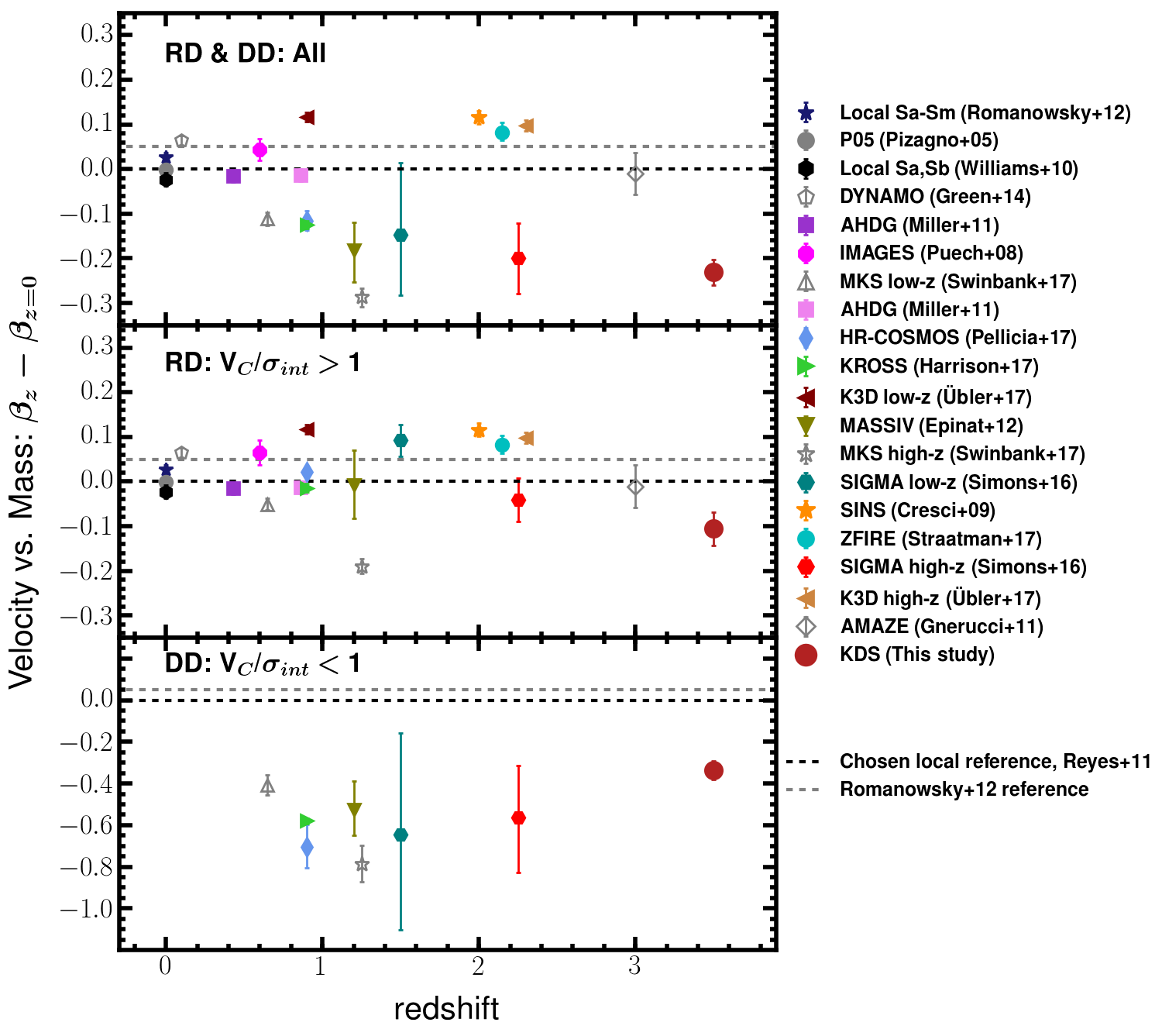}
    \caption{The velocity-normalisation offsets, $\Delta\upbeta$, from the local stellar-mass Tully-Fisher relation against redshift.
    In each panel the dashed-black horizontal line shows the \protect\cite{Reyes2011} reference velocity zero-point, and the dashed grey horizontal line shows the approximate $+0.05$ dex zero-point shift found when adopting a local reference relation recovered from fitting galaxies from \protect\cite{Romanowsky2012} (see the text and Appendix \ref{subsec:local_comparison}).
    The studies from which the comparison sample data have been taken are shown in the legend.
    Generally the median stellar mass of subsamples from the same parent sample vary between the three panels, with dispersion-dominated galaxies having somewhat lower values.
    We list these values, as well as the Tully-Fisher offsets, in Table \ref{tab:fit_results}.  
    {\it Top:} The fits to the full galaxy samples (including rotation-dominated and dispersion-dominated galaxies; see Appendix \ref{app:comparison_sample_fits}).  
    {\it Middle:} The same as the top panel but with only the rotation-dominated galaxies.
    The majority of points fall between $-0.10$ dex to $+0.10$ dex in velocity zero-point offset from the local relation (i.e. between $\simeq \pm0.35$ dex offset in stellar-mass zero-point).
    The KDS galaxies are offset to lower velocities at fixed stellar mass, suggesting that rotation velocity alone is not a good tracer of dynamical mass at ${\it z}\sim3.5$ (see \cref{sec:sigma_contribution}).     
    {\it Bottom:} The same as the top and middle panels, but with only the dispersion-dominated galaxies.
    In each survey where dispersion-dominated galaxies are available, the measured zero-points from the fits are significantly lower than the local zero-point.}
    \label{fig:velocity_evolution}
\end{figure*}

The fitting analysis described above is applied to the 5 local and 16 distant star-forming galaxy comparison samples described throughout Appendix \ref{app:comparison_samples} (and see \cref{subsec:intro_comparison_samples}).
For each of these comparison samples we make use of tabulated velocities, velocity dispersions (when available) and stellar masses and have converted all stellar mass measurements to a common \cite{Chabrier2003} IMF.
Using this information, we create rotation-dominated ($V_{\textrm{C}}/\sigma_{\textrm{int}} > 1$) and dispersion-dominated ($V_{\textrm{C}}/\sigma_{\textrm{int}} < 1$) subsamples in each of the comparison samples where this is possible (some of the samples by design contain only rotation-dominated galaxies).
The $V_{\textrm{C}}/\sigma_{\textrm{int}} > 1$ criteria is one way to pick out `disk' galaxies, i.e. those where the rotational motions exceed the random motions.
We explore how defining disk galaxies in different ways, e.g. with a stricter $V_{\textrm{C}}/\sigma_{\textrm{int}}$ cut, is connected to the observed stellar-mass Tully-Fisher evolution in \cref{subsubsec:discussion_v_over_sigma_cuts}.  
We fit the stellar-mass Tully-Fisher relation following the procedure described throughout \cref{sec:Tully-Fisher-relation}, with slope fixed to the local reference value of $\upalpha_{{\it z}=0}=0.270$, to the following samples: 
\begin{enumerate}[label=(\roman*),align=left]
\item All galaxies (RD \& DD: All)
\item Rotation-dominated only (RD: $V_{\textrm{C}}/\sigma_{\textrm{int}} > 1$)
\item Dispersion-dominated only (DD: $V_{\textrm{C}}/\sigma_{\textrm{int}} < 1$)
\end{enumerate}
The corresponding fits to the data are shown in Appendix \ref{app:comparison_sample_fits}.  
In Fig.\,\ref{fig:velocity_evolution} we compare the velocity zero-points recovered from these fits with the local velocity zero-point.
In the three panels of Fig.\,\ref{fig:velocity_evolution} the y-axis shows the difference between the fitted velocity zero-point at the redshift of the comparison sample and the local zero-point. 
Hollow, grey symbols indicate the samples for which direct comparisons to the other results are complicated due to differences in the measurement of kinematic properties (e.g. no beam-smearing corrections applied) as explained throughout Appendix \ref{subsec:distant_comparison}.
Positive values indicate that the galaxies have higher velocities at fixed stellar mass relative to the local sample, which could be interpreted as a higher ratio of dynamical to stellar mass.
Negative values indicate the opposite, however we will discuss the limitations of directly comparing dynamical mass, inferred using observed rotation velocities alone, and stellar mass in \cref{sec:sigma_contribution}. \\

The top panel of Fig.\,\ref{fig:velocity_evolution} shows the fits to the full samples.
By virtue of these samples containing both rotation-dominated and dispersion-dominated galaxies, which sit in discrepant locations in the velocity versus stellar-mass plane (see Figs \ref{fig:tf_relation} \& \ref{fig:velocity_fits}), the velocity zero-points recovered from these fits are biased low and have large uncertainties.
The middle panel of Fig.\,\ref{fig:velocity_evolution} shows the results of fitting the rotation-dominated galaxies in the comparison samples.
Note that some of the samples only contain rotation-dominated galaxies and so the point locations are unchanged between the top and middle panels.
The 12 `reliable' (filled, colour symbols in Fig.\,\ref{fig:velocity_evolution}) distant comparison samples have offsets bounded by shifts of roughly $-0.10$ dex to $+0.10$ dex in velocity zero-point from the local relation (i.e. between $\pm0.35$ dex offset in stellar-mass zero-points).
The middle panel of Fig.\,\ref{fig:velocity_evolution} does not contain any information about sample-selection effects, e.g. the correlation between the median $V_{\textrm{C}}/\sigma_{\textrm{int}}$ of each of the rotation-dominated comparison samples and the observed Tully-Fisher offset \citep[e.g.][]{Tiley2016}, or the median stellar mass of the samples.
When viewed in isolation, plots such as these provide limited insight into the evolution of scaling relations, since sample selection is a dominant caveat.
We show this explicitly in \cref{subsubsec:parent_fraction} and \cref{subsubsec:v_over_sigma_sample_selection}. \\

The bottom panel of Fig.\,\ref{fig:velocity_evolution} shows the offsets from the local stellar-mass Tully-Fisher relation for the dispersion-dominated subsamples, which are substantially below the local reference relation with velocity zero-point offsets spanning $-0.4$ to $-0.8$ dex.
The correlation between rotation velocity and stellar mass is less apparent in these subsamples, leading to large uncertainties in the recovered velocity zero-points.
It is not physically motivated to look at dispersion-dominated galaxies in the context of the stellar-mass Tully-Fisher relation, however we include the top and bottom panels here for comparison to the corresponding panels in Fig.\,\ref{fig:vtot_evolution}, which explores the evolution of the `total-velocity' (including a contribution from velocity dispersion in supporting the total mass of the systems, see \cref{subsec:vtot_m_relation}) versus stellar-mass relation.
Furthermore, the difference in offsets between the rotation-dominated and dispersion-dominated systems highlights that the galaxy sample used (in terms of the $V_{\textrm{C}}/\sigma_{\textrm{int}}$ values) is critical in determining the inferred evolution of the stellar-mass Tully-Fisher relation.
We explore this in detail in the following subsection.

\subsubsection{The importance of sample selection in the observed evolution of the stellar-mass Tully-Fisher relation}\label{subsubsec:discussion_v_over_sigma_cuts}

The aim of this subsection is predominantly to shed light on the discrepant literature results describing the evolution of the stellar-mass Tully-Fisher relation beyond the local Universe.
The carefully-selected and consistently-treated comparison samples used in this work allow the evolution to be studied across both a wide redshift range, corresponding to 12 Gyrs of cosmic time, and a wide range of galaxy properties (see Appendix \ref{subsec:distant_comparison}).
Samples are often constructed for Tully-Fisher analysis by imposing selection criteria which aim to identify star-forming galaxies that are most `disky', i.e. most kinematically evolved, and hence most representative of the spiral galaxies used to construct the local Tully-Fisher samples \citep[e.g.][]{Cresci2009,Ubler2017}.
However, at high redshift these sample-selection cuts may exclude the majority of the parent sample due to, for example, the decline in the ratio of $V_{\textrm{C}}/\sigma_{\textrm{int}}$ with increasing redshift \citep[e.g.][]{Wisnioski2015,Turner2017,Johnson2017}.
Consequently, samples resulting from strict selection criteria may not be representative of the population of typical evolving-disk galaxies at the corresponding redshift.
The key point is that the analysis of star-forming samples that are selected to be increasingly disky will lead to different conclusions than the analysis of parent population representative samples, due to differences in the kinematic properties of the two sample types at fixed stellar mass. \\

In the following subsections we show the effect of sample selection by exploring the dependence of the rotation-dominated ($V_{\textrm{C}}/\sigma_{\textrm{int}} > 1$ samples) stellar-mass Tully-Fisher offsets presented in the middle panel of Fig.\,\ref{fig:velocity_evolution} on two diagnostics of sample-selection criteria: (1) The fraction of the parent samples used in the Tully-Fisher analysis with respect to the empirically-determined rotation-dominated fraction (\cref{subsubsec:parent_fraction}); (2) The median $V_{\textrm{C}}/\sigma_{\textrm{int}}$ of the comparison samples with respect to an equilibrium model prediction for this quantity (\cref{subsubsec:v_over_sigma_sample_selection}).
We define values for these diagnostics for the typical evolving-disk population and study the link between comparison sample Tully-Fisher offsets and the departure from these values.

\subsubsection{Parent sample fraction and Tully-Fisher offsets}\label{subsubsec:parent_fraction}

\begin{figure*}
    \centering \hspace{-1.3cm}
    \begin{subfigure}[h!]{0.5\textwidth}
        \centering
        \includegraphics[height=3.8in]{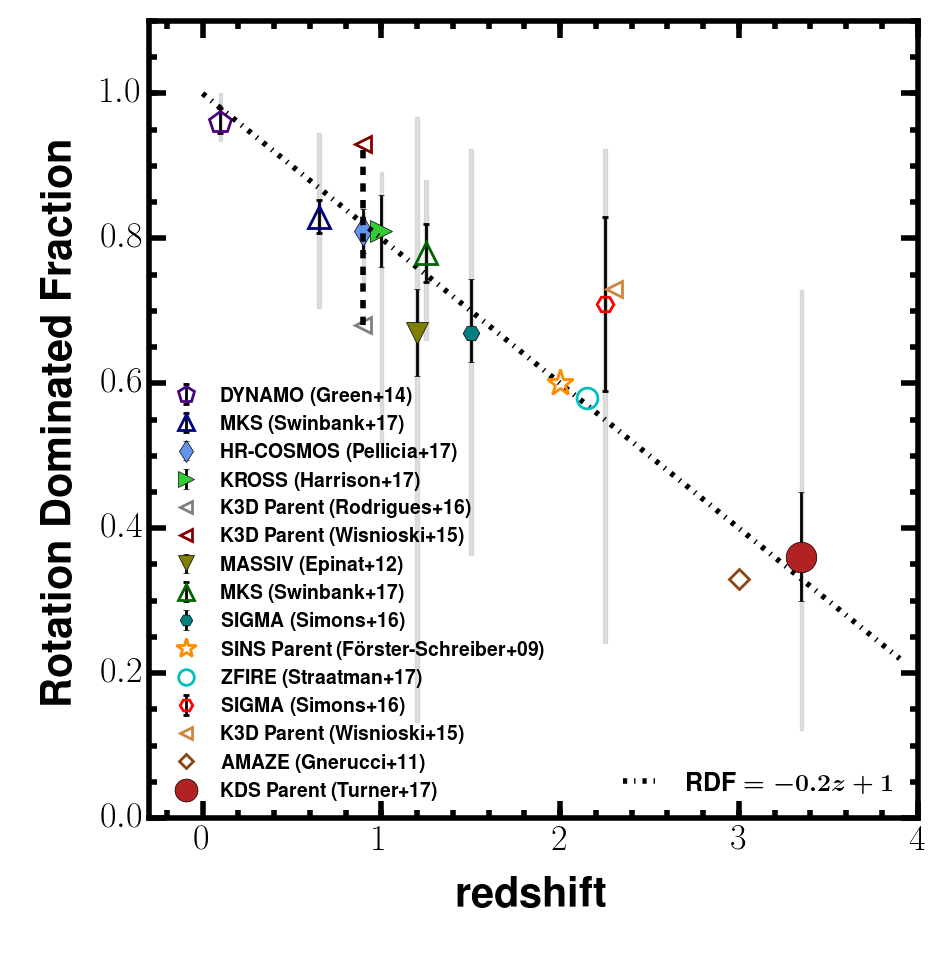}
    \end{subfigure} \hspace{0.4cm}
    \begin{subfigure}[h!]{0.5\textwidth}
        \centering
        \includegraphics[height=3.8in]{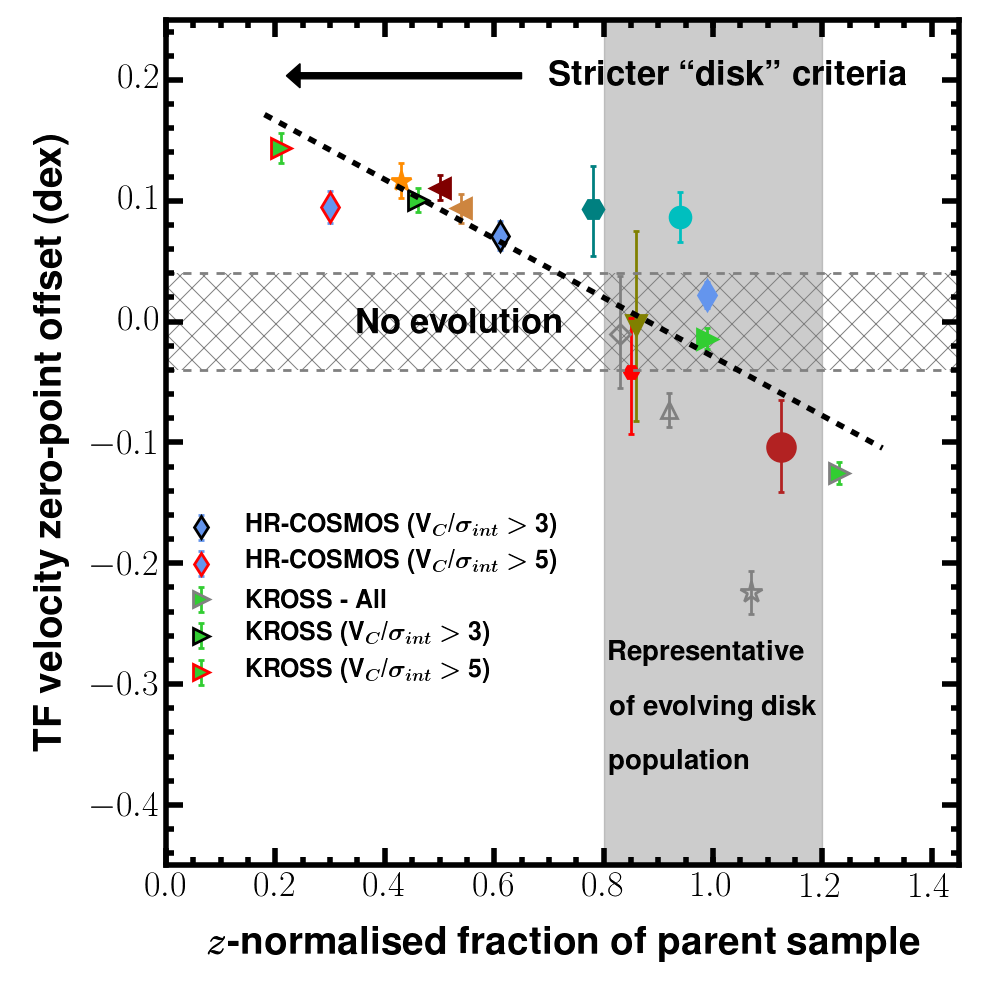}
    \end{subfigure}
    \caption{{\it Left:} The rotation-dominated fraction of star-forming galaxy samples is plotted as a function of redshift (adapted from \protect\citealt{Turner2017}), which evolves following the empirical relation RDF $= -0.2{\it z} + 1$ (dot-dashed track).
    {\it Right:} The velocity zero-point offsets from the local Tully-Fisher relation, from fitting the rotation-dominated sources (Fig.\,\ref{fig:velocity_evolution}, middle), against the fraction of the parent sample used to fit the relation.
    This fraction is redshift-normalised by dividing it by the representative rotation-dominated fraction at that redshift (dot-dashed track in the left panel).
    The studies from which the data have been collected are shown in the legend of Fig.\,\ref{fig:velocity_evolution}, and the symbols corresponding to the new subsamples created by varying the $V_{\textrm{C}}/\sigma_{\textrm{int}}$ cuts (see text) are indicated in the legend of this figure. 
    Samples in which the fraction of the parent sample used in the Tully-Fisher analysis is within $\sim20$ per cent of the representative rotation-dominated fraction are defined as `representative of evolving-disk population' (grey shaded region and see \cref{subsubsec:parent_fraction}). The black dashed line shows a linear, error-weighted fit to the data points with parameters $y = -0.24x + 0.22$. A clear trend emerges: when the fraction of the parent sample used to fit the Tully-Fisher relation is close to the representative rotation-dominated fraction at that redshift - no offsets are typically found. In contrast, when a small fraction of the parent population is used, due to increasingly-strict `disk' criteria, large, positive offsets are found.}
    \label{fig:parent_fraction_offsets}
\end{figure*}

In the following, we define the parent sample of each of the comparison samples as the number of galaxies which have been observed spectroscopically and in which the target emission line has been detected.
The parent samples are discussed explicitly in Appendix \ref{subsec:distant_comparison}.
The parent fraction is defined for each of the comparison samples as the ratio of the number of galaxies used for the stellar-mass Tully-Fisher analysis in \cref{subsec:vc_m_relation} (i.e. middle panel of Fig.\,\ref{fig:velocity_evolution}) to the number of galaxies in the parent sample, as defined above.
As a result of the decline in rotation-dominated galaxies with redshift \citep[e.g.][]{Stott2016,Turner2017}, a smaller fraction of the parent samples are considered for fitting in the rotation-velocity versus stellar-mass plane at  higher redshift.
To account for this we normalise each parent fraction using an empirically defined relation between the observed rotation-dominated fractions and redshift.
This relation is defined as RDF $= -0.2{\it z} + 1$, and is plotted in the left panel of Fig.\,\ref{fig:parent_fraction_offsets} (adapted from \citealt{Turner2017}), which also shows the rotation-dominated fraction of the parent samples against redshift.
The normalised parent fraction is defined as the ratio of the parent fraction to the empirical rotation-dominated fraction, measured at the median redshift of the samples.
We define the evolving-disk population at each redshift as all galaxies with $V_{\textrm{C}}/\sigma_{\textrm{int}} > 1$, so that the fraction of galaxies that would be used in a rotation-dominated Tully-Fisher analysis is equal to the rotation-dominated fraction, following the above empirical relation.
The normalised parent sample is small when the Tully-Fisher analysis is being applied to rarer objects which constitute only a small fraction of the full sample, and hence are not representative of the typical evolving-disk population at that epoch (i.e. moving towards smaller fractions reflects the application of increasingly-strict disk criteria). \\

In the right panel of Fig.\,\ref{fig:parent_fraction_offsets} we plot the offsets from the local Tully-Fisher relation for each comparison sample, only containing rotation-dominated ($V_{\textrm{C}}/\sigma_{\textrm{int}} > 1$) sources, against normalised parent fraction.
Using the KROSS data and the HR-COSMOS data we perform an additional sample cut of $V_{\textrm{C}}/\sigma_{\textrm{int}} > 3$ and $V_{\textrm{C}}/\sigma_{\textrm{int}} > 5$ to further explore the impact of selection criteria.
Using these subsamples we re-fit the stellar-mass Tully-Fisher relation, determining the new offsets from the local relation and the new normalised parent fractions. 
The symbols representing the $V_{\textrm{C}}/\sigma_{\textrm{int}} > 3$ and $V_{\textrm{C}}/\sigma_{\textrm{int}} > 5$ cuts are given black and red outlines respectively. \\

We highlight on Fig.\,\ref{fig:parent_fraction_offsets}, using a vertical shaded region, the location of samples which are representative of the evolving-disk population (based on the above definition).
The grey regions are lower and upper bounds (corresponding to roughly $\pm0.1$ dex, or $\sim20$ per cent tolerance) on whether the normalised parent fraction is in agreement with the expected location of the evolving-disk population (as defined above). 
We also suggest a region of no stellar-mass Tully-Fisher evolution with the horizontal hatching between $\pm 0.04$ dex\footnote{Corresponding to $\pm 0.15$ dex in stellar-mass zero-point} in velocity zero-point offset (roughly equivalent to the KDS rotation-dominated offset error).
The black arrow indicates the direction of increasingly-strict disk criteria, applied to isolate `disky' galaxies that are the closest match kinematically to the star-forming systems observed locally. \\

Clearly the samples with selection criteria designed to pick out the most disky galaxies, i.e. those with small normalised parent fractions, are those which show evolution towards higher rotation velocities at fixed stellar mass.
The most extreme example of this is for the KROSS sample with a cut of $V_{\textrm{C}}/\sigma_{\textrm{int}} > 5$, showing significant evolution towards higher rotation velocities at fixed stellar mass in comparison to the local stellar-mass Tully-Fisher relation.
Conversely, the representative samples tend not to show evolution in the Tully-Fisher relation and the high-redshift KDS galaxies appear to show evolution towards lower rotation velocities at fixed stellar mass, as a result of the decline in rotation velocities and the increased contribution of velocity dispersions to the dynamics of these systems (see \cref{sec:sigma_contribution}).
The KROSS ALL sample has a normalised parent fraction value which is greater than $1.2$, which is a consequence of the sample containing both rotation-dominated and dispersion-dominated galaxies.

We highlight the correlation between Tully-Fisher velocity offset and parent fraction by fitting a linear, error-weighted function to the datapoints in the right panel of Fig.\,\ref{fig:parent_fraction_offsets} (black dashed line) recovering the relation $y = -0.24x + 0.22$.
In \cref{subsubsec:offsets_interpretation} we interpret this correlation further.

\subsubsection{Median $V_{\textrm{C}}/\sigma_{\textrm{int}}$ and Tully-Fisher offsets}\label{subsubsec:v_over_sigma_sample_selection}

\begin{figure*}
    \centering \hspace{-1.65cm}
    \begin{subfigure}[h!]{0.5\textwidth}
        \centering
        \includegraphics[height=3.85in]{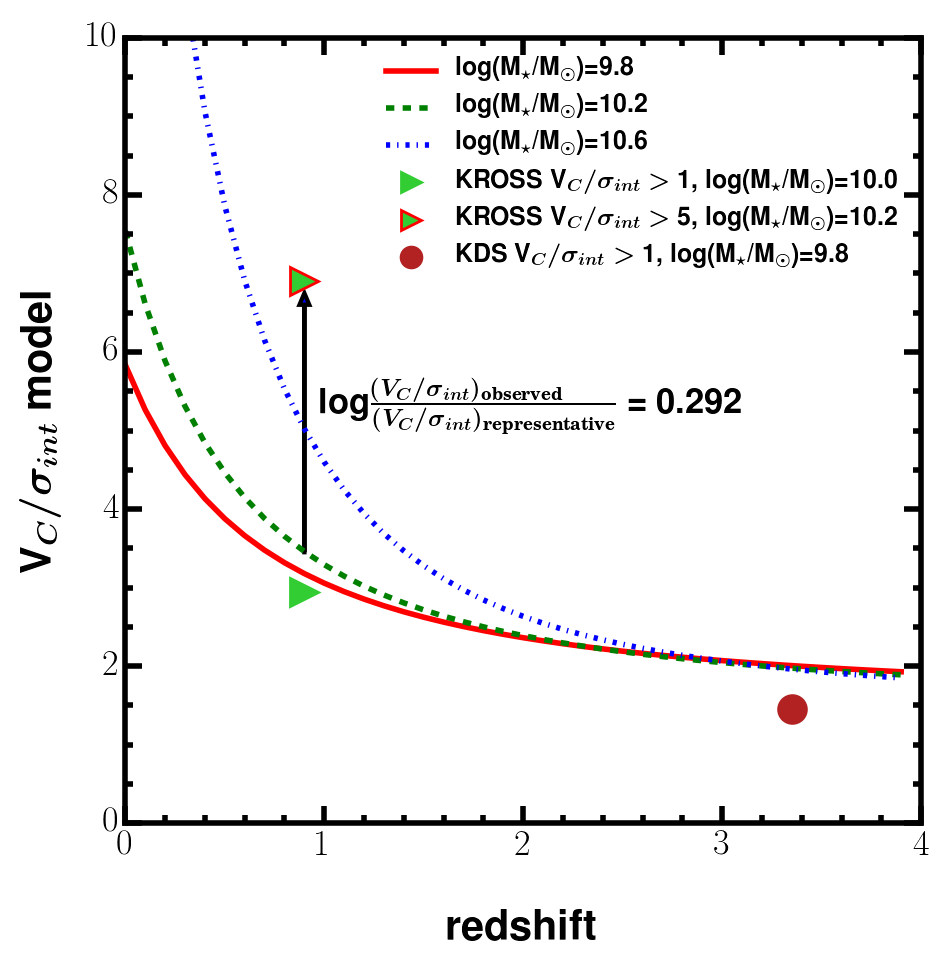}
    \end{subfigure} \hspace{0.4cm}
    \begin{subfigure}[h!]{0.5\textwidth}
        \centering
        \includegraphics[height=3.85in]{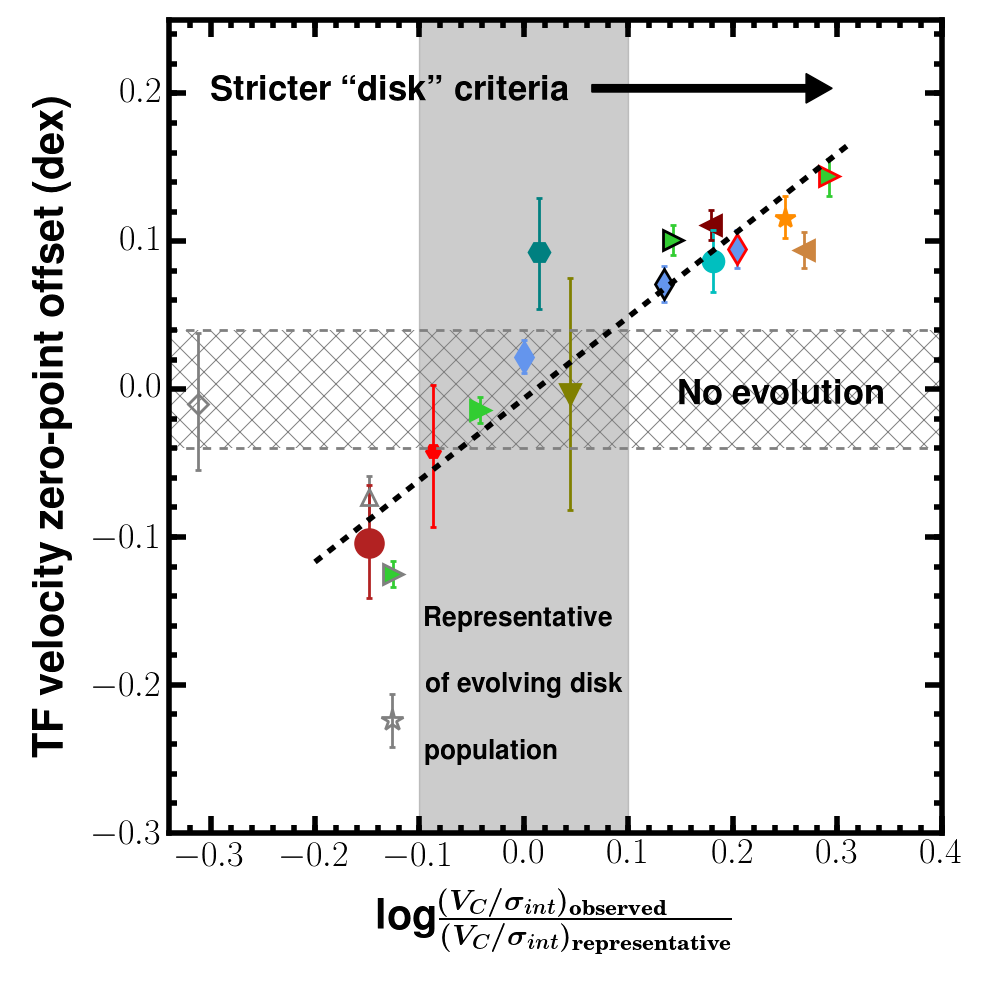}
    \end{subfigure}
    \caption{{\it Left:} $V_{\textrm{C}}/\sigma_{\textrm{int}}$ predictions computed using Equation \ref{eq:v_over_sigma_pred}, which is a function of the median stellar mass and redshift of a sample of galaxies.
    Three example curves which show the dependence of the predicted $V_{\textrm{C}}/\sigma_{\textrm{int}}$ values on mass and redshift are shown.
    For each comparison sample we compute a $\Delta \textrm{log}(V_{\textrm{C}}/\sigma_{\textrm{int}})$ value (Equation \ref{eq:log_ratio}) by comparing the median observed $V_{\textrm{C}}/\sigma_{\textrm{int}}$ value with the prediction from the model.
    We plot the KDS $V_{\textrm{C}}/\sigma_{\textrm{int}} > 1$, the KROSS $V_{\textrm{C}}/\sigma_{\textrm{int}} > 1$ and the KROSS $V_{\textrm{C}}/\sigma_{\textrm{int}} > 5$ datapoints to indicate the position of the samples relative to the model curves.    
    {\it Right:} The velocity zero-point Tully-Fisher offsets from fitting the rotation-dominated sources (Fig.\,\ref{fig:velocity_evolution}, middle) versus $\Delta \textrm{log}(V_{\textrm{C}}/\sigma_{\textrm{int}})$.
    The studies from which the data have been collected are shown in the legend of Fig.\,\ref{fig:velocity_evolution} and the symbols with coloured outlines have the same definition as the legend of Fig.\,\ref{fig:parent_fraction_offsets}.
    The black-dashed line shows the linear, error-weighted fit to the data, which has equation $y = 0.55x - 0.01$.
    The application of increasingly-strict disk criteria produces samples with median $V_{\textrm{C}}/\sigma_{\textrm{int}}$ values larger than the model predictions, and leads to increasingly-large positive velocity zero-point  offsets from the local Tully-Fisher relation.}
    \label{fig:offsets}
\end{figure*}

The median $V_{\textrm{C}}/\sigma_{\textrm{int}}$ values of the samples offers a second way to probe sample-selection criteria.
It is necessary to account for the cosmic decline and mass dependence of $V_{\textrm{C}}/\sigma_{\textrm{int}}$ \citep[e.g.][]{Wisnioski2015,Simons2017,Turner2017} in order to make direct comparisons between the median $V_{\textrm{C}}/\sigma_{\textrm{int}}$ values of samples at different redshifts. 
One way to do this is to use the simple equilibrium model proposed in \cite{Wisnioski2015}, which provides a prediction for $V_{\textrm{C}}/\sigma_{\textrm{int}}$ as a function of both mass and redshift, summarised by Equation \ref{eq:v_over_sigma_pred}:

\begin{equation}\label{eq:v_over_sigma_pred}
  \frac{V_{\textrm{C}}}{\sigma_{\textrm{int}}}(z,M_{\star}) = \frac{a}{Q_{\textrm{crit}}f_{\textrm{gas}}(z,M_{\star})}
\end{equation}

where ${\it a} = \sqrt{2}$ and $Q_{\textrm{crit}}=1.0$ for a marginally stable gas disk.
The redshift and mass dependencies are encoded in the gas fraction, the functional form of which is provided in \cite{Wisnioski2015} Equations 3-6.
Empirically, the model curves match the observed $V_{\textrm{C}}/\sigma_{\textrm{int}}$ values in the parent samples (see Fig.\,11 of \citealt{Wisnioski2015}).
Therefore, irrespective of the assumptions in the model, it provides a useful description of the evolving typical $V_{\textrm{C}}/\sigma_{\textrm{int}}$ value for star-forming galaxies as a function of redshift. 
Using the model curves we can compute a fiducial $V_{\textrm{C}}/\sigma_{\textrm{int}}$ for the comparison samples of known median stellar mass and redshift, representative of a population of typical star-forming galaxies with those properties, and determine the difference between this and the observed $V_{\textrm{C}}/\sigma_{\textrm{int}}$ for the same sample.
We use the model to define the evolving-disk population, by specifying that the median $V_{\textrm{C}}/\sigma_{\textrm{int}}$ of this population, at a given mass and redshift, is given by the model curves.  \\

The left panel of Fig.\,\ref{fig:offsets} demonstrates the redshift evolution of $V_{\textrm{C}}/\sigma_{\textrm{int}}$ for three different median stellar masses.
As an example, we show the location of the KDS $V_{\textrm{C}}/\sigma_{\textrm{int}} > 1$, the KROSS $V_{\textrm{C}}/\sigma_{\textrm{int}} > 1$ and the KROSS $V_{\textrm{C}}/\sigma_{\textrm{int}} > 5$ samples. 
We define the departure of the observed median $V_{\textrm{C}}/\sigma_{\textrm{int}}$ from the model prediction using the following log ratio:

\begin{equation}
    \Delta \textrm{log}(V_{\textrm{C}}/\sigma_{\textrm{int}}) = \textrm{log}\frac{\mkern-36mu (V_{\textrm{C}}/\sigma_{\textrm{int}})_{\textrm{observed}}}{(V_{\textrm{C}}/\sigma_{\textrm{int}})_{\textrm{representative}}}
    \label{eq:log_ratio}
\end{equation}

where $(V_{\textrm{C}}/\sigma_{\textrm{int}})_{\textrm{observed}}$ is the median observed ratio and $(V_{\textrm{C}}/\sigma_{\textrm{int}})_{\textrm{representative}}$ is the model prediction at the median stellar mass and redshift of the sample.
The black arrow in the left panel of Fig.\,\ref{fig:offsets} shows the magnitude of this ratio for the KROSS $V_{\textrm{C}}/\sigma_{\textrm{int}} > 5$ sample.
In the right panel of Fig.\,\ref{fig:offsets} we plot the stellar-mass Tully-Fisher offsets for each comparison sample against their associated $\Delta \textrm{log}(V_{\textrm{C}}/\sigma_{\textrm{int}})$ from Equation \ref{eq:log_ratio}.
The grey-shaded region is again an indication of whether the comparison samples are in agreement with the expectation for the evolving-disk population, with a tolerance of $\pm0.1$ dex.
Both the KDS and KROSS ALL samples have median $V_{\textrm{C}}/\sigma_{\textrm{int}}$ values lower than the representative region.
This is due to a combination of high velocity dispersions and low rotation velocities, which may no longer serve as a sufficient probe of the true dynamical mass (see \cref{sec:sigma_contribution}). 
We plot the same hatched, no-evolution region as in the right panel of Fig.\,\ref{fig:parent_fraction_offsets} and show the direction of increasingly-strict sample-selection criteria with the black arrow. \\

There is a clear relationship, which appears to hold throughout the comparison samples, between the stellar-mass Tully-Fisher offsets and $\Delta \textrm{log}(V_{\textrm{C}}/\sigma_{\textrm{int}})$, again with the most extreme example being the KROSS sample with a cut of $V_{\textrm{C}}/\sigma_{\textrm{int}} > 5$. 
Larger Tully-Fisher velocity zero-point offsets are observed for samples where the observed median $V_{\textrm{C}}/\sigma_{\textrm{int}}$ becomes increasingly larger than the corresponding model prediction, which we highlight using a linear, error-weighted fit to the datapoints in the right panel of Fig.\,\ref{fig:offsets} (black dashed line), which has the best-fit relation $y = 0.55x - 0.01$.
The representative samples cluster around zero Tully-Fisher offset, suggesting that the model $V_{\textrm{C}}/\sigma_{\textrm{int}}$ curves define a reference, non-evolving Tully-Fisher relation.
If an observed sample of star-forming galaxies at a particular median redshift has median $V_{\textrm{C}}/\sigma_{\textrm{int}}$ consistent with the model prediction, the standard Tully-Fisher relation fitted to those data will be in agreement with the local relation.
However it is possible to apply stricter kinematic criteria, such as the rotation-dominated galaxies being characterised by a higher $V_{\textrm{C}}/\sigma_{\textrm{int}}$ cut, which brings the median $V_{\textrm{C}}/\sigma_{\textrm{int}}$ of the new rotation-dominated sample higher at fixed stellar mass.
The response in the velocity versus stellar-mass plane is an evolution of the zero-point of the stellar-mass Tully-Fisher relation towards higher rotation velocities at fixed stellar mass in comparison to the local relation.
This is most clearly seen for the KROSS and HR-COSMOS subsamples in the right panel of Fig.\,\ref{fig:offsets}, to which several different $V_{\textrm{C}}/\sigma_{\textrm{int}}$ cuts have been applied (and see also \citealt{Tiley2016}). \\

The $V_{\textrm{C}}/\sigma_{\textrm{int}} > 1.0$ cut used to distinguish between rotation-dominated and dispersion-dominated galaxies is arbitrary and it is crucial to bear in mind that the evolution of the Tully-Fisher relation is dependent on where this boundary is placed.
Also in Tiley et al. ({\it in prep.}) the authors show that the combination of lower data quality at intermediate and high redshift, and attempts to apply corrections in order to recover intrinsic properties, results in sources scattering in and out of the $V_{\textrm{C}}/\sigma_{\textrm{int}} > 1$ bin.
This suggests that, at high redshift, the dynamical state of sources above and below this threshold can be ambiguous, due to the difficulties associated with accurately extracting kinematic properties. 
The interpretation of the relationship between $\Delta \textrm{log}(V_{\textrm{C}}/\sigma_{\textrm{int}})$ and Tully-Fisher offset is discussed further in the following subsection.

\subsubsection{Interpretation}\label{subsubsec:offsets_interpretation}

Figs \ref{fig:parent_fraction_offsets} and \ref{fig:offsets}, and the associated discussions above, describe the increased kinematic diversity of the evolving galaxy disks at high redshift.
At a particular stellar mass there exists a range of $V_{\textrm{C}}/\sigma_{\textrm{int}}$ values, which we interpret here as being an indicator of the kinematic maturity of the galaxy disks.
Fitting the stellar-mass Tully-Fisher relation to the highest $V_{\textrm{C}}/\sigma_{\textrm{int}}$ cloud leads to inferred evolution from the local relation.
Assuming that these galaxies do not have velocities which are biased high by large inclination corrections, fits to these subsamples answer the question: `what happens in the rotation-velocity versus stellar-mass plane to star-forming galaxies that are most like those disks we observe locally?'.
Furthermore, the high $V_{\textrm{C}}/\sigma_{\textrm{int}}$ samples have the advantage that rotation velocity is a better tracer of the dynamical mass.
However, with increasing redshift, these sources become increasingly rare and progressively less representative of the underlying evolving-disk population. \\ 

The results of this paper indicate that for high $V_{\textrm{C}}/\sigma_{\textrm{int}}$ samples we find Tully-Fisher velocity zero-point evolution of $+0.08$ to $+0.12$ dex (i.e. stellar-mass zero-point evolution of $-0.30$ to $-0.45$ dex) at $z \geq 1$, in agreement with previous work studying galaxies in this regime.
This evolution is consistent with a picture in which star-forming galaxies at high redshift have similar dynamical mass, but with higher gas fractions, and have simply converted less gas into stars \citep[e.g.][]{Puech2008,Ubler2017}.
Variation in the offsets is determined by sample-selection criteria, which alter the magnitude of the inferred Tully-Fisher evolution in a prescribed way as explored above.
Other factors which alter the inferred evolution are the choice of local reference relation (see Figs \ref{fig:rom_tf_relation} and \ref{fig:rom_slope_vel_evolution} in Appendix \ref{app:comparison_samples}) and the methods followed to extract kinematic parameters (which we do not attempt to correct for in this work). \\

The inferred evolution from the highest $V_{\textrm{C}}/\sigma_{\textrm{int}}$ star-forming galaxies also appears to be consistent with the predicted evolution of the velocity versus stellar-mass relation from cosmological simulations \citep[e.g.][]{Dutton2011}.
However, the model predictions for the evolution of the Tully-Fisher relation with redshift from \cite[e.g.][]{Dutton2011} or the EAGLE simulation \citep{Schaye2015} as presented in Tiley et al. ({\it in prep.}) represent an idealised scenario where all of the dynamical mass is supported by ordered rotation.
In reality this does not appear to be the case \citep[e.g.][]{Burkert2010,Ubler2017,Turner2017} and the contribution of random motions to supporting dynamical mass becomes increasingly significant with increasing redshift.
The agreement between model predictions for the evolution of the stellar-mass Tully-Fisher relation and the fits to high $V_{\textrm{C}}/\sigma_{\textrm{int}}$ samples again suggests that these galaxies are closest to being supported purely by ordered rotation.

One can instead focus on larger samples which are more representative of the typical evolving-disk population at a particular epoch, are more kinematically diverse and may have a larger contribution from random motions to supporting the dynamical mass of the systems \citep[e.g.][]{Harrison2017}.
In this case, the velocity zero-point of the stellar-mass Tully-Fisher relation does not evolve strongly and may even evolve in the opposite sense at ${\it z} > 3$, where pressure support is most significant \citep[e.g.][]{Turner2017}.
In between these extremes, the offsets from the local stellar-mass Tully-Fisher relation are mediated by both the median $V_{\textrm{C}}/\sigma_{\textrm{int}}$ of the sample and the fraction of galaxies used in the Tully-Fisher analysis relative to the parent sample as demonstrated in Figs \ref{fig:parent_fraction_offsets} and \ref{fig:offsets}.
However, it is not easy to interpret this evolution in relation to a dynamical-mass to stellar-mass ratio evolution. \\

As has been explored recently, it is necessary to account for the contribution of random motions to the dynamics of the system, especially at high redshift where star-forming galaxies appear to be highly pressurised \citep[e.g.][]{Kassin2012,Ubler2017}.
Doing so provides an opportunity to trace the true dynamical mass and to unify samples consisting of both rotation-dominated and dispersion-dominated galaxies, thus mitigating the effects of sample selection.
We explore the extent to which the rotation velocities may underestimate dynamical mass for the KDS galaxies in \cref{subsec:virial_mass_content} and consequently derive the possible form of a `total velocity', that includes a contribution from the velocity dispersion. 
Analogous to \cref{subsec:vc_m_evolution}, we then proceed to study the evolution of the total-velocity versus stellar-mass relation, using our compilation of comparison samples, throughout the following sections.

\section{Velocity dispersion contribution in tracing dynamical mass}\label{sec:sigma_contribution}

\subsection{The virial mass content of the KDS galaxies}\label{subsec:virial_mass_content}
\begin{figure*}
    \centering \hspace{-1.65cm}
    \begin{subfigure}[h!]{0.5\textwidth}
        \centering
        \includegraphics[height=3.75in]{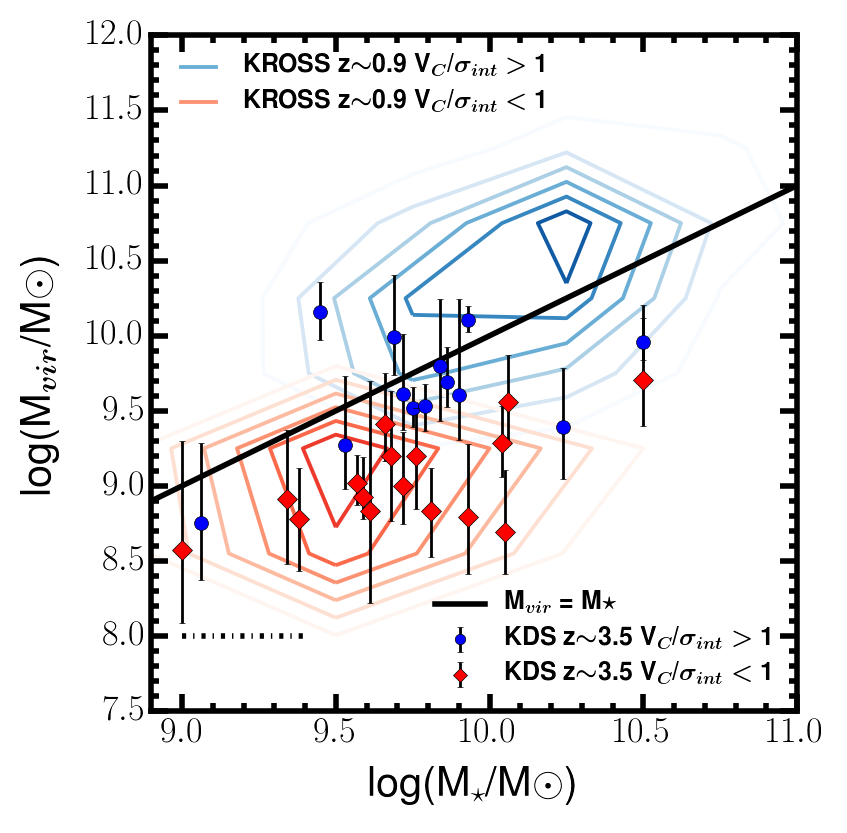}
    \end{subfigure} \hspace{+0.45cm}
    \begin{subfigure}[h!]{0.5\textwidth}
        \centering
        \includegraphics[height=3.75in]{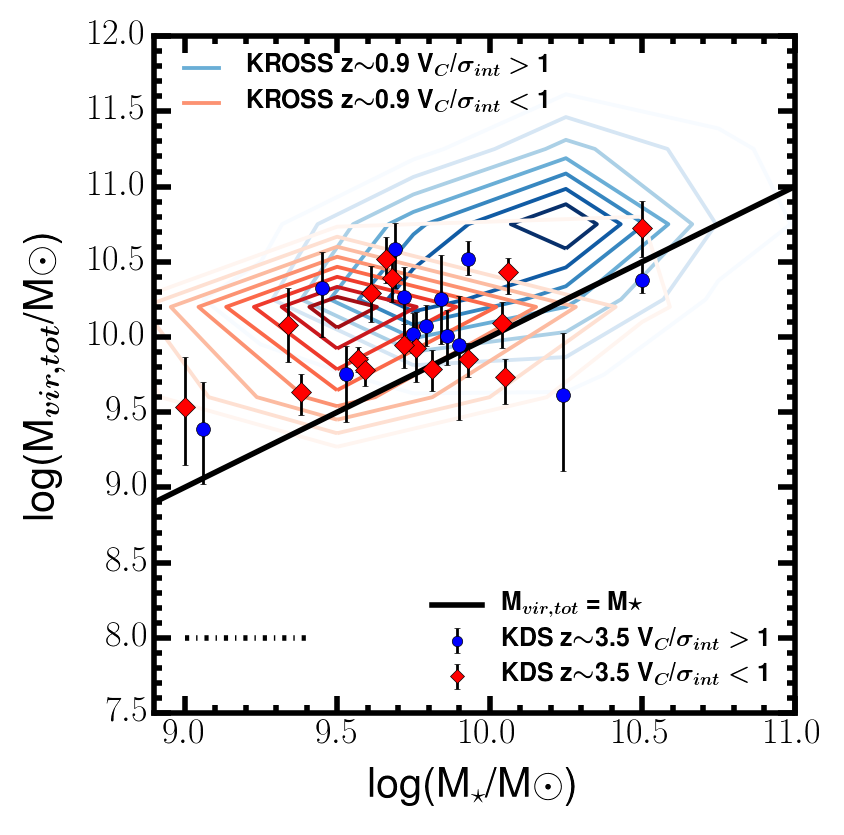}
    \end{subfigure}
    \caption{{\it Left:} Virial mass computed using only rotation velocities (Equation \protect\ref{eq:dyn_mass_kds}) versus stellar mass for the KDS isolated-field sample, with the black line indicating equality between virial mass and stellar mass.
    The blue-circular symbols show the KDS galaxies with $V_{\textrm{C}}/\sigma_{\textrm{int}} > 1$ and the red-diamond symbols show the KDS galaxies with $V_{\textrm{C}}/\sigma_{\textrm{int}} < 1$, occupying a region with lower $M_\textrm{{vir}}$ values than the rotation-dominated galaxies.    
    The majority of the points lie in the unphysical $M_{\textrm{vir}} < M_{\star}$ region, indicating that rotation velocity alone is not sufficient to provide gravitational support for the stellar mass in the systems.
    The blue and red contours show the density of rotation-dominated and dispersion-dominated galaxies from KROSS \protect\citep{Harrison2017}, both starting at 10 and increasing in increments of 10 and 3 respectively.
    {\it Right:} Total virial mass, $M_{\textrm {vir,tot}}$, computed with an additional component traced by the velocity dispersion, versus stellar mass, with the black line indicating equality between these quantities.
    The addition of this component shifts most galaxies into the $M_{\textrm {vir,tot}} > M_{\star}$ region and highlights the potential for a combination of random motions and ordered rotation to play a role in supporting the total virial mass.}
    \label{fig:dyn_masses}
\end{figure*}

In this subsection we discuss the concept of `total velocity' for the KDS galaxies, that includes a velocity-dispersion contribution ($\eta\sigma_{\textrm{int}}$), where $\eta$ can be constrained by the comparison between stellar and virial mass.
This involves making an assumption for the value of the ratio of dynamical to stellar mass for the KDS galaxies at $z\simeq3.5$, which is inherently uncertain.
However, this assumption can be informed by considering the observed gas fractions in high-redshift galaxies \citep[e.g.][]{Tacconi2013,Tacconi2017} and explored by adopting different values for the ratio in \cref{subsec:vtot_m_relation}.

The observed dynamics of a galaxy can be used to infer the total mass enclosed at different radii, which can then be compared with the stellar mass from SED fitting. 
In this way the partitioning of the total mass between baryonic components can be studied and compared with predictions.
Assuming that a galaxy is supported against gravitational collapse by ordered rotation, the rotation velocity can be used to trace the mass enclosed within radius R as follows:

\begin{equation}\label{eq:dyn_mass}
   M\left(<R\right) = \frac{RV_{\textrm{C}}(R)^{2}}{G}
\end{equation}

For the KDS isolated-field sample galaxies the rotation velocities are extracted at a radius of $2R_{1/2}$ from the intrinsic models and so the mass enclosed within this radius is given by:

\begin{equation}\label{eq:dyn_mass_kds}
   M_{\textrm{vir}} = \frac{2R_{1/2}V_{\textrm{C}}^{2}}{G}
\end{equation}

which we refer to hereafter as the virial mass, $M_{\textrm{vir}}$.
In the left panel of Fig.\,\ref{fig:dyn_masses}, we plot virial mass, computed using this simple equation, against stellar mass for the KDS isolated-field sample galaxies.
The majority of galaxies in the isolated-field sample show $M_{\textrm{vir}} < M_{\star}$ (with median value $M_{\textrm{vir}}/M_{\star}=0.32 \pm 0.23$), including those in the rotation-dominated subsample (for which the median value of $M_{\textrm{vir}}/M_{\star}=0.59 \pm 0.43$).
This is surprising because at a radius of $2R_{1/2}$ for the KDS galaxies we are tracing the bulk of the stellar mass distribution\footnote{We have verified in \cite{Turner2017} that 23/24 KDS isolated-field sample galaxies also detected in \cite{VanderWel2012} follow \Sers light profiles with $n\sim1$.
For this (exponential) distribution, a radius of $2R_{1/2}$ encloses $\sim85$ per cent of the stellar light, from which the stellar mass is inferred through SED fitting.}, and so in principle the virial mass should exceed the stellar mass if it is a measure of the total mass enclosed within $2R_{1/2}$.
Indeed at the KDS median redshift of ${\it z}\simeq3.5$, the gas fractions can be $\ge50$ per cent for typical star-forming galaxies \citep{Tacconi2013,Tacconi2017}, and dark matter is also present within the galactic disk.
If these components are gravitationally supported by rotation alone, the virial mass computed using Equation \ref{eq:dyn_mass_kds} should be substantially larger than the stellar mass of the galaxies.

To place the KDS galaxies in the context of lower redshift results, we again use galaxies from the KROSS sample.
The red and blue contours in the left panel of Fig.\,\ref{fig:dyn_masses} indicate the density of the dispersion-dominated and rotation-dominated KROSS sample galaxies respectively in the virial mass versus stellar mass plane.
The majority of the KROSS rotation-dominated galaxies show $M_{\textrm{vir}} > M_{\star}$ and almost all KROSS dispersion-dominated galaxies show $M_{\textrm{vir}} < M_{\star}$.

As shown in \cref{subsec:vc_m_relation}, the rotation-dominated KDS galaxies are $\simeq-0.10$ dex in velocity zero-point beneath the local stellar-mass Tully-Fisher relation from \cite{Reyes2011}.
The KDS galaxies also have half-light radii which are a factor of $\simeq3-4$ smaller than these local galaxies.
In order for the total mass in the more compact KDS galaxies to be supported by rotation alone, we would expect that they `spin up' to higher rotation velocities at fixed stellar mass. 
Observationally this does not appear to be the case (see e.g. \citealt{Simons2017}) suggesting that rotational motions alone are not sufficient to provide gravitational support for the total mass in the KDS galaxies, which also appears to be the case for many of the intermediate redshift KROSS galaxies.
A possible solution to the virial to stellar mass discrepancy is that random motions in the systems, as traced by the velocity dispersions, provide partial gravitational support for the total disk mass as has been previously suggested \citep[e.g.][]{Kassin2007,Burkert2010,Kassin2012,Newman2013,Ubler2017}.
This contribution becomes increasingly significant with increasing redshift as the ratio of rotation velocity to velocity dispersion decreases.
This is referred to as an `asymmetric drift' correction \citep[e.g.][]{Burkert2010}, where turbulent pressure support generated by gravitational instabilities renders the observed rotation velocity a poor tracer of the total virial mass in the galaxy.
A revised description of the total virial mass, $M_{\textrm {vir,tot}}$, is given below:

\begin{equation}\label{eq:dyn_mass_sigma}
   M_{\textrm {vir,tot}} = \frac{2R_{1/2}\left(V_{\textrm{C}}^{2} + \eta\sigma_{\textrm{int}}^{2}\right)}{G}
\end{equation}

which includes a contribution from the velocity dispersion of the galaxies.
Here we do not seek to derive a precise value for $\eta$, rather to show that a significant contribution to the dynamical mass from random motions appears necessary in order to provide gravitational support for the expected baryonic material within $2R_{1/2}$.
We use the value $\eta=4.0$, which generates a median ratio of total virial mass to stellar mass in the KDS isolated-field sample equal to 2 (i.e. median $M_{\textrm {vir,tot}}/M_{\star} = 2$\footnote{The $\eta$ values required for individual galaxies to have $M_{\textrm {vir,tot}}/M_{\star} = 2$ vary widely, with $\eta_{\textrm{min}}=0.25$ and $\eta_{\textrm{max}} = 37$. }).
This implies that, on average across the KDS sample and on the scales traced by the observations, the dynamical mass should be twice as large as the stellar mass.
This value of $\eta$ is comparable to the value found for an exponential mass distribution ($\eta=3.4$, e.g. \citealt{Burkert2010,Newman2013}) and for a non-rotating spherical mass distribution of constant density ($\eta=5$).
The adopted value of $\eta$ is somewhat arbitrary, and so the impact on the results when varying this parameter between a minimum of $\eta=2.0$ (corresponding to median $M_{\textrm {vir,tot}}/M_{\star} \sim 1.1$) and a maximum of $\eta=6.0$ (corresponding to median $M_{\textrm {vir,tot}}/M_{\star} \sim 2.7$) is explored throughout \cref{subsec:vtot_m_relation}. \\

In the right panel of Fig.\,\ref{fig:dyn_masses} we plot total virial mass, computed using Equation \ref{eq:dyn_mass_sigma}, versus stellar mass for the KDS galaxies.
By design, with this additional virial mass component sourced by the velocity dispersions, most of the isolated-field sample galaxies shift to the $M_{\textrm {vir,tot}} > M_{\star}$ region and the discrepancy between dispersion-dominated and rotation-dominated galaxies no longer remains.
We again plot the rotation-dominated and dispersion-dominated KROSS sample galaxies with the blue and red contours respectively in this plane, with $M_{\textrm {vir,tot}}$ computed using the same equation.
The dispersion-dominated galaxies from the KROSS sample also shift into the $M_{\textrm{vir}} > M_{\star}$ region, showing similar values to the galaxies from the KDS sample. 

From Equation \ref{eq:dyn_mass_sigma}, we can also define a `total velocity', which is given by: 

\begin{equation}\label{eq:tot_velocity}
   V_{\textrm{tot}} = \sqrt{V_{\textrm{C}}^{2} + 4.0\sigma_{\textrm{int}}^{2}}
\end{equation}

This is similar to the $S_{0.5} = \sqrt{0.5V_{\textrm{C}}^{2} + \sigma_{\textrm{g}}^{2}}$ relation \citep{Kassin2007}, which uses the combination of observed rotation velocity and integrated velocity dispersion as a better tracer of dynamical mass (and Cf. also the circular velocity given by equation 1 of \citealt{Ubler2017}).
One important difference between $V_{\textrm{tot}}$ and $S_{0.5}$ is the use of intrinsic rather than integrated gas velocity dispersions and also the direct addition in quadrature of the contribution from velocity dispersions to the observed rotation velocities, in an attempt to find a substitute velocity which traces the total dynamical mass.
We proceed to explore the derived total velocities in the context of the total-velocity versus stellar-mass relation throughout the following sections. 

\subsection{The total-velocity versus stellar-mass relation for the KDS galaxies}\label{subsec:vtot_m_relation}

\begin{figure*}
\centering \hspace{-1.13cm}
\includegraphics[width=\textwidth]{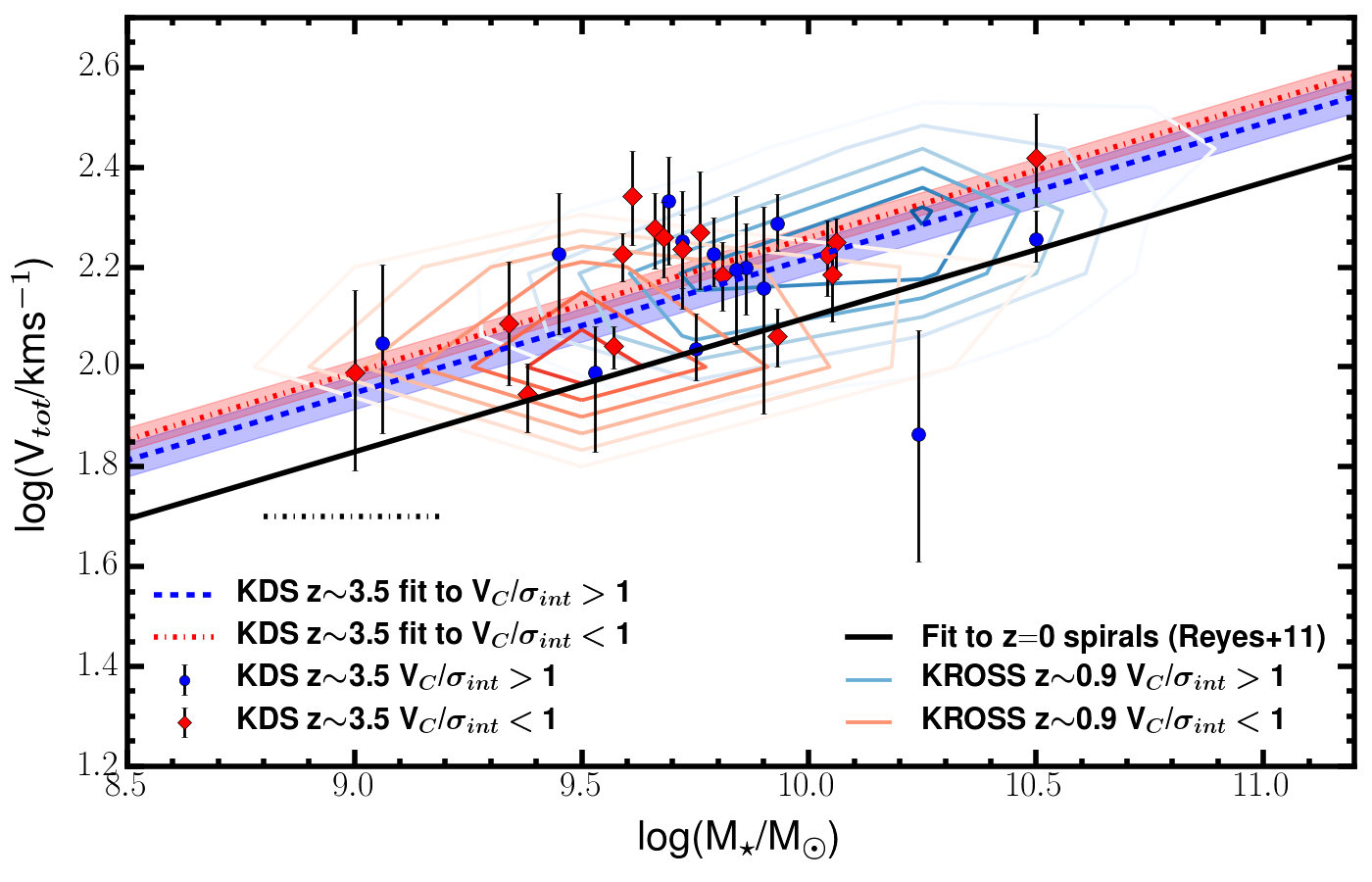}
\caption{Total-velocity versus stellar-mass for both the KDS and KROSS galaxies, with the same symbol convention, fit-lines and local reference relation as Fig.\,\ref{fig:tf_relation}.
As discussed in the text, the total velocity is likely a better tracer of the true dynamical mass than the rotation velocity alone.  
    In contrast to Fig.\,\ref{fig:tf_relation}, the rotation-dominated and dispersion-dominated galaxies from both KDS and KROSS fall on the same sequence when using the total velocity rather than the rotation velocity.
    This suggests that sample-selection effects, which aim to distinguish between these subsamples, are less important when studying the total-velocity versus stellar-mass relation.
    The relation $\textrm{log}({\it V_{\textrm{tot}}}) = \upbeta + \upalpha[\textrm{log}({\it M_{\star}}) - 10.1]$ is fitted to the combined samples of rotation-dominated and dispersion-dominated galaxies in KDS and KROSS, returning zero-point offsets which are in agreement, and roughly $+0.1$ dex in total-velocity zero-point above the local zero-point.}
\label{fig:tot_v_mass_relation}
\end{figure*}

Given the sample-selection caveats discussed in \cref{subsubsec:discussion_v_over_sigma_cuts}, it would be extremely useful to measure the evolution of the connection between dynamical and stellar mass following a method which is insensitive to sample-selection effects.
By extension, this would allow fits to full galaxy samples (i.e. both rotation-dominated, defined in whatever way, and dispersion-dominated) regardless of the observed kinematic properties of the individual galaxies. 
We do this by including pressure support (following Equation \ref{eq:tot_velocity}) to the rotation velocities of the KDS galaxies and plotting the total-velocity versus stellar-mass relation in the right panel of Fig.\,\ref{fig:tot_v_mass_relation}.
We assume that the contribution of pressure to the rotation velocities of local galaxies is negligible, due to the large $V_{\textrm{C}}/\sigma_{\textrm{int}}$ ratios, and continue to compare higher redshift fits with the fiducial relation recovered from the fit to the \cite{Reyes2011} galaxies.
This approach is verified by adding a constant velocity dispersion of 20kms$^{-1}$, appropriate for local spiral galaxies \citep[e.g.][]{Epinat2008a}, and calculating the total velocity of the \cite{Reyes2011} galaxies using Equation \ref{eq:tot_velocity}.
The normalisation difference between the total-velocity versus stellar-mass and velocity versus stellar-mass relations is $+0.02$ dex, which is small in comparison to other systematics such as the choice of $\eta$ parameter and local reference relation. \\

\begin{figure*}
\centering
\includegraphics[width=\textwidth]{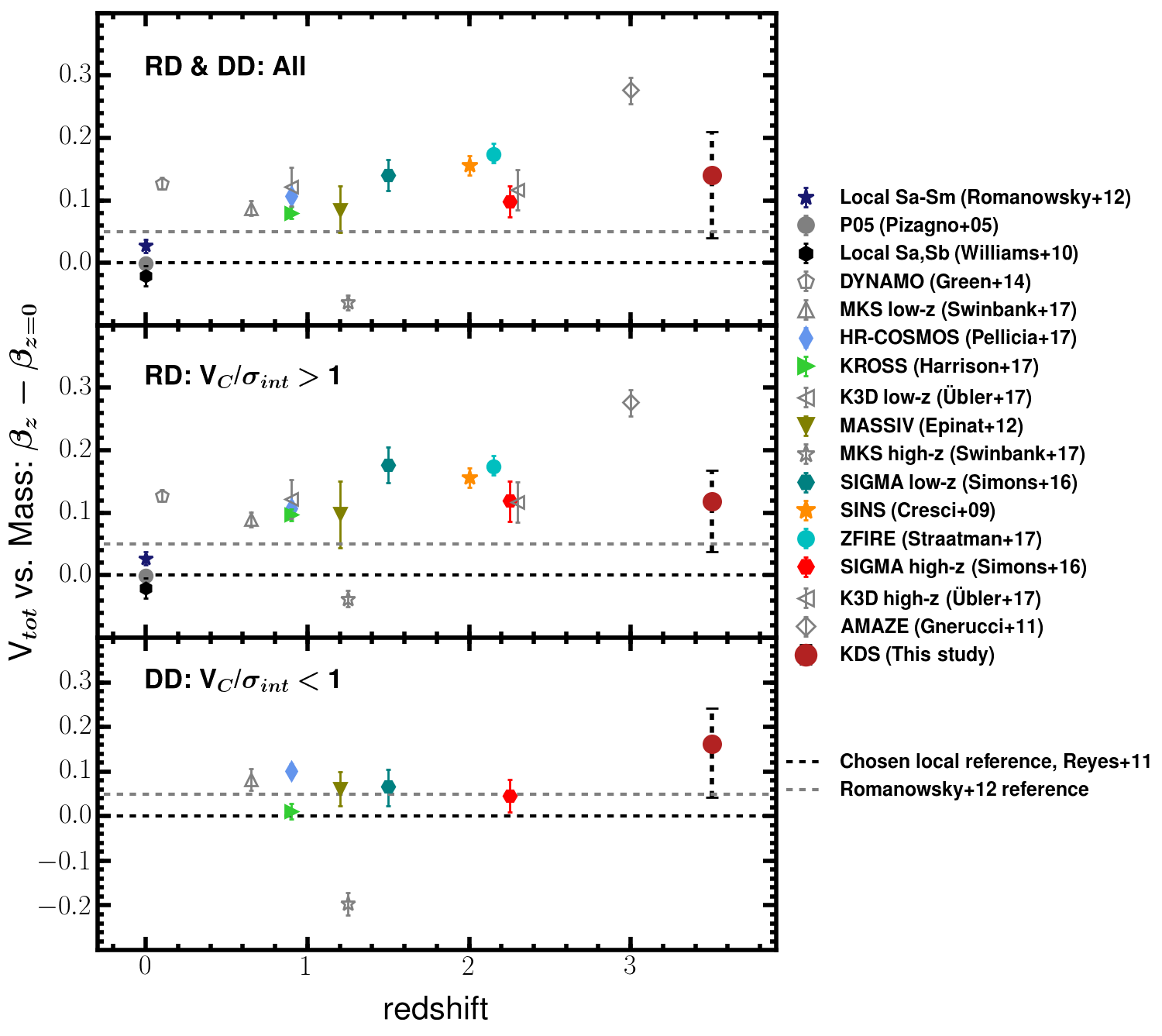}
    \caption{Total velocity offsets from the local stellar-mass Tully-Fisher relation for each of the comparison samples, plotted against redshift.
    Each panel is the same as in Fig.\,\ref{fig:velocity_evolution} but with the Tully-Fisher offsets computed using the total velocity from Equation \ref{eq:tot_velocity}.
    As per Fig.\,\ref{fig:velocity_evolution} we indicate the zero-point shift of $+0.05$ dex found when using the \protect\cite{Romanowsky2012} reference relation with the grey-dashed line.
    Also indicated with the black-dashed error bar on the KDS datapoints are the lower and upper limits on the KDS total velocity offsets when using the values $\eta=2$ and $\eta=6$ respectively (see text).
    The dispersion-dominated galaxies sit on almost the same relationship as the rotation-dominated galaxies, so that the fits to the full galaxy samples no longer average over two subsamples in different regions of the plane.
    The fits to the full samples and the rotation-dominated galaxies with added pressure support suggest a fairly constant shift in total-velocity zero-point of between $+0.08$ to $+0.15$ dex ($-0.30$ to $-0.55$ dex in stellar-mass zero-point) at ${\it z}\geq1$.}
    \label{fig:vtot_evolution}
\end{figure*}

\begin{figure*}
\centering
\includegraphics[width=\textwidth]{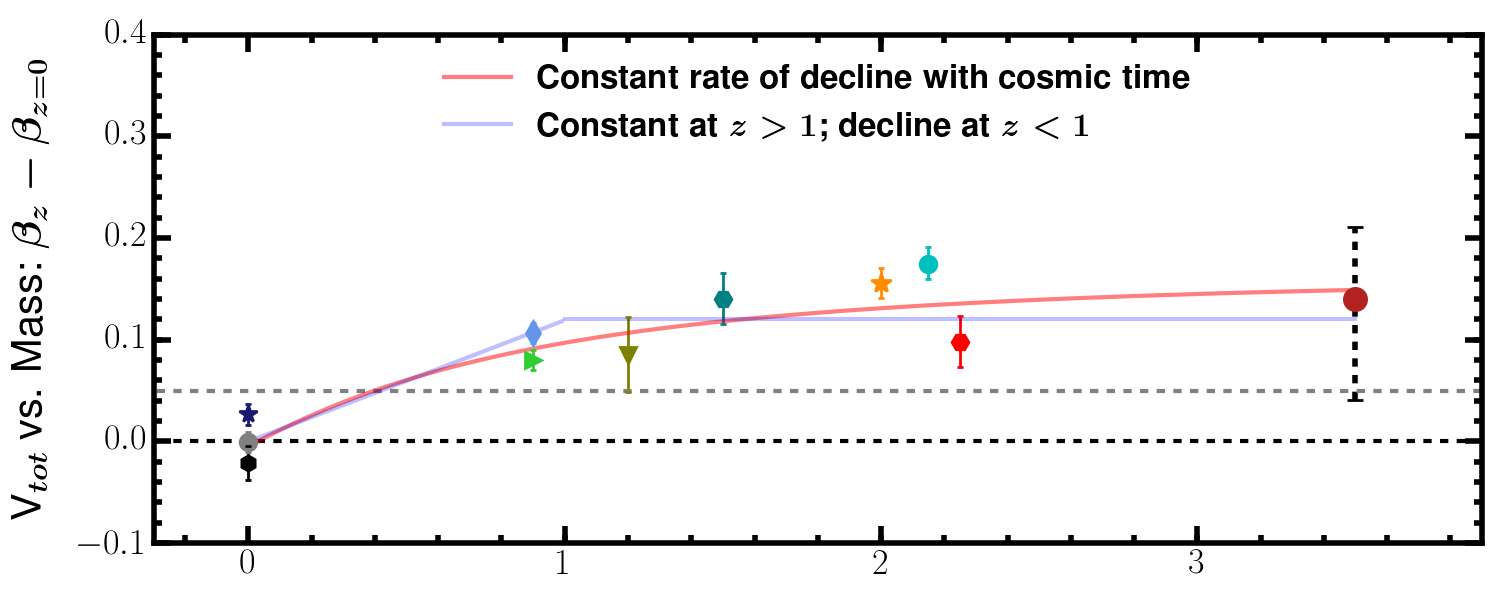}
    \caption{We replicate the top panel of Fig.\,\ref{fig:vtot_evolution}, plotting the full-sample total-velocity offsets against redshift, with the grey-hollow symbols omitted.
    The red solid line shows an offset which is declining constantly with time, starting at an offset of $+0.15$ dex (see text).
    This reflects a constant decline in the ratio of dynamical to stellar mass throughout star-forming galaxy samples as the age of the Universe increases.
    The blue solid line shows a constant offset of $+0.12$ dex at ${\it z} > 1$, followed by a constant decline at ${\it z} < 1$.
    On the basis of these data, we cannot distinguish between these two scenarios.} 
    \label{fig:vtot_evolution_lines}
\end{figure*}

In the total-velocity versus stellar-mass plane, the discrepancy between rotation-dominated and dispersion-dominated galaxies disappears for the KDS sample and is greatly reduced in the KROSS sample.
The same fitting procedure as described above is applied in turn to the full KDS sample, the rotation-dominated subsample and the dispersion-dominated subsample, returning the values $\upbeta_{\textrm{tot,all},{\it z}=3.5} = 2.269\pm0.020$, $\upbeta_{\textrm{tot,rot},{\it z}=3.5} = 2.245\pm0.033$ and $\upbeta_{\textrm{tot,disp},{\it z}=3.5} = 2.286\pm0.023$.
Fitting the same subsamples of KROSS galaxies returns the values values $\upbeta_{\textrm{tot,all},{\it z}=0.9} = 2.207\pm0.010$, $\upbeta_{\textrm{tot,rot},{\it z}=0.9} = 2.223\pm0.010$ and $\upbeta_{\textrm{tot,disp},{\it z}=0.9} = 2.139\pm0.019$.
The fits to the full KDS and KROSS samples suggest total-velocity zero-point offsets of $\simeq+0.14$ dex and $\simeq+0.08$ dex towards higher total velocities at fixed stellar mass respectively from the local relation.
Crucially, because of the high velocity dispersions observed throughout the KDS sample \citep{Turner2017}, the zero-point offsets shift substantially to move above the local relation, bringing the rotation-dominated and dispersion-dominated subsamples into agreement.  
This balance between the increased random motions and decreased rotational motions in the KDS galaxies suggests that $V_{\textrm{tot}}$ is a better tracer of the dynamical mass. \\

For this analysis we chose $\eta=4.0$ in Equation \ref{eq:tot_velocity}, corresponding to the value required for the KDS sample median $M_{\textrm {vir,tot}}/M_{\star} = 2$.
In \cite{Epinat2009} the value $\eta=1.35$ is adopted and in \cite{Newman2013} $\eta=3.4$ is quoted for an exponential mass distribution.
If instead we adopt $\eta=2.0$ and fit the full sample of KDS galaxies in the total-velocity versus stellar-mass plane using the above relation, we recover the normalisation $\upbeta_{\textrm{tot,all},\eta=2.0} = 2.162\pm0.021$ and with $\eta=6.0$ we find $\upbeta_{\textrm{tot,all},\eta=6.0} = 2.337\pm0.021$.
Taking these values as encompassing the range of possible total-velocity zero-points leads to large errors on the quoted zero-point above, so that it reads $\upbeta_{\textrm{tot,all},{\it z}=3.5} = 2.269^{+0.07}_{-0.10}$ (i.e. a zero-point offset from the local relation of $+0.14^{+0.07}_{-0.10}$ dex).
However at the lower end of the $2.0 < \eta < 6.0$ range, the discrepancy between the expected virial and stellar mass (see \cref{subsec:virial_mass_content}) is still present and towards the upper end the relationship between virial mass and stellar mass flattens, suggesting that the velocity dispersion term is too large.
The precise choice of $\eta$ has a significant impact on the extent to which the KDS total-velocity versus stellar-mass relationship is observed to evolve.
However, for the reasons above we believe $\eta=4.0$ is a reasonable choice.  

\subsubsection{Evolution of the total-velocity versus stellar-mass relation out to $z\sim4$}\label{subsubsec:vtot_m_evolution}

To explore these ideas over a wider redshift baseline, we calculate the total velocities of the comparison sample galaxies, where possible, and apply the same fitting method as described in \cref{subsec:vc_m_relation}.
We again use a fixed slope of $\upalpha_{{\it z}=0}=0.270$, in order to measure the total-velocity zero-point offsets from the local relation.
This allows us to compare with the offsets measured from fitting the standard stellar-mass Tully-Fisher relation throughout \cref{subsec:vc_m_evolution}.
In Fig.\,\ref{fig:vtot_evolution} we plot the total-velocity versus stellar-mass offsets against redshift for each of the comparison samples in which velocity dispersion measurements were available.
The most dramatic difference between Figs \ref{fig:velocity_evolution} and \ref{fig:vtot_evolution} is the velocity zero-point shift in the bottom panel, for the dispersion-dominated galaxies, which move up to almost the same position in the total-velocity versus stellar-mass plane as the rotation-dominated galaxies (see Fig.\,\ref{fig:v_tot_fits} also). \\

The offsets computed from fits to the full samples and to the rotation-dominated subsamples are now almost indistinguishable, with the comparison samples at ${\it z}\geq1$ showing offsets in the range $+0.08$ to $+0.15$ dex in total-velocity zero-point ($-0.30$ to $-0.55$ dex in stellar-mass zero-point) above the local relation.
These results are subject to several systematics.
For example setting $\eta=2.0$ and $\eta=6.0$ in Equation \ref{eq:tot_velocity} as described above leads to large errors on the recovered KDS total-velocity offsets.
We show these errors, associated with the uncertainty in the value of $\eta$, with the black-dashed error bars in each of the panels of Fig.\,\ref{fig:vtot_evolution}.
We stress that this error is not so severe for the lower redshift comparison samples in which the velocity dispersions are smaller.
Adopting a different local reference relation can change the KDS offset by $-0.05$ dex (see Fig.\,\ref{fig:rom_slope_vel_evolution}), represented in Fig.\,\ref{fig:vtot_evolution} by the grey-dashed zero-point lines.
The impact of these effects adds uncertainty to the extent of the inferred evolution of the total-velocity versus stellar-mass relation.      
However, the total-velocity offsets are consistently positive amongst the star-forming galaxy comparison samples at $z\geq1$.
This suggests an evolving ratio of dynamical to stellar mass and a transition between the magnitude of the dynamical support provided by ordered and random motions, due to the steady rise in the intrinsic velocity dispersions of star-forming galaxies with increasing redshift \citep[e.g.][]{Wisnioski2015,Turner2017}.
We focus on the interpretation of this result in the following section.

\subsection{The addition of velocity dispersion is required to trace the galaxy potential wells}\label{subsec:pressure_support_unification}

As was discussed in \cref{subsubsec:vtot_m_evolution} and in Fig.\,\ref{fig:vtot_evolution}, a more complete way to study the evolution of the relationship between dynamical and stellar mass is to attempt to account for the effects of pressure support in the galaxies.
This reduces the kinematic diversity observed in high-redshift galaxies by including the `missing' dynamical component traced by velocity dispersions, bringing rotation-dominated and dispersion-dominated galaxies into better agreement in the velocity versus stellar-mass plane (see the fits in Fig.\,\ref{fig:v_tot_fits}) and allowing us to fit the Tully-Fisher relation to the full samples of galaxies.
This avoids the problematic issue of choosing criteria to define a Tully-Fisher sample, which, as we have shown throughout \cref{subsubsec:discussion_v_over_sigma_cuts}, entirely determine the extent to which the relation is observed to evolve. \\

The fits suggest that the pressure corrected samples all have positive total-velocity versus stellar-mass relation offsets, with a mean value of roughly $+0.12$ dex in total-velocity zero-point ($-0.45$ dex in stellar-mass zero-point) from the local stellar-mass Tully-Fisher relation.
This is similar in magnitude to the offsets of $-0.44$ dex at ${\it z}\simeq0.9$ and $-0.42$ dex at ${\it z}\simeq2.3$ in stellar-mass zero-point quoted in \cite{Ubler2017}, in which the effects of pressure support have been included.
This is interpreted as a decrease in stellar mass relative to gas mass, as well as an increasing baryonic to dark matter fraction with redshift on the scales traced by the ionised gas emission.
The combined impact of these effects would maintain a relatively constant ratio of dynamical to stellar mass on the disk scale above ${\it z}\simeq1.0$, which is traced by the stellar-mass Tully-Fisher relation.
In Fig.\,\ref{fig:vtot_evolution_lines} we plot the total-velocity offsets measured from fitting the combined rotation-dominated and dispersion-dominated galaxies throughout the comparison samples, (i.e. a reproduction of the top panel of Fig.\,\ref{fig:vtot_evolution}).
The blue line shows an offset that is at a constant level of $+0.12$ dex at ${\it z} > 1$ and declines to the local relation at a constant rate (relative to redshift) at ${\it z} < 1$.  \\

An alternative scenario is that populations of star-forming galaxies have been gradually drifting onto the local Tully-Fisher relation by maintaining constant total velocity and growing in stellar mass at a rate which is roughly constant with time. 
To demonstrate this, we plot an offset of $+0.15$ dex at ${\it z} = 3.5$ (corresponding to 12 Gyrs in the past) which declines constantly with time onto the local relationship at ${\it z} = 0$, with the red line in Fig.\,\ref{fig:vtot_evolution_lines}.
Both scenarios imply an important period of stellar mass assembly within star-forming galaxy populations at ${\it z} < 1$ which starts to bring them onto the local stellar-mass Tully-Fisher relation. 
The blue and red lines, representing the increased influence of dark matter on disk scales \citep[e.g.][]{Ubler2017,Lang2017,Genzel2017} and a constant decline in the ratio of dynamical to stellar mass with time respectively, both appear to provide an adequate description of the data.
However the data do not allow us to further distinguish between these two scenarios. \\

What is clear is that the trend to observe positive velocity zero-point offsets across all the comparison samples is not seen unless a velocity dispersion term is taken into account, as a direct consequence of sample-selection effects and the incomplete dynamical evolution of star-forming galaxies at intermediate and high redshift.
This is especially true for the KDS sample, in which the large observed velocity dispersions suggest that accounting for pressure support of this type is especially important at ${\it z}\simeq3.5$, where the interstellar medium is increasingly turbulent and gas rich.
It will be intriguing to follow up these observations in the future with increased senstitivity and higher spatial resolution, i.e. in the {\it JWST} era, in order to compare stellar and gaseous velocity dispersions and to further test the role of motions traced by the gaseous velocity dispersion in supporting mass in the systems against gravitational collapse.

\section{CONCLUSIONS}\label{sec:conclusion}
We have used rotation velocities, $V_{\textrm{C}}$, and velocity dispersions, $\sigma_{\textrm{int}}$ from the KMOS Deep Survey \citep{Turner2017}, along with measurements from several carefully-selected comparison samples spanning $0 < {\it z} < 3$ (see Appendix \ref{app:comparison_samples}), to investigate the evolution of the stellar-mass Tully-Fisher relation.
To explain discrepant literature results, we have explored the connection between sample-selection effects and Tully-Fisher evolution, finding a strong correlation between two tracers of sample-selection criteria and velocity zero-point offsets from the local reference relation.  
We have also studied the impact of adding pressure support in the derivation of rotation velocities as a way to both trace the true dynamical mass and to mitigate the effects of sample-selection criteria. 
The main conclusions of this work are summarised as follows:

\begin{itemize}
    
    \item We fit the stellar-mass Tully-Fisher relation, $\textrm{log}({\it V_{\textrm{C}}}) = \upbeta + \upalpha[\textrm{log}({\it M_{\star}}) - 10.1]$, to rotation-dominated galaxies from the KMOS Deep Survey using a fixed slope of $\upalpha=0.270$, determined from fitting the same relation to local reference data from \cite{Reyes2011}.
    The recovered velocity zero-point, $\upbeta_{\textrm{rot},{\it z}=3.5} = 2.02\pm0.04$, is offset by $-0.10$ dex ($+0.37$ dex in stellar-mass zero-point) from the $z=0$ reference relation velocity zero-point, suggesting lower rotation velocities at fixed stellar mass in the KDS galaxies (see Fig.\,\ref{fig:tf_relation}). \\

    \item We fit the same fixed-slope relation to data from 16 distant comparison samples spanning $0 < {\it z} < 4$, divided into rotation-dominated and dispersion-dominated subsamples.
    The fits to the rotation-dominated subsamples show a variety of offsets from the local relation (on, above and below), in agreement with the discrepancies quoted in the literature and with no clear correlation between the offsets and redshift (see Fig.\,\ref{fig:velocity_evolution}).
    Increasingly-strict `disky' sample-selection criteria result in larger inferred velocity offsets at fixed stellar mass with respect to the local Tully-Fisher relation.
    In contrast, no offset is generally found for samples which are representative of the evolving-disk population - defined as either: (1) a population of $V_{\textrm{C}}/\sigma_{\textrm{int}} > 1$ star-forming galaxies, so that the fraction used in a Tully-Fisher analysis is in agreement with the evolving rotation-dominated fraction (Fig.\,\ref{fig:parent_fraction_offsets}); (2) a population where the average $V_{\textrm{C}}/\sigma_{\textrm{int}}$ value agrees with the prediction from a simple equilibrium model (Fig.\,\ref{fig:offsets}).
    The strong connection between sample-selection criteria and Tully-Fisher offsets highlights the kinematic diversity of high-redshift galaxies and demonstrates that previous, discrepant results in the literature for the evolution of the stellar-mass Tully-Fisher relation can be explained by taking sample-selection criteria into account.\\

    \item We show using a comparison of the KDS virial mass and stellar mass that a contribution from velocity dispersion is likely required to trace dynamical mass and consequently define a `total velocity' of the form $V_{\textrm{tot}} = \sqrt{V_{\textrm{C}}^{2} + 4.0\sigma_{\textrm{int}}^{2}}$ (Fig.\,\ref{fig:dyn_masses}).
    Using this relation, rotation-dominated and dispersion-dominated galaxies lie on the same sequence in the total-velocity versus stellar-mass plane (Fig.\,\ref{fig:tot_v_mass_relation}), in contrast to in the rotation-velocity versus stellar-mass plane.
    This allows us to fit the full KDS sample (both rotation-dominated and dispersion-dominated) consistently in the total-velocity versus stellar-mass plane without imposing any sample-selection criteria, finding an offset of $+0.14$ dex in total-velocity zero-point from the local relation. \\

    \item Using the total velocity also unifies the rotation-dominated and dispersion-dominated galaxies throughout the comparison samples.
    We explore the evolution of the total-velocity versus stellar-mass relation, which is independent of selection criteria, finding a mean total-velocity zero-point offset from the local relation of $\sim+0.12$ dex ($-0.45$ dex in stellar-mass zero-point) at $z\geq1$ (see Fig.\,\ref{fig:vtot_evolution}).
    The evolutionary trend throughout the total-velocity offsets suggests a constant decline in the ratio of dynamical to stellar mass with cosmic time at $z<4$, reflecting the accumulation of stellar mass and the kinematic evolution of star-forming galaxies.
    However, the data do not allow us to distinguish this scenario from one in which the ratio of dynamical to stellar mass stays constant in the range $1 < z < 4$ and declines steadily thereafter (Fig.\,\ref{fig:vtot_evolution_lines}).
\end{itemize}

It is crucial to consider the dynamical maturity of galaxies when determining whether the stellar-mass Tully-Fisher relation evolves with redshift.
Physically interpreting Tully-Fisher evolution as tracing evolution of the ratio of dynamical to stellar mass requires that rotation velocity is a good tracer of dynamical mass.
This is most likely to be true for high $V_{\textrm{C}}/\sigma_{\textrm{int}}$ samples, within which the total mass of the galaxies is closest to being supported entirely by ordered rotation, however these samples become less representative of the evolving-disk population at high redshift.
The galaxies in samples with lower median $V_{\textrm{C}}/\sigma_{\textrm{int}}$ have made less progress towards forming a stable, rotating disk and have lower velocities at fixed stellar mass, with the magnitude of the velocity dispersion tracing the `missing' dynamical mass component. 
Adding in a pressure support term to the velocities resolves the discrepancy in the rotation-velocity versus stellar-mass plane between star-forming galaxies at a particular epoch which span a wide range in $V_{\textrm{C}}/\sigma_{\textrm{int}}$, and allows a single relation to be fitted to full samples without imposing sample cuts which potentially bias the results.
The evolutionary trend in the pressure-corrected stellar-mass Tully-Fisher relation with redshift suggests a scenario in which gas rich galaxies at high redshift have yet to form the bulk of their stars and may have a smaller baryon to dark matter fraction on the disk scale.

\section*{Acknowledgements}

OJT acknowledges the financial support of the Science and Technology Facilities Council through a studentship award. 
MC and OJT acknowledge the KMOS team and all the personnel of the European Southern Observatory Very Large Telescope for outstanding support during the KMOS GTO observations.
AMS and ALT acknowledge the Science and Technology Facilities Council through grant code ST/L00075X/1.
JSD acknowledges the contribution of the EC FP7 SPACE project ASTRODEEP (Ref.No: 312725).
AMS acknowledges the Leverhulme Foundation.
We acknowledge the hard work of each of the teams who measured the properties of the galaxies in the comparison samples, without which this work would not be possible. 


\clearpage 
\bibliographystyle{mnras}
\bibliography{library.bib}
\clearpage


\appendix

\section{Comparison samples}\label{app:comparison_samples}
When attempting to assess the evolution of scaling relations by comparing the results from different surveys across cosmic time, it is essential to monitor sample-selection criteria and the differing methods used to compute the physical properties of the galaxies.
We carefully chose several local and distant comparison samples which have studied the dynamical properties of star-forming galaxies of varying median stellar mass, and paid close attention to the measurement of stellar mass (correcting to a \citealt{Chabrier2003} IMF where appropriate), intrinsic rotation velocities and velocity dispersions.
We list the details of these measurements for each of the local and distant comparison samples in the following subsections.
We did not however seek to correct the different measurements of rotation velocity, e.g. extraction from the rotation curves at different radii, to a common standard, which is a caveat of this analysis.  

\begin{figure}
\centering \hspace{-1.13cm}
\includegraphics[width=0.49\textwidth]{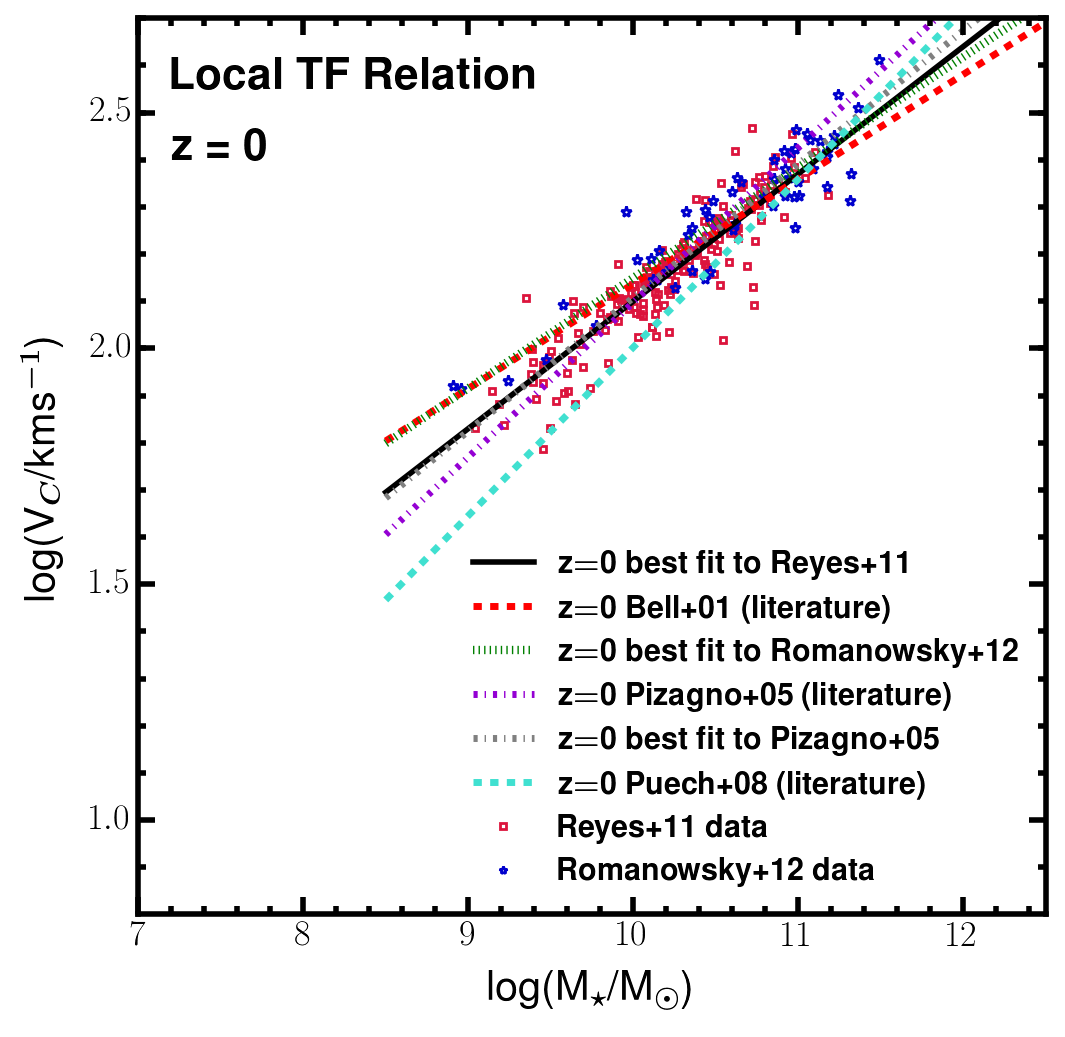}
\caption{189 spiral galaxies from \protect\cite{Reyes2011} are plotted with the red squares and 57 Sa-Sm type spiral galaxies from \protect\cite{Romanowsky2012} are plotted with the blue stars in the rotation-velocity versus stellar-mass plane.
The relation $\textrm{log}({\it V_{\textrm{C}}}) = \upbeta + \upalpha[\textrm{log}({\it M_{\star}}) - 10.1]$ was fitted to both samples, finding best fit parameters $\upbeta_{{\it z}=0} = 2.173$ and $\upalpha_{{\it z}=0} = 0.234$ for the \protect\cite{Romanowsky2012} sample (green-dashed line) and $\upbeta_{{\it z}=0} = 2.127$ and $\upalpha_{{\it z}=0} = 0.270$ for the \protect\cite{Reyes2011} sample (black-solid line).
To put these fits in context we also plot the most commonly-used local stellar-mass Tully-Fisher relations taken directly from the literature (i.e. without re-fitting the data points) and our best-fit to the \protect\cite{Pizagno2005} galaxies.
We used the \protect\cite{Reyes2011} relation as our local reference, as this has been used commonly in other studies of the stellar-mass Tully-Fisher relation, has been recovered from a fit to a large number of galaxies and is intermediate in slope between the relations plotted in this figure.
As described in the main text (\cref{subsec:vc_m_evolution}), when adopting the \protect\cite{Romanowsky2012} relation as a reference, the velocity zero-point offsets inferred for the comparison samples shift by $-0.05$ dex, however the evolutionary trends remain the same.  
Further information on these local comparison samples is given throughout Appendix \ref{subsec:local_comparison}.}
\label{fig:rom_tf_relation}
\end{figure}

We chose the spiral galaxies presented in \cite{Reyes2011} as our local reference sample to assess evolution in the rotation-velocity and total-velocity versus stellar-mass planes.
As described in the main-text, we fitted the stellar-mass Tully-Fisher relation to this sample to recover the local slope and velocity zero-point.
By fitting fixed-slope relations to the local and distant comparison samples, and comparing with the reference sample zero-point, we found the data points used in the evolution plots throughout \cref{sec:Tully-Fisher-relation} (i.e. Figs \ref{fig:velocity_evolution} and \ref{fig:vtot_evolution}). 
We also list 4 alternative measurements of the local stellar-mass Tully-Fisher relation (details in Appendix \ref{subsubsec:Reyes_2011} - \ref{subsubsec:Bell_2001}) and plot these in Fig.\,\ref{fig:rom_tf_relation}, to assess the impact of adopting a different local reference relation on the conclusions of this work. 

\subsection{Local comparison samples}\label{subsec:local_comparison}

\subsubsection{Reyes et al. 2011 spiral galaxies (${\it z}\simeq0$)}\label{subsubsec:Reyes_2011}
\cite{Reyes2011} presented a detailed study of the stellar-mass Tully-Fisher relation using a representative sample of 189 disk galaxies from the Sloan Digital Sky Survey (SDSS).
Rotation velocities, $V_{80}$, were extracted by fitting an arctangent function to the H$\upalpha$ position-velocity profiles and reading off at the radius containing $80$ per cent of the {\it i}-band light (i.e. roughly $3R_{\textrm{d}}$).
Stellar masses were computed using the method of \cite{Bell2003}, which uses colour-dependent mass-to-light ratios, and we converted from the \cite{Kroupa2002} IMF adopted in \cite{Reyes2011} to a Chabrier IMF (i.e. we apply a $\simeq0.06$ dex shift towards higher stellar mass).
The median stellar mass is $\textrm{log}   (M_{\star}/M_{\odot})=10.2$.
We fitted the stellar-mass Tully-Fisher relation to the galaxies in this survey using the relation $\textrm{log}({\it V_{\textrm{C}}}) =  \upbeta + \upalpha[\textrm{log}({\it M_{\star}}) - 10.10]$, finding $\upbeta_{{\it z}=0} = 2.127 \pm 0.005$ and $\upalpha_{{\it z}=0} = 0.270 \pm 0.009$, in good agreement with the values $\upbeta_{{\it z}=0} = 2.142 \pm 0.004$ and $\upalpha_{{\it z}=0} = 0.278 \pm 0.010$ quoted in \cite{Reyes2011}, with the small normalisation discrepancy a result of the change in IMF.
Subsequently, we study the evolution of the stellar-mass Tully-Fisher relation (\cref{subsec:vc_m_evolution}) and the total-velocity versus stellar-mass relation (\cref{subsubsec:vtot_m_evolution}) by holding the slope fixed to this local value and determining the velocity zero-points of the distant samples (see Appendix \ref{subsec:distant_comparison}).
This has been verified in \cref{subsec:vtot_m_relation} by re-fitting the total velocities of the \cite{Reyes2011} galaxies, assuming velocity dispersion values typical for the local Universe \citep[e.g.][]{Epinat2008}, finding negligible difference in the zero-point and slope recovered from the velocity versus stellar-mass fit.

\subsubsection{Romanowsky \& Fall 2012 spiral galaxies (${\it z}\simeq0$)}\label{subsubsec:romanowsky_spirals}
\cite{Romanowsky2012} carried out a detailed analysis of the kinematic properties of 64 spiral galaxies and 40 early-type galaxies in a study of the specific angular momentum of galaxies in the local Universe.
In \cite{Romanowsky2012}, these spiral galaxies were taken from the compilation of \cite{Kent1986,Kent1987,Kent1988}, which covers a wide range of morphological types from Sa-Sm and with gas rotation curve measurements from optical emission lines \citep{Rubin1980,Rubin1982,Rubin1985} and \ion{H}{i} measurements (various literature sources).
The rotation velocity was extracted from these rotation curves, which extend to several effective radii, at the point of flattening.
No velocity dispersion measurements were presented for the spiral galaxies in this sample and so we assumed that the galaxies are rotation-dominated.
We corrected the stellar masses presented in \cite{Romanowsky2012}, computed by assuming a fixed mass-to-light ratio for all spiral galaxies, using Equation 1 of \cite{Fall2013}, which asserts that the mass-to-light ratio is a function of extinction corrected $({\it B-V})_{0}$ colour.
The $({\it B-V})_{0}$ colours were taken from the HYPERLEDA catalogue \citep{Paturel2003}, in which 7 of the 64 spiral galaxies do not have a measurement, leaving a reference sample size of 57 galaxies.
The median stellar mass for this reference sample is $\textrm{log} (M_{\star}/M_{\odot})=10.8$.

\subsubsection{Williams et al. 2009 early-type spirals (${\it z}\simeq0$)}\label{subsubsec:williams_spirals}
\cite{Williams2009a,Williams2010} described kinematic measurements of 14 local early-type spiral galaxies (Sa,Sb) and 14 S0 type galaxies.
We focus solely on the early-type spirals as a comparison sample throughout this work.
We adopted the gas velocity, available for 10/14 galaxies, derived from fitting the flat region of the [\ion{N}{ii}] position-velocity diagram and giving values which are $\simeq-0.10$ dex lower on average than the stellar velocities, leaving 10 early-type spiral galaxies for analysis.
Velocity dispersion measurements were not presented for these galaxies and so we do not compute or plot $V_{\textrm{tot}}$ and assumed the galaxies that form this comparison sample are dominated by ordered rotation.
The {\it Ks}-band mass-to-light ratio was computed for each galaxy by leaving it as a free parameter in the dynamical modelling, and was then used to find the stellar mass of the galaxies.
The mean mass-to-light ratio adopted in \cite{Williams2009a} is a factor of 1.71 higher than for the early-type spirals in \cite{Romanowsky2012}.
We applied this mean correction factor to the stellar masses presented in \cite{Williams2010} for consistency with \cite{Romanowsky2012}, and hence the other comparison samples.
After applying this correction the median stellar mass for this sample is $\textrm{log}(M_{\star}/M_{\odot})=11.0$.

\subsubsection{Bell and de Jong 2001 spiral galaxies (${\it z}\simeq0$)}\label{subsubsec:Bell_2001}
\cite{Bell2001} used a sample of local spiral galaxies to examine the stellar-mass Tully-Fisher relation, with stellar masses computed using colour-dependent mass-to-light ratios and a `diet Salpeter' IMF.
This resulted in stellar masses $\simeq0.08$ dex larger than the Chabrier IMF \citep{Cresci2009}.
The velocities were taken from \cite{Verheijen1997}, which used the flat part of the \ion{H}{i} rotation curve for galaxies in the Ursa Major Cluster. 
The stellar-mass Tully-Fisher relation presented in \cite{Bell2001}: $\textrm{log}({\it V_{\textrm{C}}}) = (2.159 \pm 0.009) + (0.222 \pm 0.013)[\textrm{log}({\it M_{\star}}) - 10.1]$ (having converted to the formalism used in this study and added $+0.024$ dex to the velocity zero-point in the conversion from diet Salpeter to Chabrier IMF), has been used frequently as a local comparison relation throughout the literature.
This is in good agreement with our fit to the \cite{Romanowsky2012} spirals and we make use of this relation in Fig.\,\ref{fig:rom_tf_relation}, which presents our comparison of local reference relations.

\subsubsection{Pizagno et al. 2005 SDSS spirals (${\it z}\simeq0$)}\label{subsubsec:pizagno_2005}
Another popular local reference sample is \cite{Pizagno2005}, in which the stellar-mass Tully-Fisher relation was fitted to a collection of 81 spiral galaxies from the Sloan Digital Sky Survey (SDSS), with stellar masses measured using colour dependent mass-to-light ratios following \cite{Bell2003}, assuming a diet Salpeter IMF.
The rotation velocities were extracted from the velocity profiles at $2.2R_{\textrm{d}}$.
We converted their fit result to our formalism, using the Chabrier IMF, giving $\textrm{log}({\it V_{\textrm{C}}}) = (2.130 \pm 0.033) + (0.328 \pm 0.013)[\textrm{log}({\it M_{\star}}) - 10.1]$, which has slope marginally steeper than our fit to the \cite{Reyes2011} spirals.
We have used the tabulated data in \cite{Pizagno2005}, with stellar masses corrected to a Chabrier IMF, to refit the stellar-mass Tully-Fisher relation with slope and zero-point free to vary, and found the best fit $\textrm{log}({\it V_{\textrm{C}}}) = (2.136 \pm 0.007) + (0.283 \pm 0.011)[\textrm{log}({\it M_{\star}}) - 10.1]$.
This is in slight tension with the original fit, which has a steeper slope, and in better agreement with the \cite{Reyes2011} reference relation (of which the data in \cite{Pizagno2005} is a subset). \\

\subsubsection{Hammer et al. 2007, Puech et al. 2008 (${\it z}\simeq0$)}\label{subsubsec:hammer_steep_slope}
In \cite{Hammer2007}, the {\it K}-band Tully-Fisher relation was constructed for a compilation of galaxies from 3 different studies \citep{Courteau1997,Verheijen2001,Pizagno2007} containing spiral galaxies with Hubble types Sa-Sm.
This was converted to the stellar-mass Tully-Fisher in \cite{Puech2008} by applying colour dependent mass-to-light ratios as per \cite{Bell2003}, with a diet Salpeter IMF.
The best-fit relation, after correcting to a Chabrier IMF, is $\textrm{log}({\it V_{\textrm{C}}}) = 2.038 + 0.357[\textrm{log}({\it M_{\star}}) - 10.1]$ which again has a significantly steeper slope than in \cite{Bell2001}.
Whilst carrying out this fit, the authors restricted themselves to the high-mass end of the Tully-Fisher relations by only considering galaxies with $\textrm{log}({\it V_{\textrm{C}}}) > 2.2$.
It is possible that this resulted in a steeper slope in the velocity versus stellar-mass plane, since a narrow mass range provides little constraint on the slope of the fit. \\

\subsubsection{Summary of local comparison samples}

\begin{figure*}
    \centering \hspace{-1.65cm}
    \begin{subfigure}[h!]{0.48\textwidth}
        \centering
        \includegraphics[height=5in]{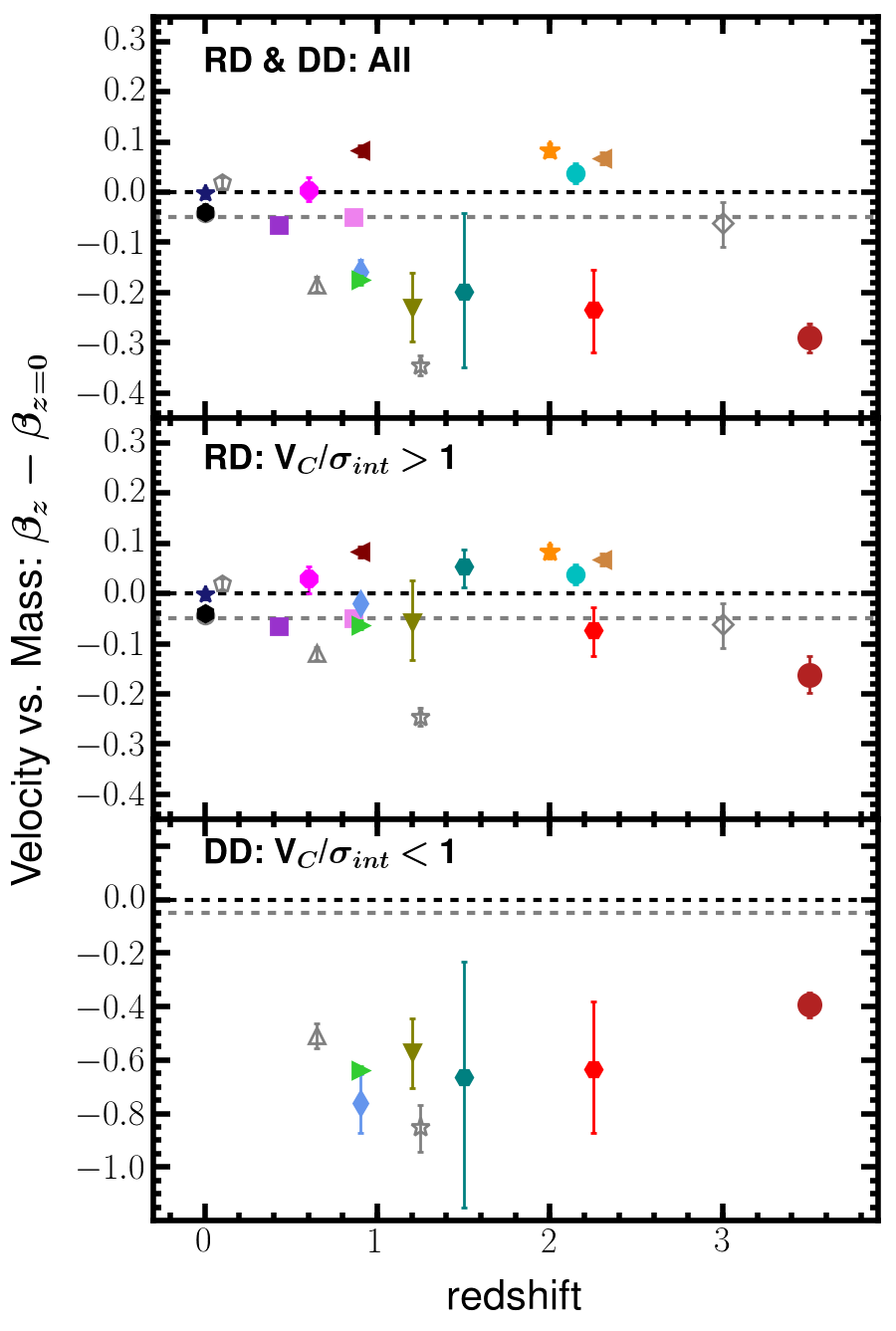}
    \end{subfigure} \hspace{+0.45cm}
    \begin{subfigure}[h!]{0.48\textwidth}
        \centering
        \includegraphics[height=5in]{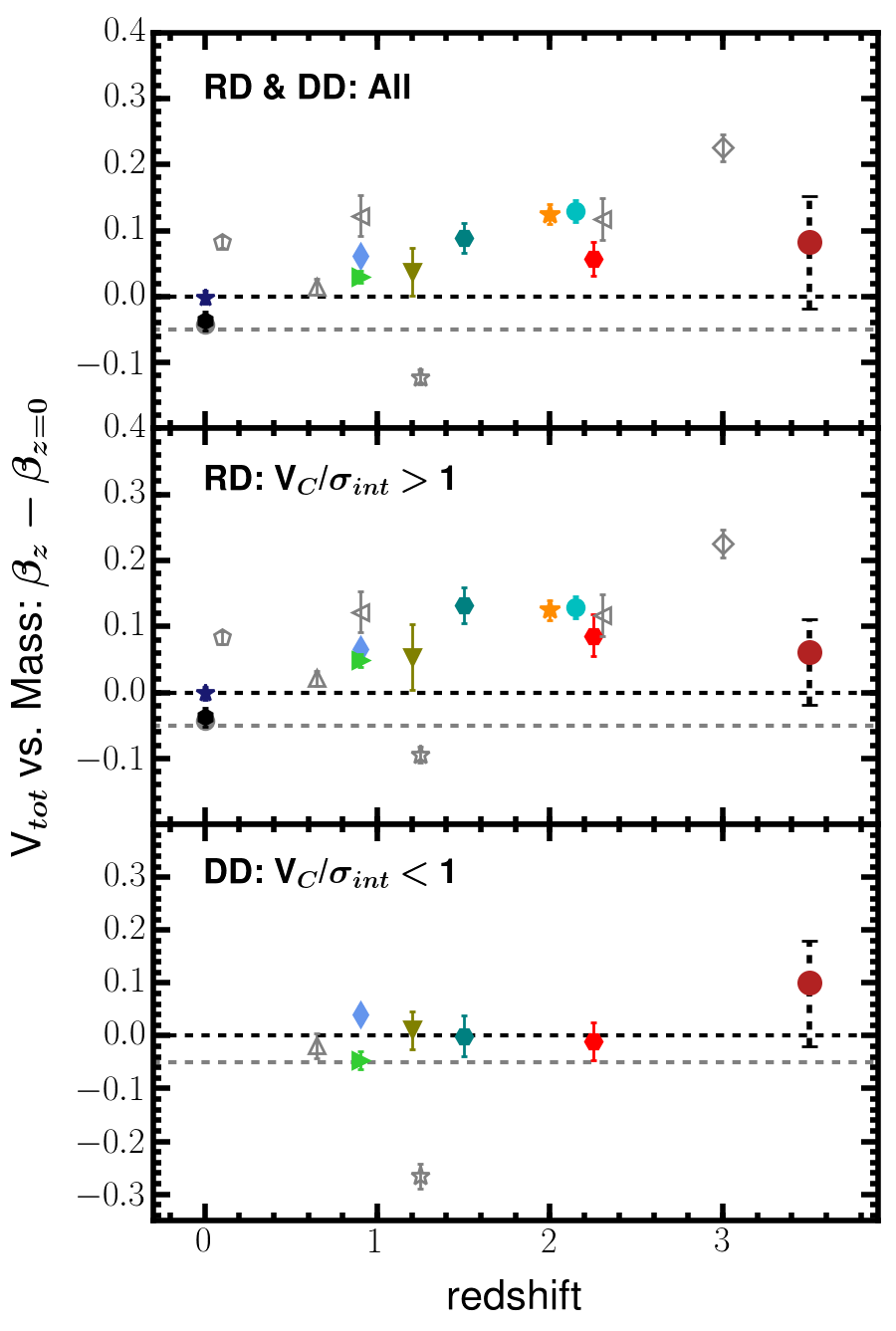}
    \end{subfigure}
    \caption{{\it Left:} The stellar-mass Tully-Fisher offsets against redshift, as per Fig.\,\ref{fig:velocity_evolution}, using a local stellar-mass Tully-Fisher relation defined from fitting the \protect\cite{Romanowsky2012} galaxies.
    This has both a shallower slope and a higher velocity zero-point than the \protect\cite{Reyes2011} relation and has almost the same functional form as the commonly adopted \protect\cite{Bell2001} relation.
    The result when adopting this relation is a shift towards lower velocity zero-point offsets throughout the comparison samples by $\simeq-0.05$ dex, although the trends remain unchanged.
    We show the approximate position of the \protect\cite{Reyes2011} zero-point with the grey-dashed horizontal line in each of the three panels.
    {\it Right:} The total-velocity versus stellar-mass offsets against redshift, as per Fig.\,\ref{fig:vtot_evolution}, using the local relation from fitting the \protect\cite{Romanowsky2012} galaxies.
    Using this local relation shifts the total-velocity zero-points downwards by approximately $\simeq-0.05$ dex, although the trend for all comparison samples to be offset to higher total velocities at fixed stellar mass is still observed.
    The position of the \protect\cite{Reyes2011} zero-point is again shown with the grey-dashed horizontal line and the black-dashed error bar on the KDS datapoint indicates the range of total-velocity zero-point offset differences found when varying the $\eta$ parameter over the range $2.0 < \eta < 6.0$ as per Fig.\,\ref{fig:vtot_evolution}.}
    \label{fig:rom_slope_vel_evolution}
\end{figure*}

In Fig.\,\ref{fig:rom_tf_relation} we plot the stellar-mass Tully-Fisher relations recovered from fitting the data in \cite{Pizagno2005,Reyes2011,Romanowsky2012}, as well as the literature relations from \cite{Bell2001,Pizagno2005,Puech2008} in an attempt to understand any discrepancy between these.
We plot also the data from \cite{Romanowsky2012} with the blue stars and the data from \cite{Reyes2011} with the red squares.
Generally these are in good agreement over the range $10.0 < \textrm{log} (M_{\star}/M_{\odot}) < 11.0$, where the density of the reference galaxies is highest, but diverge when extrapolated to lower stellar masses as a result of differences in the slope of the relations.
The chosen local reference relation will therefore impact the inferred evolution of the stellar-mass Tully-Fisher relation when carrying out the fixed-slope fitting procedure described in \cref{subsec:vc_m_evolution}.
Throughout this work we used the fit to the \cite{Reyes2011} data as our local reference relation, which has slope intermediate between the two extremes explored in Fig.\,\ref{fig:rom_tf_relation}.
We also carried out the analysis of \cref{subsec:vc_m_evolution} using the fit to the \cite{Romanowsky2012} data as our local comparison sample (which is almost equivalent to the \cite{Bell2001} relation), and found the same evolutionary trends but with the velocity zero-points shifted by roughly $-0.05$ dex towards lower values due to a combination of the shallower slope and higher velocity normalisation in the \cite{Romanowsky2012} galaxies.
In Fig.\,\ref{fig:rom_slope_vel_evolution} we plot the velocity and total-velocity offsets against redshift (analogous to Figs \ref{fig:velocity_evolution} and \ref{fig:vtot_evolution}) recovered from adopting the \cite{Romanowsky2012} stellar-mass Tully-Fisher relation as a local reference.
This $-0.05$ dex shift is a significant fraction of the evolution of the stellar-mass Tully-Fisher relation inferred for the `diskiest' galaxies in \cref{subsubsec:discussion_v_over_sigma_cuts} ($+0.12$ dex) and if the \cite{Romanowsky2012} relation was adopted we would infer that there had been less evolution out to $z\sim4$.
We stress however that the same evolutionary trends are recovered in our test of the two different reference relations. \\

\subsection{Distant comparison samples}\label{subsec:distant_comparison}
We now list the details of the distant comparison samples, noting the methods used to extract the kinematic parameters and the number of galaxies in each parent sample. 
We discuss the properties which are deemed not comparable to the other samples (grey hollow symbols throughout \cref{sec:Tully-Fisher-relation} and \cref{sec:sigma_contribution} and grey cells in Table \ref{tab:fit_results}).
In some of the studies from which the data have been drawn, the authors make specific reference to the evolution of the stellar-mass Tully-Fisher relation.
We compare the conclusions presented in the original studies with those that we find here using the same data.  

\subsubsection{DYNAMO (${\it z}\simeq0.1$)}\label{subsubsec:dynamo_comp}
\cite{Green2014} described H$\upalpha$ IFU measurements of 67 star-forming galaxies, with half the sample selected as ${\it z}\simeq2$ analogues on the basis of high specific star-formation rates and gas fractions.
For this reason, the sample does not represent `typical' star-forming galaxies at ${\it z}\simeq0.1$ and we use hollow symbols for the DYNAMO data points throughout the figures in \cref{sec:Tully-Fisher-relation} and \cref{sec:sigma_contribution}.
The velocity values were extracted at a fixed radius from an arctangent model fitted to the velocity fields (2.7$R_{\textrm{d}}$, where $R_{\textrm{d}}$ is the exponential scale radius, here measured in the r-band) and the velocity dispersion was taken as the luminosity-weighted average of the beam-smearing corrected dispersion field.
The galaxies were classified using the scheme described in \cite{Flores2006}.
We removed galaxies flagged to have `complex kinematics' and those with $V_{\textrm{C}}/\sigma_{\textrm{int}} < 1$, to leave only the rotation-dominated subsample of 51 galaxies.
The stellar masses were computed in \cite{Kauffmann2003}, using model mass-to-light ratios and dust corrected {\it z}-band luminosities, with the sample having median stellar mass of $\textrm{log}(M_{\star}/M_{\odot})=10.3$.
Throughout \cite{Green2014} the stellar-mass Tully-Fisher relation was explored, finding an evolution of $\simeq +0.09$ dex (no error) in velocity zero-point towards higher velocities at fixed stellar mass in comparison to the local relation of \cite{Bell2001}.
This is slightly larger than the offset of $+0.07\pm0.01$ dex in velocity zero-point we find in comparison to the \cite{Reyes2011} reference relation and significantly larger than the evolution inferred in comparison to the \cite{Romanowsky2012} relation (which is close to the \citealt{Bell2001} relation, see Fig. \ref{fig:rom_tf_relation}).
This suggests some tension between the fit results for the DYNAMO sample presented in \cite{Green2014} and the results inferred from the same data throughout this study.
However, in both the original results and from the fitting analysis in this paper, there appears to be evolution of the stellar-mass Tully-Fisher relation in the DYNAMO galaxies at ${\it z} \simeq 0.1$ (roughly 1 Gyr in the past).
This is possibly a consequence of the sample selection, in which many of the galaxies in the sample have unusually high gas content, and hence a large ratio of dynamical to stellar mass, for this redshift. 

\subsubsection{AHDG (${\it z}\simeq0.4$ and ${\it z}\simeq0.8$)}\label{subsubsec:ahdg_low_comp}
In this study, \cite{Miller2011} described photometric and kinematic measurements of 129 disk-like galaxies with a broad morphological selection criteria that included irregular systems, systems which displayed signs of interactions and bulge-dominated disks.
The broad selection criteria was chosen to reduce potential bias towards selecting only symmetric spirals, which represent the end point of isolated evolution.
In this work we split the full sample into a low redshift subsample with $0.2 < {\it z} < 0.6$, with median ${\it z}=0.4$ and containing 67 galaxies and a high-redshift subsample with $0.6 < {\it z} < 1.3$, with median ${\it z}=0.8$ and containing 62 galaxies.
Velocity measurements were computed from Keck~{\sc II} DEIMOS spectra, using an arctangent fit to the position-velocity diagrams of various optical emission lines and extracting at a fiducial radius of 2.2$R_{\textrm{d}}$.
No velocity dispersion measurements were made and so again we do not measure $V_{\textrm{tot}}$ and assumed, given the disk-like selection criteria, that all galaxies in both redshift ranges are dominated by ordered rotation.
Stellar masses were computed from SED fitting using a Chabrier IMF, with the authors then using the mass enclosed within 2.2$R_{\textrm{d}}$.
The remainder of the comparison samples used the total stellar mass whilst fitting the stellar-mass Tully-Fisher relation, and so we added $+0.187$ dex (as stated in \citealt{Miller2012}) to the quoted stellar masses to account for this. 
Following this correction, the median stellar mass values are $\textrm{log}(M_{\star}/M_{\odot})=10.0$ and $\textrm{log}   (M_{\star}/M_{\odot})=10.3$ for the low and high-redshift subsamples respectively. \\

Throughout \cite{Miller2011,Miller2012} the stellar-mass Tully-Fisher relation was explored in comparison to local relations presented in \cite{Bell2001} and \cite{Pizagno2005}, finding no evidence for evolution in the zero-point of the fit to the data.
This is consistent with the lack of evolution found in our analysis for the same data.
In \cite{Miller2011,Miller2012} the authors did not attempt to pick out ${\it z}=0$ disk analogues by applying stricter sample-selection criteria and so the star-forming galaxies are likely to be representative of the evolving-disk population at the redshifts studied.
The broad sample selection criteria and subsequent lack of inferred evolution in the stellar-mass Tully-Fisher relation for these galaxies is consistent with the findings in \cref{subsubsec:discussion_v_over_sigma_cuts}, despite the lack of reported velocity dispersions preventing us from plotting datapoints in Figs \ref{fig:parent_fraction_offsets} and \ref{fig:offsets} to show this explicitly.  

\subsubsection{IMAGES (${\it z}\simeq0.6$)}\label{subsubsec:images_comp}

In \cite{Puech2008} the authors evaluated the stellar-mass Tully-Fisher relation using 63 star-forming galaxies from the IMAGES large programme \cite{Yang2008}, which made use of the GIRAFFE integral-field spectrograph.
The velocity measurements were made by fitting PSF-convolved thin-disk models with arctangent velocity fields to the observed [\ion{O}{ii}] velocity fields, and extracting at the flat region of the rotation curve.
Velocity dispersion values were not reported in \cite{Puech2008} or in the companion IMAGES survey papers \citep[e.g.][]{Neichel2008,Yang2008}.
The authors provided a kinematic classification for each galaxy following \cite{Flores2006}, and we removed all those with `complex kinematics' to leave a sample of 28 galaxies.
We also constructed a sample containing 14 galaxies that are classed as `rotation-dominated', plotted in the $V_{\textrm{C}}/\sigma_{\textrm{int}} > 1$ panels throughout \cref{sec:Tully-Fisher-relation} and \cref{sec:sigma_contribution}, although there is only a very small difference in the velocity zero-point offsets between these and the sample of 28 galaxies that also contain `perturbed rotators'. 
Stellar masses were computed using {\it K}-band luminosities and a colour dependent mass-to-light ratio following the method described in \cite{Bell2003}.
This method assumed a diet Salpeter IMF, which we corrected to Chabrier by reducing the stellar masses by a factor 1.19 \citep{Cresci2009}.
The median stellar mass value amongst the full sample is $\textrm{log} (M_{\star}/M_{\odot})=10.3$.
In \cite{Puech2008} the authors studied the stellar-mass Tully-Fisher relation and found an evolution of $+0.13\pm0.06$ dex in velocity zero-point in comparison with the local relation described in Appendix \ref{subsubsec:hammer_steep_slope}.
When fitting the same data with the slope fixed to the shallower value found in the \cite{Reyes2011} relation, which also has higher velocity zero-point, we found an evolution of $+0.04\pm0.03$ dex.
This highlights the importance of the choice of local reference relation in determining the extent of the evolution, which is much less extreme in comparison to the \cite{Reyes2011} relation.
The galaxies to which the relation is fitted are the `diskiest' in the \cite{Puech2008} sample.
Observing evolution in the relation when fitting these galaxies is therefore consistent with the discussion throughout \cref{subsubsec:discussion_v_over_sigma_cuts}, although we cannot plot the \cite{Puech2008} data point in Figs \ref{fig:parent_fraction_offsets} and \ref{fig:offsets} to show this explicitly due to the lack of reported velocity dispersions. 

\subsubsection{MUSE+KMOS - MKS (${\it z}\simeq0.65$ and ${\it z}\simeq1.25$)}\label{subsubsec:mks_low_comp}
\cite{Swinbank2017} measured the kinematics of $\simeq150$ main-sequence star-forming galaxies spanning $0.28 < {\it z} < 1.65$ using Multi-Unit Spectroscopic Explorer (MUSE) observations of [\ion{O}{ii}] emitters and KMOS observations of H$\upalpha$ emitters. 
In this work we split the sample into two redshift ranges; the lower redshift subsample with $0.28 < {\it z} < 1.0$, median ${\it z}=0.65$ and containing 107 galaxies and a higher redshift subsample with $1.0 < {\it z} < 1.65$, median $z=1.25$ and containing 43 galaxies.
The velocity was measured by fitting disk+halo dynamical models to the observed data and extracting at $3.0R_{\textrm{d}}$.
These measurements have not had beam-smearing corrections applied and so throughout \cref{sec:Tully-Fisher-relation} and \cref{sec:sigma_contribution} the points are plotted with hollow symbols. 
The velocity dispersions were measured from the observed linewidths and corrected for beam-smearing following the method described in \cite{Johnson2017}.
We used the combination of the two measurements to construct rotation-dominated and dispersion-dominated subsamples in each redshift range on the basis of $V_{\textrm{C}}/\sigma_{\textrm{int}} > 1$.
Stellar masses were computed using the {\scriptsize HYPER-Z} SED fitting code with a Chabrier IMF, with a median stellar mass of $\textrm{log}  (M_{\star}/M_{\odot})=9.3$ across the lower redshift subsample and $\textrm{log}  (M_{\star}/M_{\odot})=9.9$ across the higher redshift subsample.
The stellar mass Tully-Fisher relation is not studied in \cite{Swinbank2017}. 

\subsubsection{HR-COSMOS (${\it z}\simeq0.9$)}\label{subsubsec:hr_cosmos_comp}
\cite{Pelliccia2017} described kinematic measurements for a subsample of 82 galaxies from the HR-COSMOS survey, which made use of slit spectroscopy using the Visible Multi-Object Spectrograph (VIMOS).
Various optical emission lines were targeted and velocity measurements were extracted by fitting PSF-convolved exponential disk, flat and arctangent velocity field models to the data.
The results from these different models are consistent within the uncertainties and in most cases the rotation velocity was extracted at $2.2R_{\textrm{d}}$ from the exponential disk model.
The velocity dispersion was also constrained as a parameter in the model, giving beam-smearing corrected measurements of both velocity and velocity dispersion for the sample.
This allowed for classification into rotation-dominated and dispersion-dominated subsamples.
Stellar masses were computed from SED fits to 30-band UV-IR photometry in the COSMOS field using the {\scriptsize LE PHARE} software \citep{Arnouts2002,Ilbert2006} with a Chabrier IMF.
The rotation-dominated subsample have $\textrm{log}(M_{\star}/M_{\odot})=10.2$, whilst the dispersion-dominated subsample have $\textrm{log}   (M_{\star}/M_{\odot})=9.7$.
Throughout \cite{Pelliccia2017} the authors studied the evolution of the stellar-mass Tully-Fisher relation over the range $0 < {\it z} < 1.2$ using several datasets from the literature, concluding that there is no significant evolution in their dataset or when studying all datasets together.
This is consistent with the results we find when fitting the HR-COSMOS data, which we conclude are representative of the evolving-disk population at ${\it z}\simeq0.9$ due to the lack of additional sample cuts which aim to isolate the most disky galaxies.  

\subsubsection{KROSS (${\it z}\simeq0.9$)}\label{subsubsec:kross_comp}
\cite{Stott2016} described the first kinematic measurements from the KMOS Redshift One Spectroscopic Survey, with the full sample and derived values presented in \cite{Harrison2017}.
Velocity and velocity dispersion measurements were made from IFU observations of the H$\upalpha$ emission line for $\simeq600$ galaxies as explained below, providing a very large and diverse statistical sample at ${\it z}\simeq1.0$.
Exponential disk models were fitted to the observed 2D velocity fields to provide a smoothly varying 1D profile, and measurements were extracted at $3.4R_{\textrm{d}}$.  
Velocity dispersion measurements were extracted either at the outskirts when the data extend to large enough radii, or using a median of the observed values when they do not.
The observed velocity and velocity dispersion measurements were then corrected for the effects of beam-smearing using the methods discussed in \cite{Johnson2017}.
We only used sources with quality flags 1-3, with no sign of AGN and with inclination angles $\theta_{\textrm{im}} > 25^{\circ}$ from the catalogue presented in \cite{Harrison2017}, in order to minimise the uncertainties on the kinematic parameters, which left a total of 475 galaxies.
Using the velocity and velocity dispersion measurements we defined rotation-dominated (382/475, 80 per cent) and dispersion-dominated (93/475, 20 per cent) subsamples.
Stellar masses were estimated using a fixed mass-to-light ratio applied to the {\it H}-band magnitudes, with median value $\textrm{log} (M_{\star}/M_{\odot})=10.1$ for the rotation-dominated subsample and $\textrm{log}   (M_{\star}/M_{\odot})=9.7$ for the dispersion-dominated subsample.
The analysis presented in \cite{Harrison2017} suggestsed that rotation-dominated galaxies ($V_{\textrm{C}}/\sigma_{\textrm{int}} > 1$) at ${\it z}\simeq0.9$ lie on the \cite{Reyes2011} stellar-mass Tully-Fisher relation.
This is consistent with the results we found when fitting the KROSS sample.
In \cite{Tiley2016}, an analysis of the stellar mass Tully-Fisher relation for the KROSS sample, it was reported that applying stricter $V_{\textrm{C}}/\sigma_{\textrm{int}}$ cuts led to inferred evolution towards higher velocities at fixed stellar mass in the stellar-mass Tully-Fisher relation.
We reproduced this trend throughout the analysis presented in this paper, interpreting these cuts as selecting galaxies that are more kinematically evolved and closer to tracing dynamical mass using rotation velocities alone.

\clearpage
    \thispagestyle{empty}
    \begin{landscape}

\begin{table}
\centering
\begin{threeparttable}
\caption{We list the zero-points, $\upbeta$, recovered from fitting the function $\textrm{log}({\it V}) = \upbeta + \upalpha[\textrm{log}({\it M_{\star}}) - 10.1]$ to the velocity versus stellar-mass and total-velocity versus stellar-mass relations for the local and distant comparison samples listed in Appendices \ref{subsec:local_comparison} and \ref{subsec:distant_comparison} respectively.
The slope, $\upalpha$, is held fixed throughout the fitting to the value 0.270 recovered from fitting the local galaxy sample presented in \protect\cite{Reyes2011}.
For reference, the slopes and zero-points of popular local comparison samples from the literature are presented in Appendix \ref{subsec:local_comparison}.
The grey cells show the zero-point parameters which are not deemed directly comparable to the others, as explained throughout Appendix \ref{subsec:distant_comparison}, and are correspondingly plotted with grey hollow symbols in Figs \ref{fig:velocity_evolution}, \ref{fig:parent_fraction_offsets}, \ref{fig:offsets}, \ref{fig:vtot_evolution} and \ref{fig:rom_slope_vel_evolution}.}
\label{tab:fit_results}
\begin{tabular}{*{5}{l}@{\hskip 0.2in}*{3}{c}@{\hskip 0.2in}*{3}{c}@{\hskip 0.2in}*{3}{c}} 

\toprule
\multirow{3}{*}{\textbf{Survey}} & \multirow{3}{*}{\textbf{z}} & \multicolumn{3}{c}{} & \multicolumn{3}{c}{} & \multicolumn{6}{c}{\textbf{$\upbeta$ values from fits to the data}} \\[1ex]
 &  & \multicolumn{3}{c}{\textbf{M$\bmath{_{\star}}$}} & \multicolumn{3}{c}{\textbf{Number}} & \multicolumn{3}{c}{\textbf{V$\bmath{_{\textrm{C}}}$ vs. M$\bmath{_{\star}}$}, $\upalpha=0.270$} & \multicolumn{3}{c}{\textbf{V$\bmath{_{\textrm{tot}}}$ vs. M$\bmath{_{\star}}$}, $\upalpha=0.270$} \\ \addlinespace
 &  & \textbf{All} & \textbf{RD} & \textbf{DD} & \textbf{All} & \textbf{RD} & \textbf{DD} & \textbf{All} & \textbf{RD} & \textbf{DD} & \textbf{All} & \textbf{RD} & \textbf{DD} \\ \addlinespace
 \midrule
\textbf{Local Samples} & & & & & & & & & & & & & \\
\midrule
Reyes+11 (REFERENCE) & 0.0 & 10.2 & 10.2 & - & 16 & 16 & 0 & $2.127\pm0.010$ & $2.127\pm0.010$ & - & - & - & -  \\[1ex]
Romanowsky+12 & 0.0 & 10.8 & 10.8 & - & 16 & 16 & 0 & $2.154\pm0.012$ & $2.154\pm0.012$ & - & - & - & -  \\[1ex]
Pizagno+05  & 0.0 & 10.3 & 10.3 & - & 81 & 81 & 0 & $2.126\pm0.009$ & $2.126\pm0.009$ & - & - & - & -  \\[1ex]
Williams+10 & 0.0 & 11.2 & 11.2 & - & 10 & 10 & 0 & $2.104\pm0.015$ & $2.104\pm0.015$ & - & - & - & -  \\[1ex]
\midrule
\textbf{Distant Samples} & & & & & & & & & & & & &  \\
\midrule
DYNAMO & 0.1 & 10.3 & 10.3 & - & 51 & 51 & 0 & \cellcolor{lightgray}$2.191\pm0.011$ & \cellcolor{lightgray}$2.191\pm0.011$ & - & \cellcolor{lightgray}$2.254\pm0.010$ & \cellcolor{lightgray}$2.254\pm0.010$ & -  \\[1ex]
AHDG (low-z) & 0.4 & 10.0 & 10.0 & - & 67 & 67 & 0 & $2.111\pm0.012$ & $2.111\pm0.012$ & - & - & - & -  \\[1ex]
$^{*}$IMAGES & 0.6 & 10.3 & 10.3 & - & 28 & 14 & 0 & $2.170\pm0.023$ & $2.192\pm0.028$ & - & - & - & -  \\[1ex]
MKS (low-z) & 0.7 & 9.3 & 9.4 & 8.8 & 106 & 86 & 20 & \cellcolor{lightgray}$2.015\pm0.016$ & \cellcolor{lightgray}$2.075\pm0.014$ & \cellcolor{lightgray}$1.720\pm0.045$ & \cellcolor{lightgray}$2.214\pm0.012$ & \cellcolor{lightgray}$2.216\pm0.012$ & \cellcolor{lightgray}$2.209\pm0.022$  \\[1ex]
AHDG (high-z) & 0.8 & 10.3 & 10.3 & - & 62 & 62 & 0 & $2.115\pm0.013$ & $2.115\pm0.013$ & - & - & - & -  \\[1ex]
HR-COSMOS & 0.9 & 10.1 & 10.2 & 9.7 & 80 & 65 & 15 & $2.015\pm0.023$ & $2.148\pm0.011$ & $1.428\pm0.111$ & $2.233\pm0.009$ & $2.235\pm0.010$ & $2.230\pm0.015$  \\[1ex]
KROSS & 0.9 & 10.0 & 10.1 & 9.7 & 475 & 328 & 93 & $2.002\pm0.008$ & $2.112\pm0.009$ & $1.549\pm0.013$ & $2.207\pm0.010$ & $2.223\pm0.010$ & $2.139\pm0.019$  \\[1ex]
KMOS$^{3D}$ (low-z) & 0.9 & 10.5 & 10.5 & - & 65 & 65 & 0 & $2.239\pm0.008$ & $2.239\pm0.008$ & - & \cellcolor{lightgray}2.249 & \cellcolor{lightgray}2.249 & -  \\[1ex]
MASSIV & 1.2 & 9.9 & 10.1 & 9.8 & 46 & 30 & 16 & $1.940\pm0.065$ & $2.119\pm0.076$ & $1.615\pm0.125$ & $2.212\pm0.036$ & $2.226^{+0.052}_{-0.047}$ & $2.188\pm0.037$  \\[1ex]
MKS (high-z) & 1.3 & 9.9 & 10.0 & 9.3 & 43 & 35 & 8 & \cellcolor{lightgray}$1.841\pm0.021$ & \cellcolor{lightgray}$1.938\pm0.017$ & \cellcolor{lightgray}$1.347^{+0.082}_{-0.094}$ & \cellcolor{lightgray}$2.064\pm0.012$ & \cellcolor{lightgray}$2.089\pm0.012$ & \cellcolor{lightgray}$1.930\pm0.025$  \\[1ex]
SIGMA (low-z) & 1.5 & 10.1 & 10.2 & 9.5 & 27 & 18 & 9 & $1.981\pm0.150$ & $2.221\pm0.037$ & $1.524^{+0.427}_{-0.498}$ & $2.267\pm0.025$ & $2.307\pm0.028$ & $2.187\pm0.039$  \\[1ex]
SINS & 2.0 & 10.6 & 10.6 & - & 16 & 16 & 0 & $2.242\pm0.014$ & $2.242\pm0.014$ & - & $2.283\pm0.016$ & $2.283\pm0.016$ & -  \\[1ex]
ZFIRE & 2.2 & 10.2 & 10.2 & - & 21 & 21 & 0 & $2.212\pm0.020$ & $2.212\pm0.020$ & - & $2.302\pm0.016$ & $2.302\pm0.016$ & -  \\[1ex]
SIGMA (high-z) & 2.3 & 10.0 & 10.2 & 10.0 & 17 & 12 & 5 & $1.924^{+0.081}_{-0.077}$ & $2.086\pm0.048$ & $1.540^{+0.247}_{-0.258}$ & $2.223\pm0.024$ & $2.246\pm0.033$ & $2.170\pm0.037$  \\[1ex]
KMOS$^{3D}$ (high-z) & 2.3 & 10.5 & 10.5 & - & 46 & 46 & 0 & $2.222\pm0.012$ & $2.222\pm0.012$ & - & \cellcolor{lightgray}2.199 & \cellcolor{lightgray}2.199 & -   \\[1ex]
AMAZE & 3.0 & 10.0 & 10.0 & - & 5 & 5 & 0 & \cellcolor{lightgray}$2.116\pm0.047$ & \cellcolor{lightgray}$2.116\pm0.047$ & - & \cellcolor{lightgray}$2.402\pm0.022$ & \cellcolor{lightgray}$2.402\pm0.022$ & - \\[1ex]
\textbf{KDS} & 3.5 & 9.8 & 9.8 & 9.7 & 29 & 13 & 16 & $1.900\pm0.030$ & $2.026\pm0.036$ & $1.790^{+0.042}_{-0.045}$ & $2.269\pm0.020$ & $2.245\pm0.033$ & $2.286\pm0.023$ \\[1ex]

\hline
\end{tabular}
\begin{tablenotes}
      \small
       \item $^{*}$ The categories `All' and `RD' in this case correspond to the combination of the perturbed rotators and rotation-dominated galaxies and the rotation-dominated galaxies on their own respectively (see Appendix \ref{subsubsec:images_comp}).
       None of the galaxies in these categories are classified as dispersion-dominated.
    \end{tablenotes}
  \end{threeparttable}
  \end{table}

      \end{landscape}
    \clearpage

\subsubsection{KMOS$^{3D}$ (${\it z}\simeq0.9$ and ${\it z}\simeq1.25$)}\label{subsubsec:kmos_3d_comp}
The KMOS$^{3D}$ survey presented in \cite{Wisnioski2015} described KMOS H$\upalpha$ observations of $\simeq600$ massive SFGs clustered around ${\it z}\simeq0.9$ and ${\it z}\simeq2.3$.
We made use of data presented in a recent, thorough study of the evolution of the stellar-mass Tully-Fisher relation over the range $0.9 < {\it z} < 2.3$ from \cite{Ubler2017}, using 316 KMOS$^{3D}$ galaxies with detected and spatially resolved H$\upalpha$ emission.
The dynamical modelling of the data was presented in \cite{Wuyts2016b}, in which exponential mass models were fitted simultaneously to one-dimensional extractions along the kinematic axis of the velocity and velocity dispersion fields.
The rotation velocity was extracted as the maximum of the model rotation curve.
Various cuts were made to the parent sample to remove merger candidates and to ensure high signal-to-noise and main-sequence sampling, leaving 240 galaxies.
Further cuts were made to remove galaxies where the peak velocity was not constrained, where the velocity dispersion peak did not coincide with the galaxy centre and where $V_{\textrm{C}}/\sigma_{\textrm{int}} < \sqrt{4.4}$, in order to build a Tully-Fisher sample of 65 galaxies at ${\it z}\simeq0.9$ and 46 at ${\it z}\simeq2.3$.
To calculate the parent fractions we assumed that the parent sample of 316 galaxies is divided evenly between the two redshift slices.  
The extracted stellar masses were computed following the procedure described in \cite{Wuyts2011a}, which uses a Chabrier IMF.
The median stellar mass in both redshift slices is $\textrm{log}   (M_{\star}/M_{\odot})=10.5$. \\

Since tabulated data was not provided, the rotation velocity values and stellar masses were extracted from Fig.\,6 of \cite{Ubler2017} using WebPlotDigitiser \citep{rohatgi17}.
In \cite{Ubler2017} the authors found velocity zero-point offsets, in comparison to the \cite{Reyes2011} local relation, of $+0.10$ dex and $+0.07$ dex for the ${\it z}\simeq0.9$ and ${\it z}\simeq2.3$ subsamples respectively (errors not quoted).
These are in agreement with the values $+0.11\pm0.01$ dex and $+0.09\pm0.01$ dex we found when fitting the values extracted from the plots.
The observed evolution from the local relation at both redshift slices is expected following the discussion of \cref{subsubsec:discussion_v_over_sigma_cuts}, since the sample selection described in \cite{Ubler2017} was designed to pick out the most `disky' galaxies with high $V_{\textrm{C}}/\sigma_{\textrm{int}}$ values. \\

Throughout \cite{Ubler2017} the authors also considered a {\it circular velocity}, which contains a velocity dispersion term, to account for the non-negligible contribution of pressure to the gravitational support of the systems.
The measured offsets from the local relation in the circular velocity versus stellar-mass plane increased to $+0.12$ dex at both redshift ranges.
We do not have access to velocity dispersion measurements for these galaxies and so cannot compute the total velocity in the same way as the other samples.
Consequently, we plot the quoted circular velocity offsets from \cite{Ubler2017} with grey-hollow symbols in Fig.\,\ref{fig:vtot_evolution}.

\subsubsection{MASSIV (${\it z}\simeq1.2$)}\label{subsubsec:massiv_comp}
The Mass Assembly Survey with SINFONI in VVDS (MASSIV) \citep{Vergani2012,Contini2012,Epinat2012} utilised the Spectrograph for INtegral-Field Observations in the Near Infrared (SINFONI) to collect H$\upalpha$ emission line observations for 46 star-forming galaxies over the range $0.9 < {\it z} < 1.6$.
The velocity and velocity dispersion measurements were extracted after fitting a PSF-convolved model arctangent function to the data to counter the effects of beam-smearing.
The rotation velocity was extracted at $\simeq3R_{\textrm{d}}$ from the intrinsic model and the velocity dispersion is taken as the average of the beam-smearing corrected velocity dispersion map.
We defined rotation-dominated and dispersion-dominated subsamples of 30 and 16 galaxies respectively using the ratio $V/\sigma_{\textrm{int}} > 1$.
The stellar masses were computed in \cite{Contini2012} using SED fits to the photometry with an assumed Salpeter IMF \citep{Salpeter1955}, which we converted to a Chabrier IMF by dividing by a factor 1.8.
The median stellar mass is $\textrm{log}   (M_{\star}/M_{\odot})=10.1$ in the rotation-dominated subsample and $\textrm{log}  (M_{\star}/M_{\odot})=9.8$ in the dispersion-dominated subsample.
In \cite{Vergani2012} the authors studied the evolution of the stellar-mass Tully-Fisher relation since ${\it z}\simeq1.2$ using the MASSIV sample, making comparisons with both the \cite{Bell2001} and \cite{Pizagno2007} local relations.
For the rotation-dominated MASSIV galaxies, velocity zero-point evolution of $+0.11$ dex was found in comparison to the \cite{Pizagno2007} local relation, whereas no evolution was found in comparison with the \cite{Bell2001} relation, consistent with our findings for the MASSIV sample.
This stresses further the importance of choosing a consistent reference relation when making comparisons between different intermediate and high-redshift studies.

\subsubsection{SIGMA (${\it z}\simeq1.5$ and ${\it z}\simeq2.25$)}\label{subsubsec:sigma_low_comp}
\cite{Simons2016} presented the Keck/MOSFIRE Survey in the near-Infrared of Galaxies with Multiple position Angles (SIGMA), which was a study of the internal kinematics of star-forming galaxies at ${\it z}\simeq2$.
We split the full sample into 27 galaxies with $1.3 < {\it z} < 1.8$ and 17 galaxies with $2.0 < {\it z} < 2.5$ with measured velocities and velocity dispersions.
These come from parent samples of 33 and 25 galaxies at the respective redshift intervals, assuming that 9 galaxies which were cut from the sample on the basis of emission line extent are divided equally between the two ranges.
The velocity and velocity dispersion were measured using {\scriptsize ROTCURVE} to fit the 2D longslit spectra (containing high S/N detections of either the H$\upalpha$ or [\ion{O}{iii}]$\lambda$5007 emission line), which takes into account the effects of beam-smearing.
Stellar masses were computed using SED fits with a Chabrier IMF with median values of $\textrm{log} (M_{\star}/M_{\odot})=10.1$ and $\textrm{log}(M_{\star}/M_{\odot})=10.0$ for the samples at lower and higher redshift respectively. \\

In \cite{Simons2016} the stellar-mass Tully-Fisher was explored, with the location of galaxies in the velocity versus stellar-mass plane found to depend strongly on the ratio of $V_{\textrm{C}}/\sigma_{\textrm{int}}$.
For 12 massive, rotation-dominated galaxies from the full sample (both redshift ranges) an evolution of $-0.44$ dex in $\textrm{log}   (M_{\star}/M_{\odot})$ zero-point offset was reported, in comparison to the \cite{Reyes2011} relation, corresponding to $+0.12$ dex in velocity zero-point offset.
After splitting the sample into two redshift ranges we found velocity zero-point offsets of $+0.09\pm0.04$ dex at ${\it z}\sim1.5$ and $-0.04\pm0.05$ dex at ${\it z}\sim2.25$, although there is a large degree of scatter in the velocity versus stellar-mass plane amongst the higher redshift galaxies.
Given the normalised parent fraction and the median $V_{\textrm{C}}/\sigma_{\textrm{int}}$ of the lower redshift sample, there is a discrepancy between the observed and expected stellar mass Tully-Fisher offsets in the right panels of Figs \ref{fig:parent_fraction_offsets} and \ref{fig:offsets}.
This suggests higher observed rotation velocities at fixed stellar mass relative to the other comparison samples. 

\subsubsection{SINS (${\it z}\simeq2.0$)}\label{subsubsec:sins_comp}
The Spectroscopic Imaging survey in the near-infrared (SINS) was presented in \cite{ForsterSchreiber2009}, describing SINFONI H$\upalpha$ observations of 80 massive star-forming galaxies.
\cite{Cresci2009} described dynamical modelling of a subsample of 18 galaxies from the SINS parent sample, selected due to the prominence of ordered rotational motions.   
We concentrated solely on this subsample, since the dynamical modelling included steps to correct the velocity and velocity dispersion fields for the effects of beam-smearing.
16 of the 18 galaxies have reliable stellar mass measurements and 11 have intrinsic velocity dispersion measurements provided in \cite{Cresci2009}, which we used as comparison samples in the velocity versus stellar-mass and total-velocity versus stellar-mass planes respectively. \\

The velocities were extracted in \cite{Cresci2009} as the best fit parameter from the IDL code {\scriptsize DYSMAL}, which derives rotation curves given an input radial mass distribution.
This fitting procedure also constrained the intrinsic velocity dispersions of the disks, corrected for beam-smearing and instrumental resolution effects.
Stellar masses were computed using SED fitting with a Chabrier IMF, with median value for the 16 galaxies of $\textrm{log}  (M_{\star}/M_{\odot}) = 10.6$.
In \cite{Cresci2009} the authors constructed the ${\it z}\simeq2$ stellar-mass Tully-Fisher relation and found a $\textrm{log}(M_{\star}/M_{\odot})$ zero-point offset of $+0.09$ dex in comparison to the \cite{Bell2001} local relation.
This is in agreement with the value of $+0.11\pm0.01$ dex that we found when fitting the \cite{Cresci2009} galaxies, consistent with evolution of the stellar-mass Tully-Fisher relation since ${\it z}\simeq2$.
In the context of the discussion throughout \cref{subsubsec:discussion_v_over_sigma_cuts}, we expected to find evolution for this sample due to the nature of the sample selection criteria, which aimed to isolate the highest $V_{\textrm{C}}/\sigma_{\textrm{int}}$ galaxies from the SINS sample.

\subsubsection{ZFIRE (${\it z}\simeq2.15$)}\label{subsubsec:zfire_comp}

In \cite{Straatman2017}, the authors described Keck/MOSFIRE longslit spectroscopic observations of 22 star-forming galaxies over the range $2.0 < {\it z} < 2.5$.
The initial sample consisted of 38 galaxies from which the 22 were selected as the best candidates for accurate kinematic modelling on the basis of the S/N of the emission lines.
Rotation velocities and velocity dispersions were measured by fitting an arctangent function to the spectra in 2D, assuming an exponential disk profile for the emission line intensity (which is either H$\upalpha$ or [\ion{O}{iii}]$\lambda$5007).
The fitting procedure included a beam-smearing correction so that the velocities, extracted at $2.2R_{\textrm{d}}$, and velocity dispersions used in fitting the stellar-mass Tully-Fisher relation were intrinsic properties.
Only one galaxy in the sample shows $V_{\textrm{C}}/\sigma_{\textrm{int}} < 1$, which we removed to focus solely on the 21 rotation-dominated galaxies.
The stellar masses were computed using the {\scriptsize FAST} SED fitting code with Chabrier IMF, with median value for the sample $\textrm{log}(M_{\star}/M_{\odot})=10.2$. 
The authors fitted the stellar-mass Tully-Fisher relation to these galaxies and recovered the best-fit relation $\textrm{log}({\it V_{\textrm{C}}}) = (2.20 \pm 0.05) + (0.193 \pm 0.108)[\textrm{log}({\it M_{\star}}) - 10.1]$ when allowing both the slope and zero-point to vary.
When fixing the slope to the value described in \cite{Reyes2011} the authors inferred a velocity zero-point evolution of $+0.07$ dex.
This is consistent with the value $+0.08\pm0.02$ dex that we find when fitting the same galaxies and consistent with a moderate evolution of the stellar-mass Tully-Fisher relation since ${\it z}\simeq2.15$.
Given that no special selection criteria have been applied, the ZFIRE datapoint is an outlier in the right panel of Fig.\,\ref{fig:parent_fraction_offsets}, since the normalised parent fraction for the sample is $\sim1$.
However, the observed Tully-Fisher velocity offset agrees with the expectation in the right panel of Fig.\,\ref{fig:offsets}, as the sample has high median $V_{\textrm{C}}/\sigma_{\textrm{int}}$ in comparison to the model prediction. 

\subsubsection{AMAZE (${\it z}\simeq3.0$)}\label{subsubsec:amaze_comp}
\cite{Gnerucci2011} presented the resolved dynamical properties of the Assessing the Mass-Abundance redshift Evolution (AMAZE) sample, using SINFONI [\ion{O}{iii}]$\lambda$5007 measurements for 33 galaxies.
11 of these galaxies were judged to be rotation-dominated by assessing the deviations from a planar fit to the position-velocity diagram, with the other 22 discarded in the analysis presented in \cite{Gnerucci2011}.
Rotation velocities and intrinsic velocity dispersions were measured by fitting model rotation curves, derived from exponential mass distributions, to the observed velocity fields, with the extracted $V_{\textrm{C}}$ value taken as the large radius limit of the rotation curve and the $\sigma_{\textrm{int}}$ as the maximum of the difference in quadrature between the $\sigma_{\textrm{obs}}$ map and the $\sigma_{\textrm{model}}$ map (which also takes into account instrumental resolution and beam-smearing; see their Equation 8).

As explained in \cite{Turner2017}, we constructed a `clean' sample of 5 galaxies that have constrained velocity measurements, and velocity dispersion measurements consistent with being greater than zero.
Stellar masses were derived from SED fitting using a Chabrier IMF, with median value for the clean sample of $\textrm{log}(M_{\star}/M_{\odot})=10.0$.
Due to the small sample size with large intrinsic scatter, we plot all comparison points for the AMAZE sample with hollow symbols.
In \cite{Gnerucci2011} the authors fitted the stellar-mass Tully-Fisher relation at ${\it z}\simeq3$ to the full sample of 11 rotation-dominated galaxies.
A velocity zero-point evolution of $+0.286$ dex is claimed in comparison to the \cite{Bell2001} relation, far higher than in any of the other comparison samples and requiring rapid evolution of the relation over the range $2 < {\it z} < 3$.
In our clean sample, many of the uncertain high-velocity galaxies are omitted and we find a velocity zero-point evolution of $-0.01$ dex, consistent with a picture in which the large velocity dispersions observed throughout this sample become a significant component of the dynamical mass budget.

\subsubsection{Summary of distant comparison samples}

\begin{figure}
\centering \hspace{-1.13cm}
\includegraphics[width=0.49\textwidth]{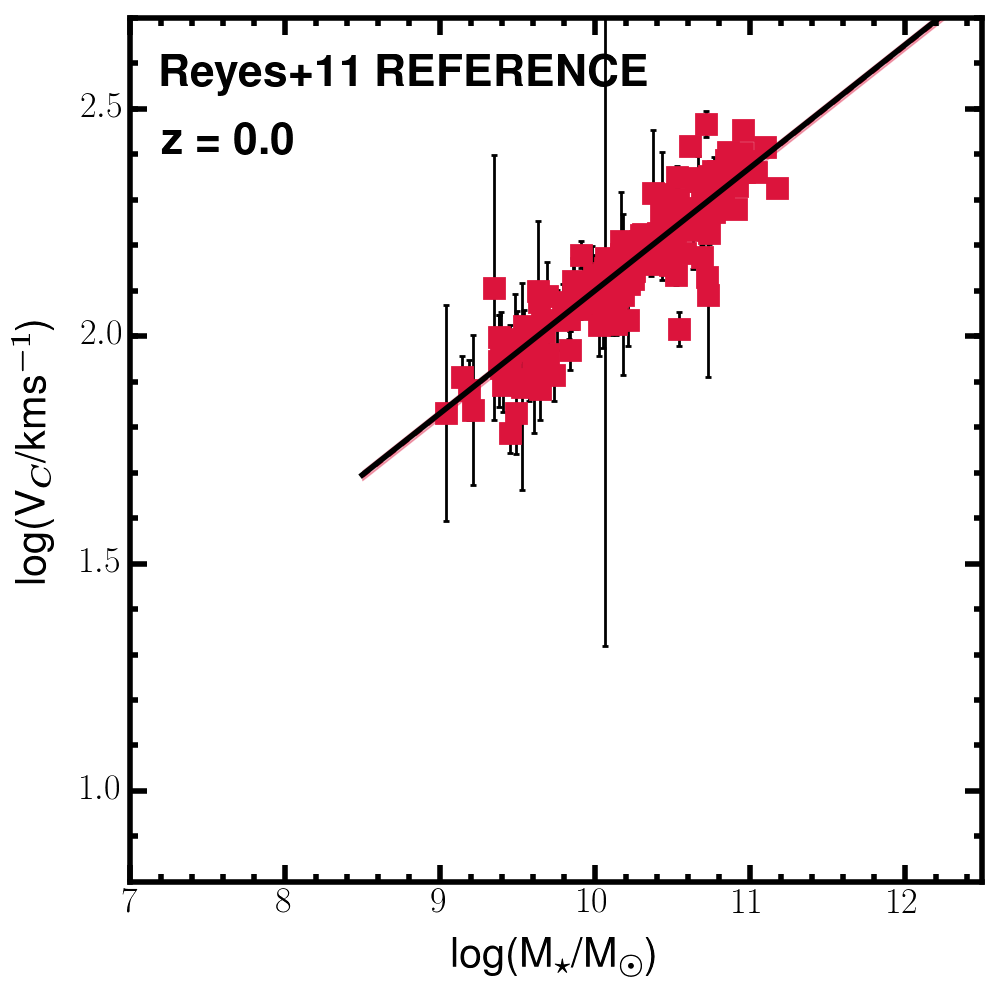}
\caption{As an enlarged example of the subplots shown throughout Figs \ref{fig:velocity_fits} and \ref{fig:v_tot_fits}, we plot the 189 spiral galaxies from \protect\cite{Reyes2011} with the solid red squares with best-fit relation $\textrm{log}({\it V_{\textrm{C}}}) = 2.127 + 0.270[\textrm{log}({\it M_{\star}}) - 10.1]$ (solid-black line).}
\label{fig:app_reyes_fit}
\end{figure}

We have endeavoured to collect a large number of star-forming galaxy samples with kinematic measurements, covering a wide range in physical properties, redshift and sample-selection criteria.
This is to provide an unbiased perspective of the evolution of the stellar mass Tully-Fisher relation over cosmic time. 
Given the numerous methodologies which have been followed in the different studies to compute and extract physical properties, there are unavoidable systematics associated with making use of published measurements.
Keeping this in mind as an important caveat, and to summarise the results of \cref{sec:Tully-Fisher-relation} and \cref{sec:sigma_contribution}, the normalisation of the Tully-Fisher relation does appear to evolve with cosmic time for the diskiest subsamples of galaxies, and the range of normalisation shifts quoted in the literature are quantifiable through an understanding of sample-selection criteria.
As the sample sizes and data quality continue to increase over the range $0 < z < 4$, it will be possible to further understand the physical processes which are driving this evolution.

\section{Fits to comparison sample data}\label{app:comparison_sample_fits}

\begin{figure*}
    \centering \hspace{-1.65cm}
    \begin{subfigure}[h!]{0.246\textwidth}
        \centering
        \includegraphics[height=1.825in]{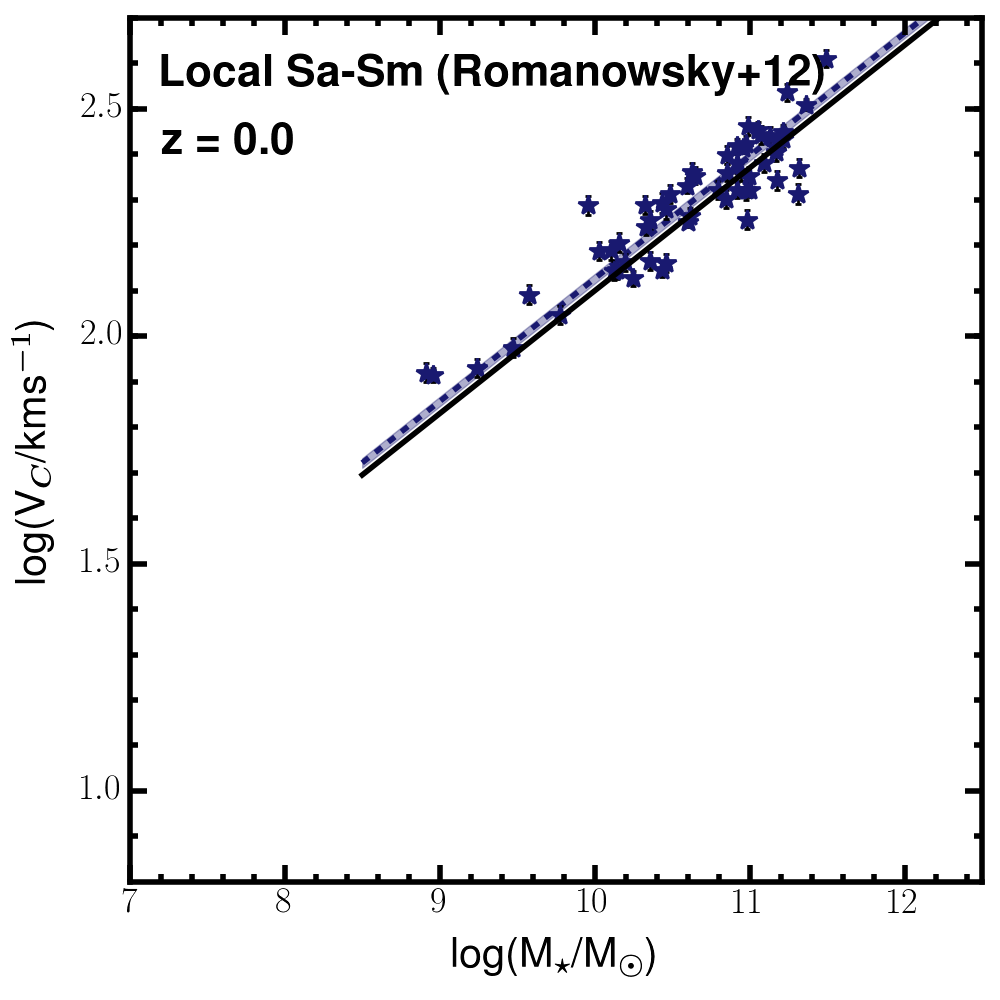}
    \end{subfigure} \hspace{0.15cm}
    \begin{subfigure}[h!]{0.246\textwidth}
        \centering
        \includegraphics[height=1.825in]{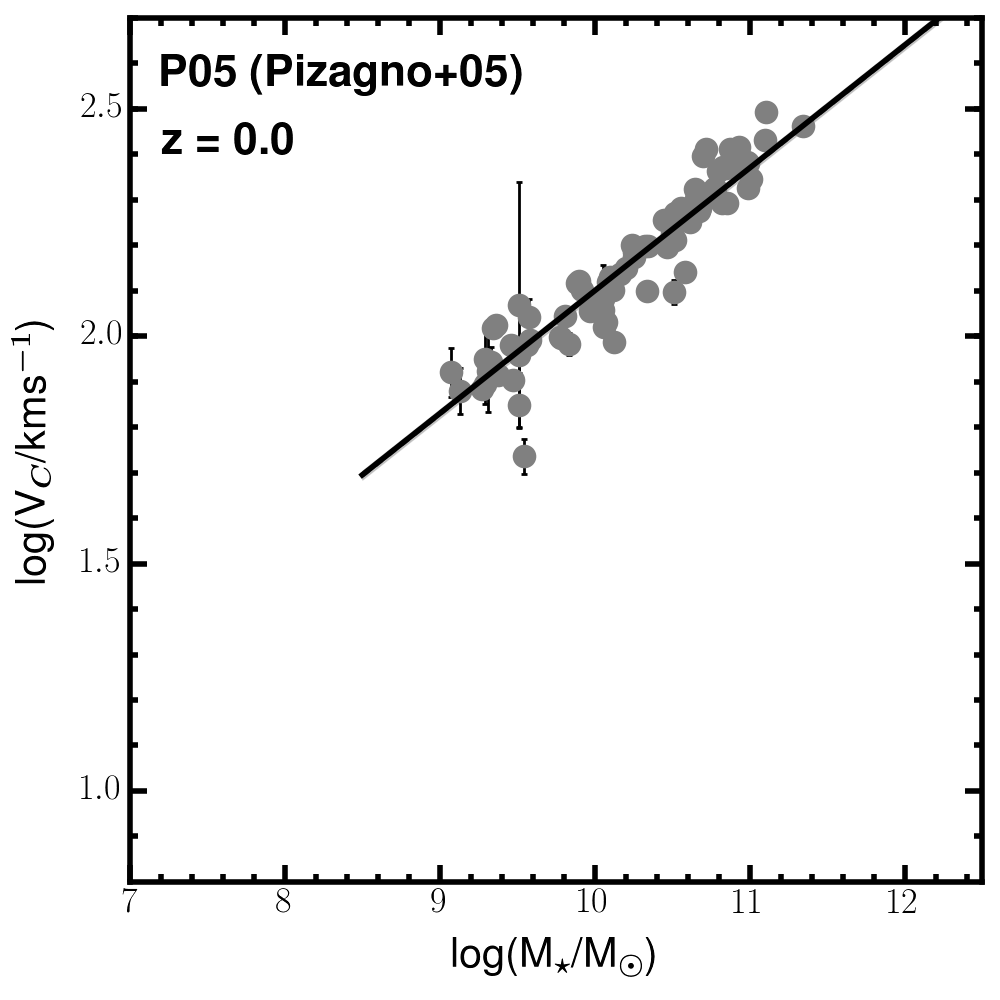}
    \end{subfigure} \hspace{0.15cm}
    \begin{subfigure}[h!]{0.246\textwidth}
        \centering
        \includegraphics[height=1.825in]{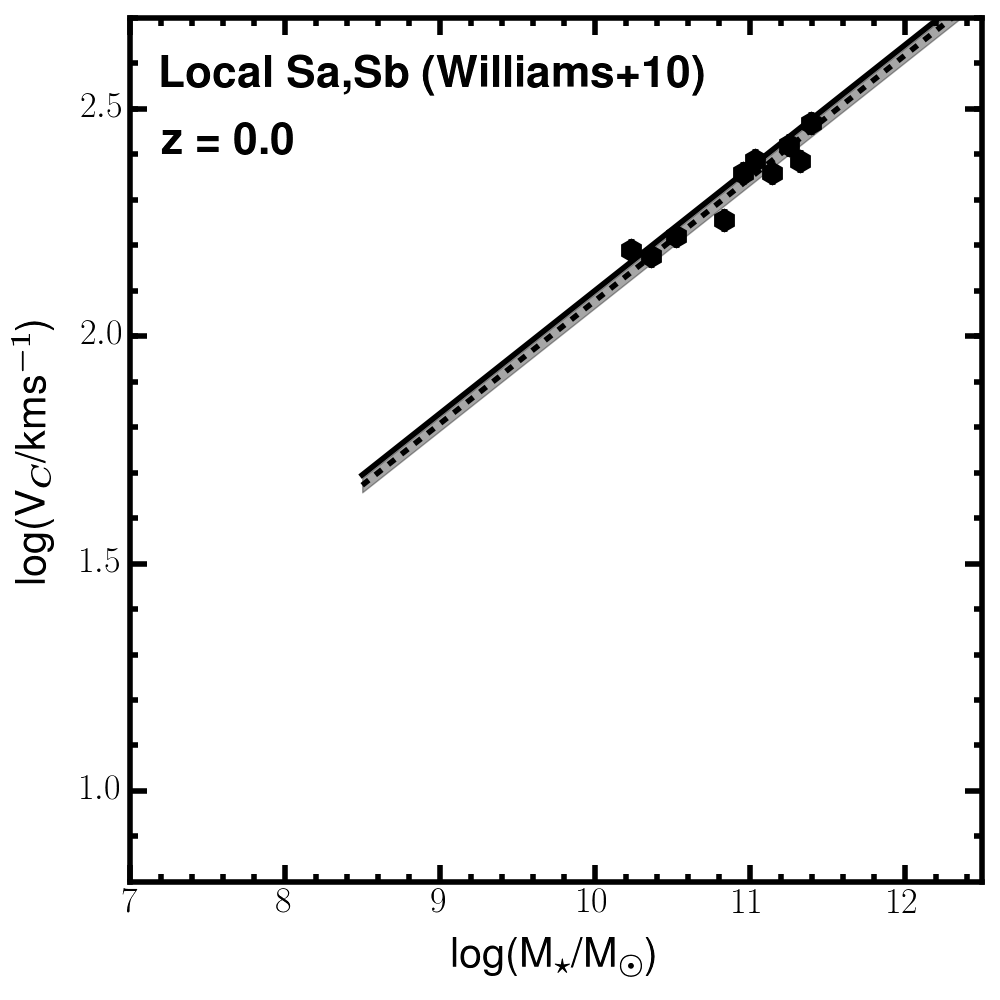}
    \end{subfigure} \hspace{0.15cm}
    \begin{subfigure}[h!]{0.246\textwidth}
        \centering
        \includegraphics[height=1.825in]{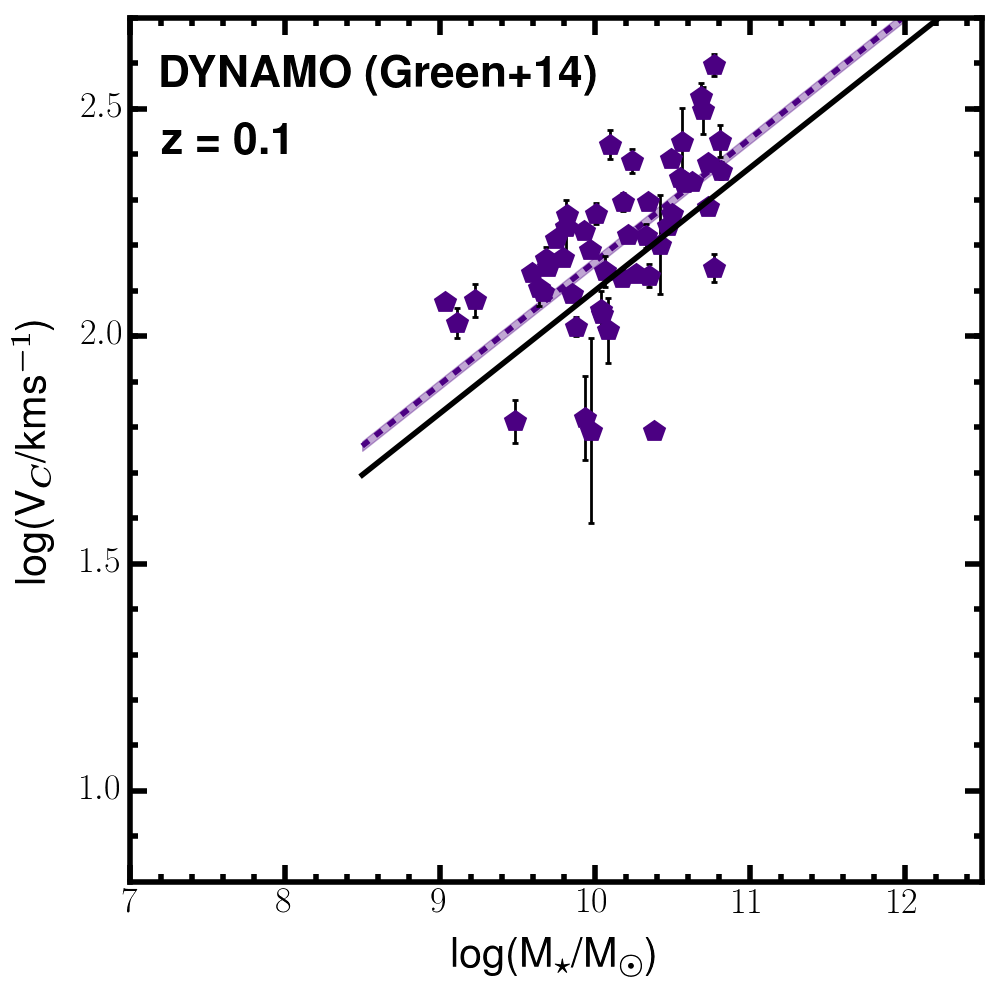}
    \end{subfigure} \hspace{0.15cm}
    \begin{subfigure}[h!]{0.246\textwidth}
        \centering \hspace{-2.85cm}
        \includegraphics[height=1.825in]{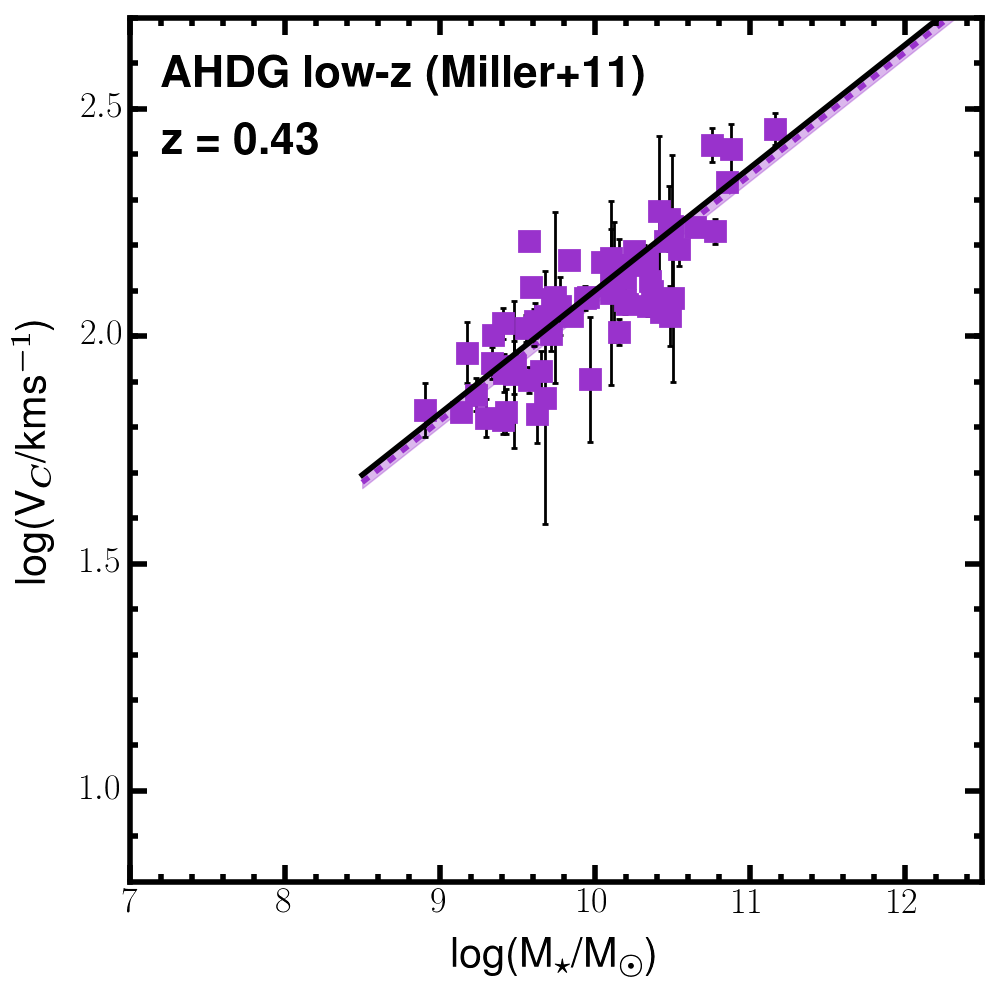}
    \end{subfigure} \hspace{-1.4cm}
    \begin{subfigure}[h!]{0.246\textwidth}
        \centering 
        \includegraphics[height=1.825in]{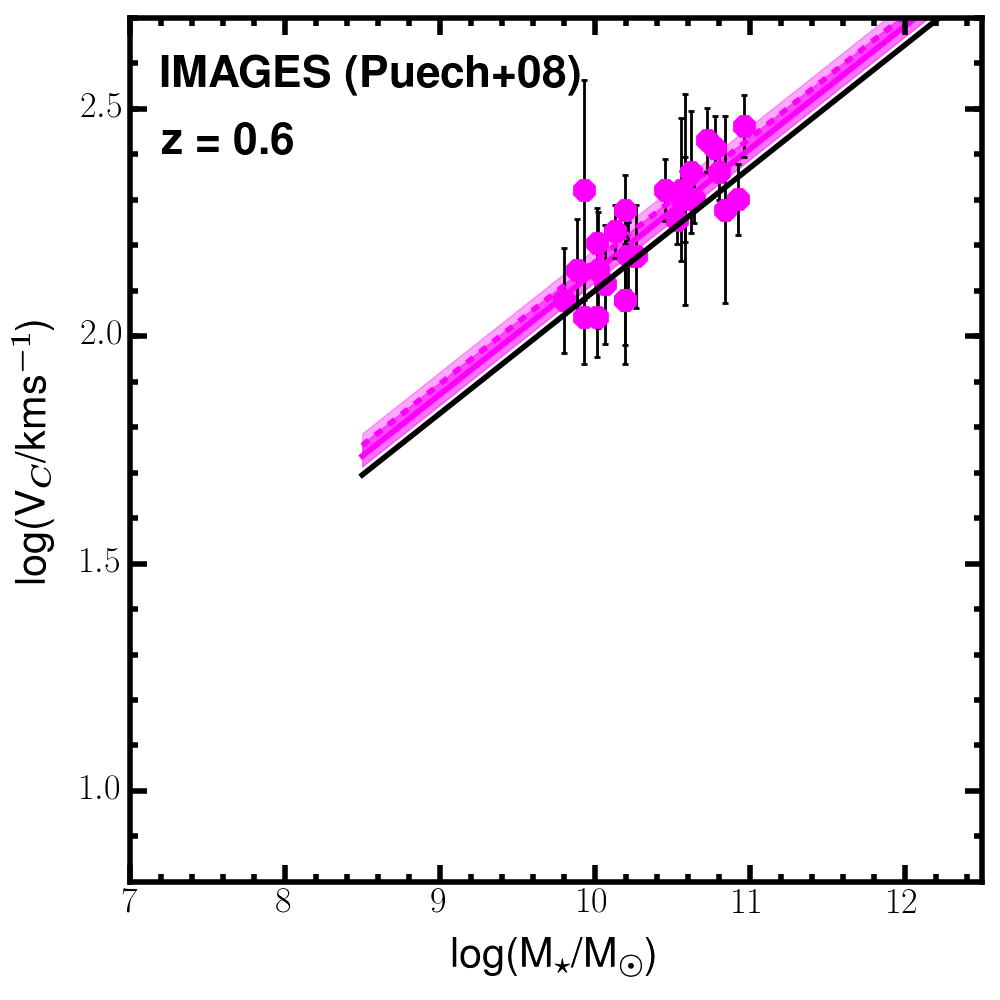}
    \end{subfigure}  \hspace{0.15cm}
    \begin{subfigure}[h!]{0.246\textwidth}
        \centering 
        \includegraphics[height=1.825in]{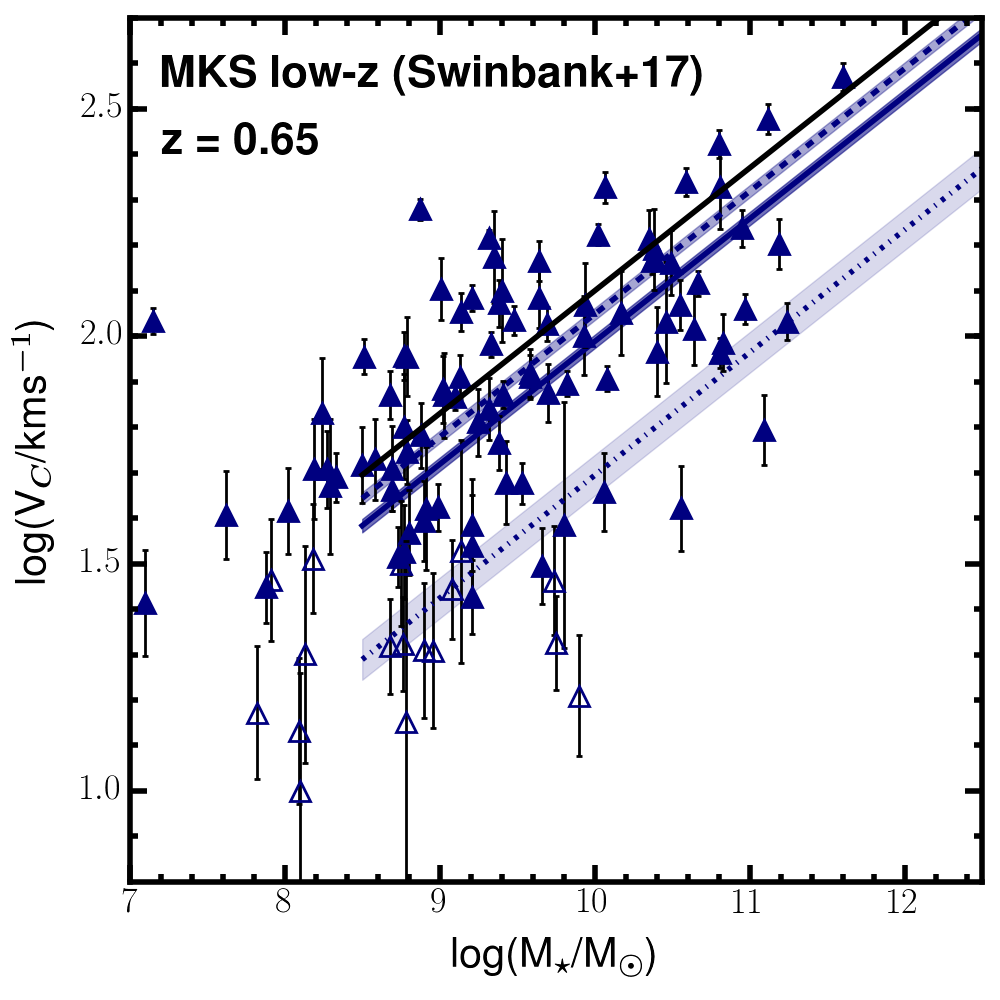}
    \end{subfigure} \hspace{0.15cm}
    \begin{subfigure}[h!]{0.246\textwidth}
        \centering 
        \includegraphics[height=1.825in]{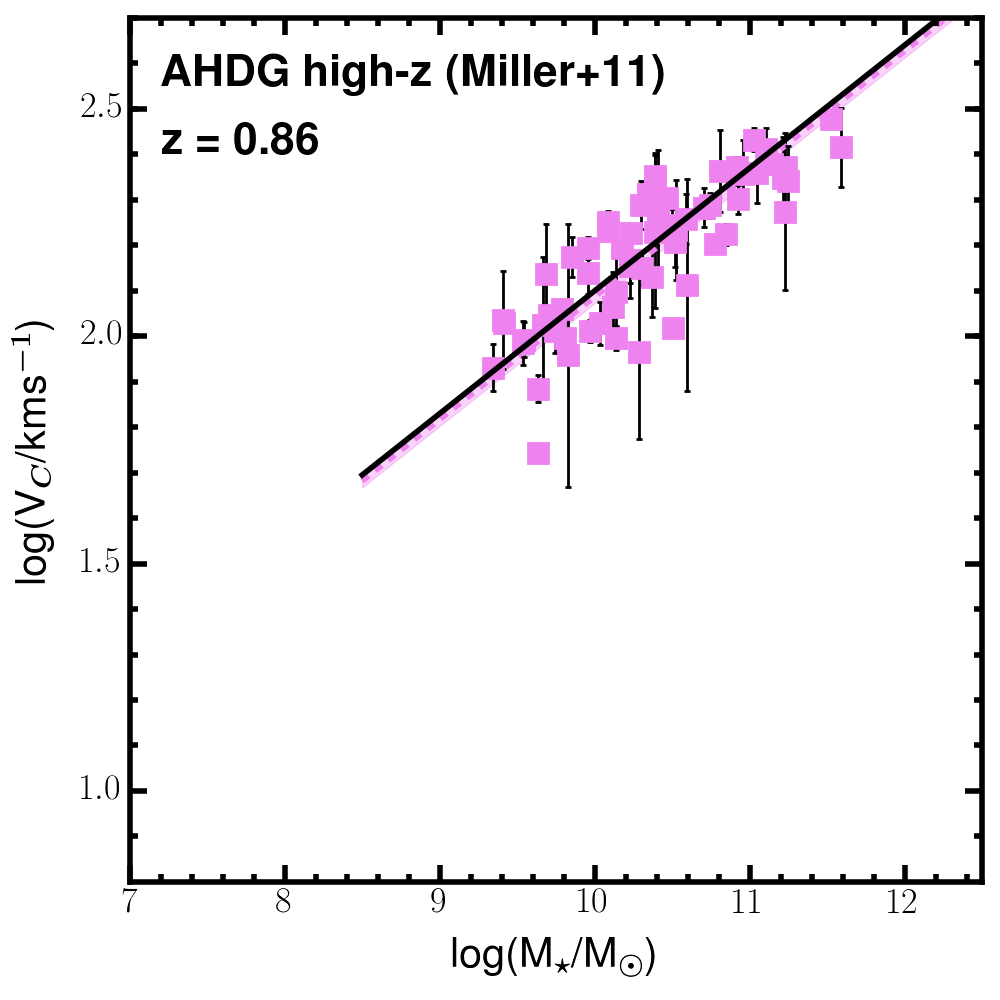}
    \end{subfigure} 
    \begin{subfigure}[h!]{0.246\textwidth}
        \centering \hspace{-2.85cm}
        \includegraphics[height=1.825in]{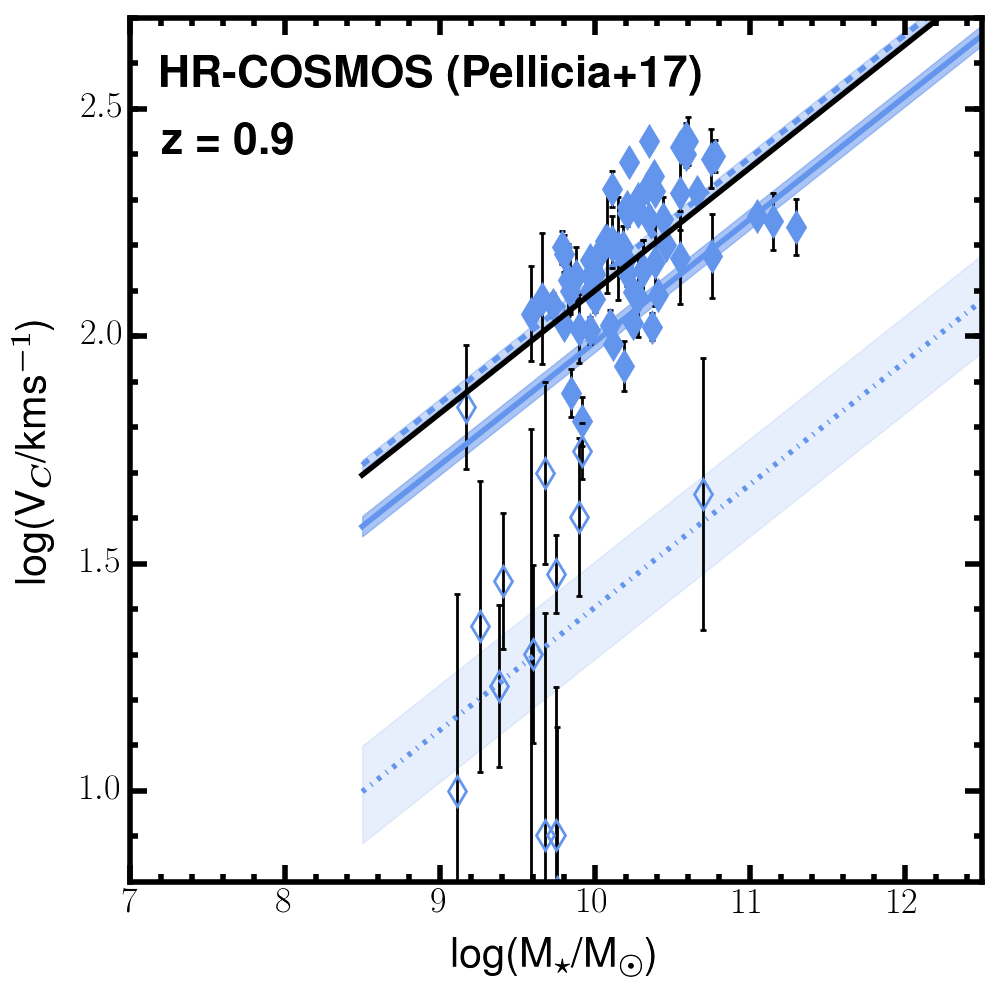}
    \end{subfigure} \hspace{-1.4cm}
    \begin{subfigure}[h!]{0.246\textwidth}
        \centering 
        \includegraphics[height=1.825in]{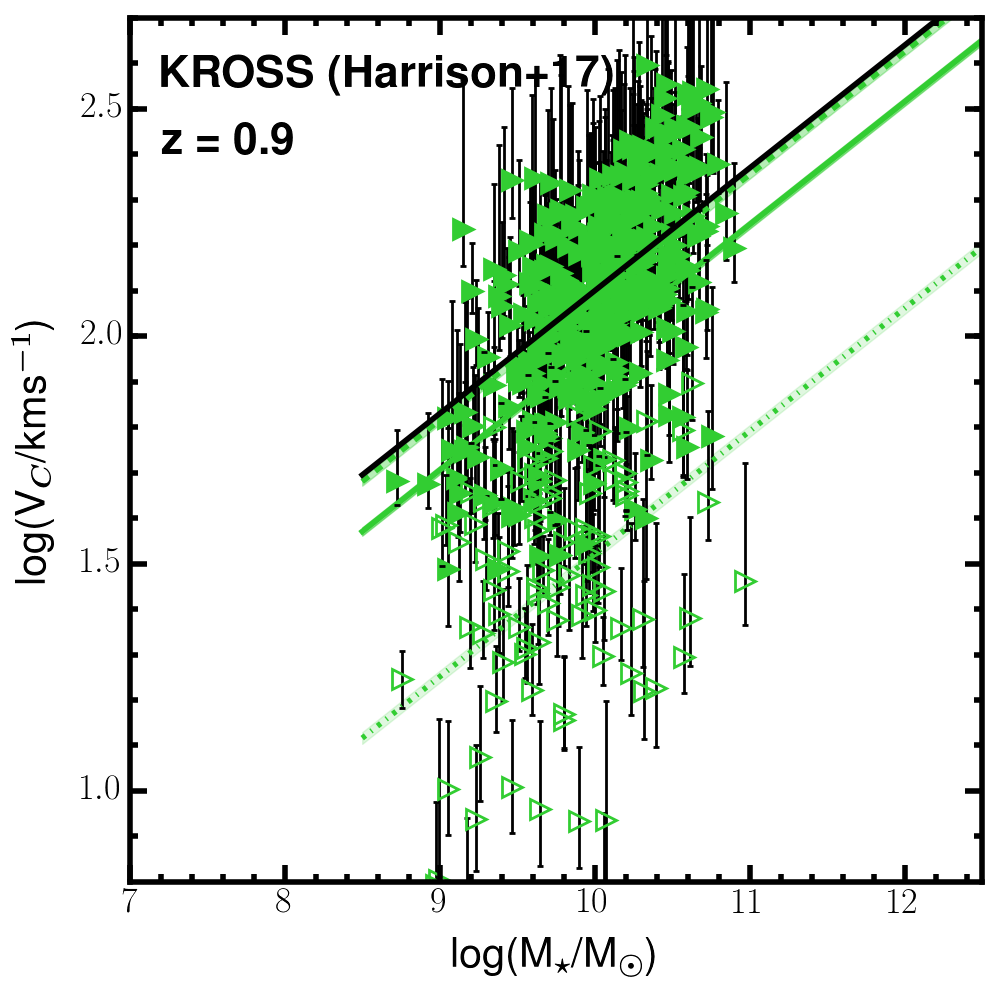}
    \end{subfigure} \hspace{0.15cm}
    \begin{subfigure}[h!]{0.246\textwidth}
        \centering
        \includegraphics[height=1.825in]{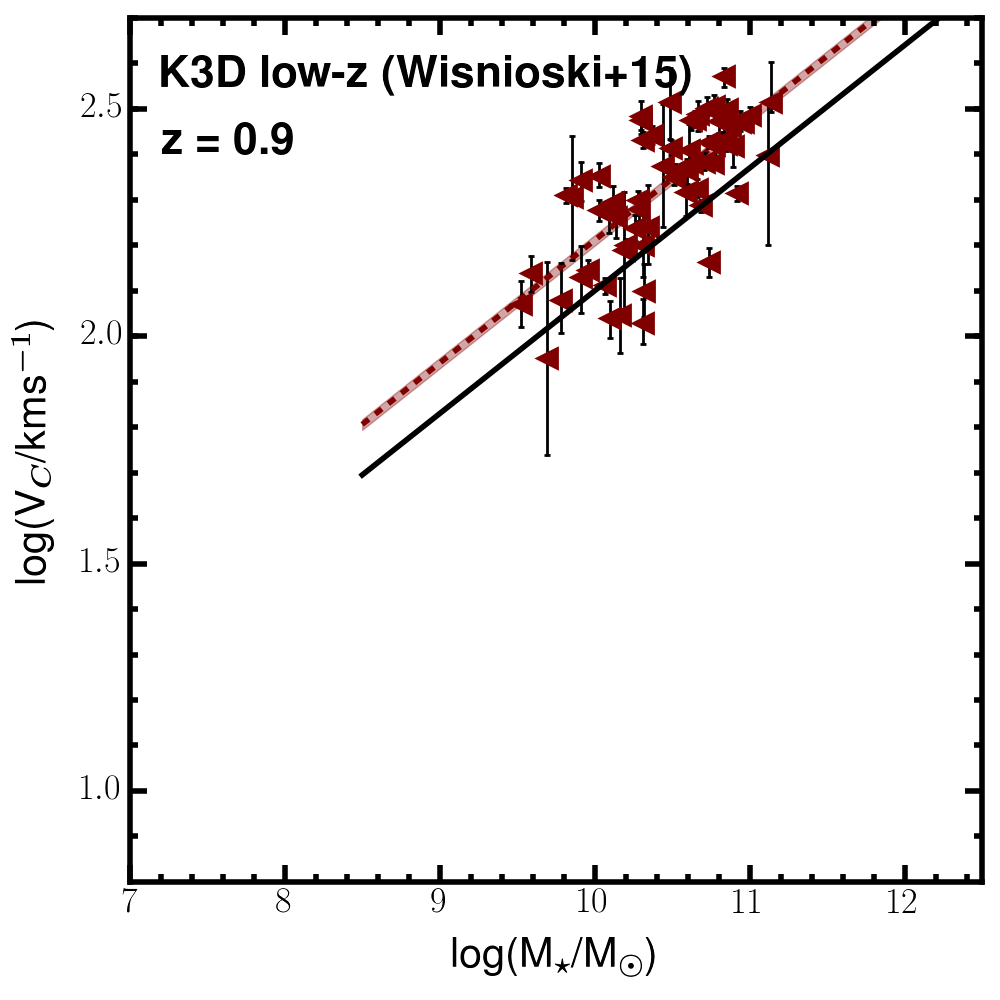}
    \end{subfigure} \hspace{0.15cm}
    \begin{subfigure}[h!]{0.246\textwidth}
        \centering
        \includegraphics[height=1.825in]{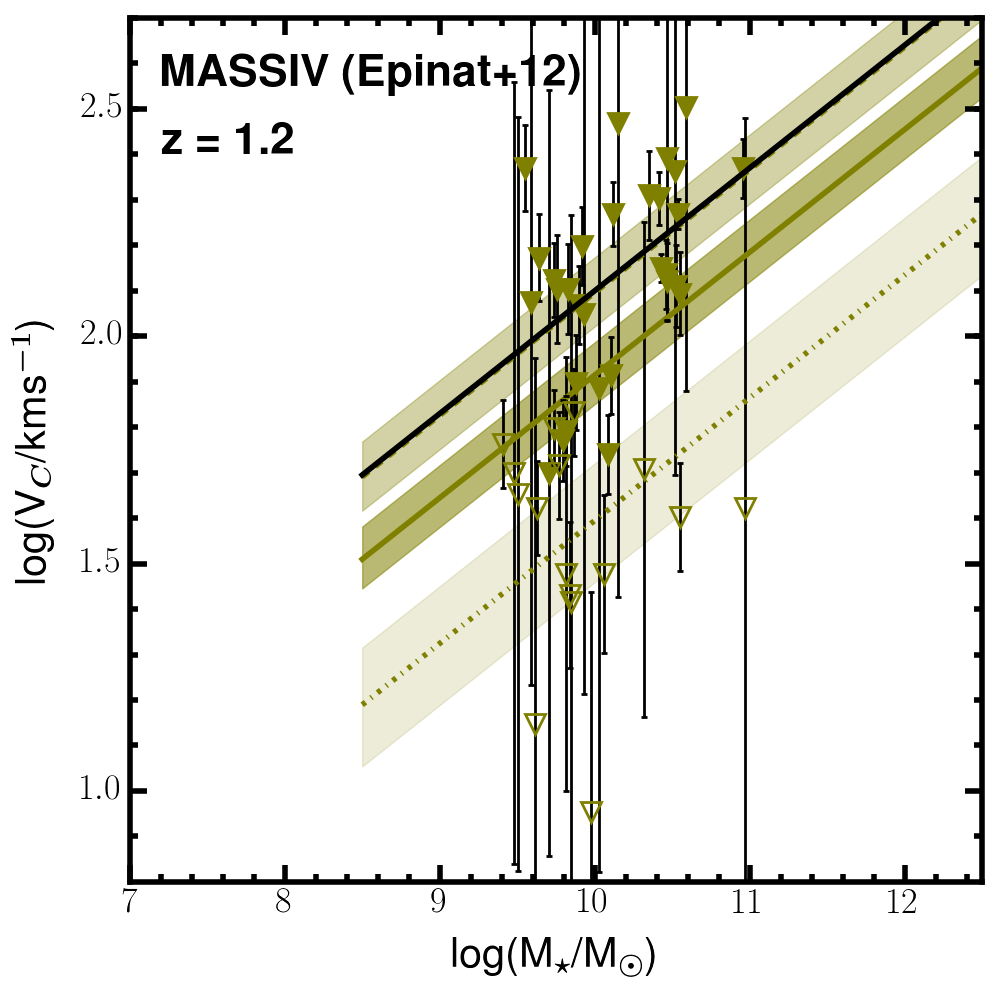}
    \end{subfigure}
    \begin{subfigure}[h!]{0.246\textwidth}
        \centering \hspace{-2.85cm}
        \includegraphics[height=1.825in]{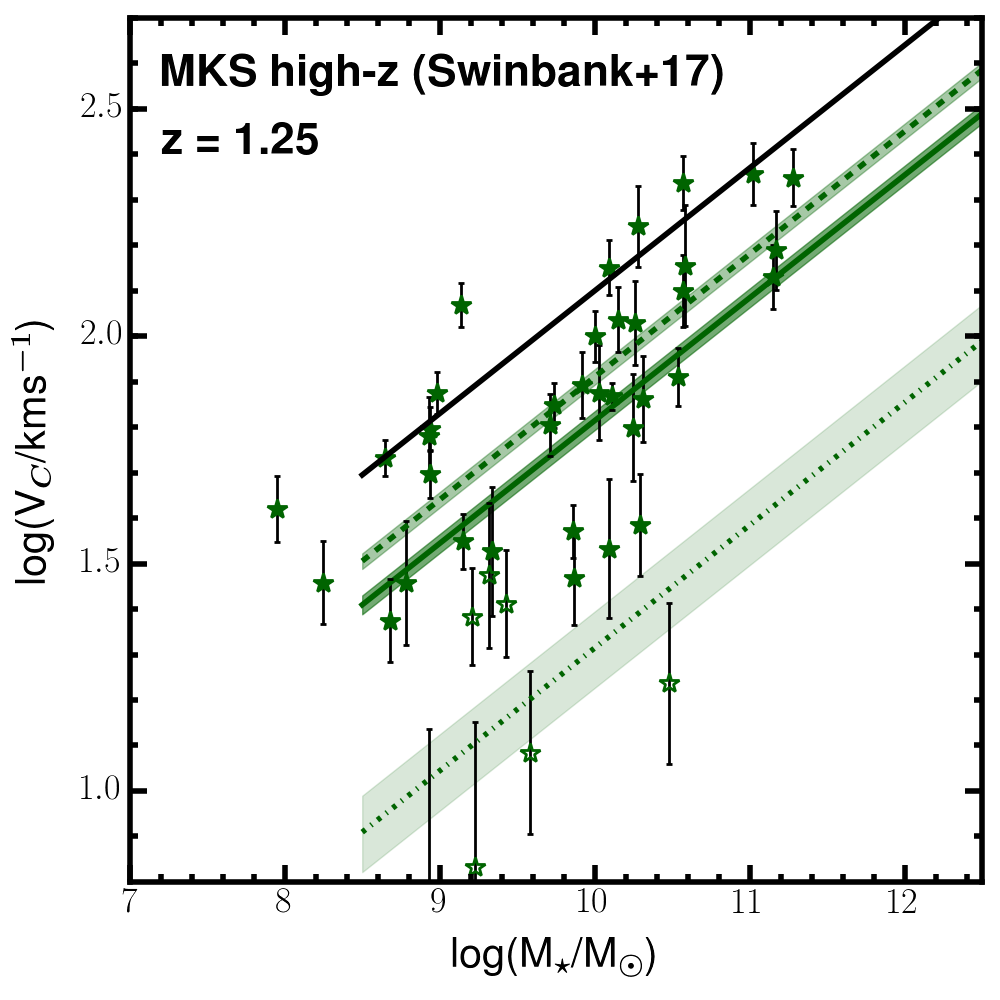}
    \end{subfigure} \hspace{-1.4cm}
    \begin{subfigure}[h!]{0.246\textwidth}
        \centering 
        \includegraphics[height=1.825in]{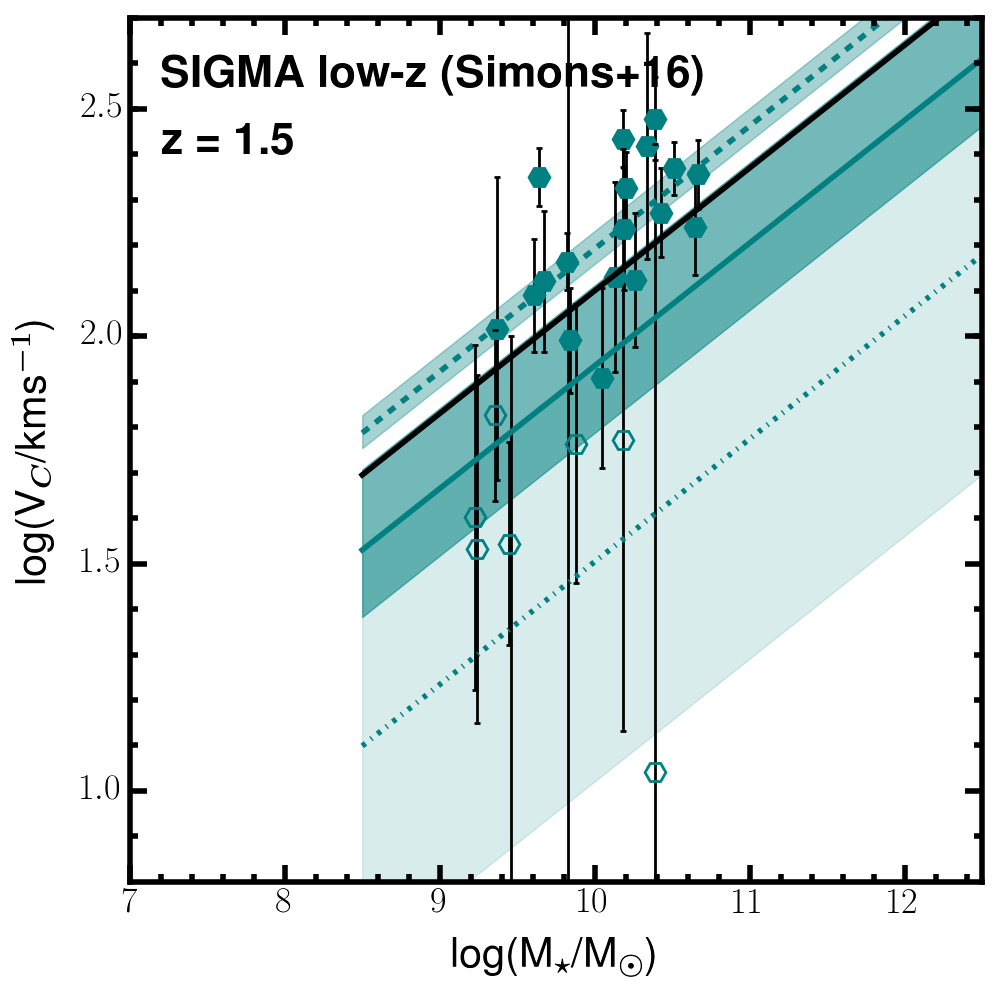}
    \end{subfigure} \hspace{0.15cm}
    \begin{subfigure}[h!]{0.246\textwidth}
        \centering
        \includegraphics[height=1.825in]{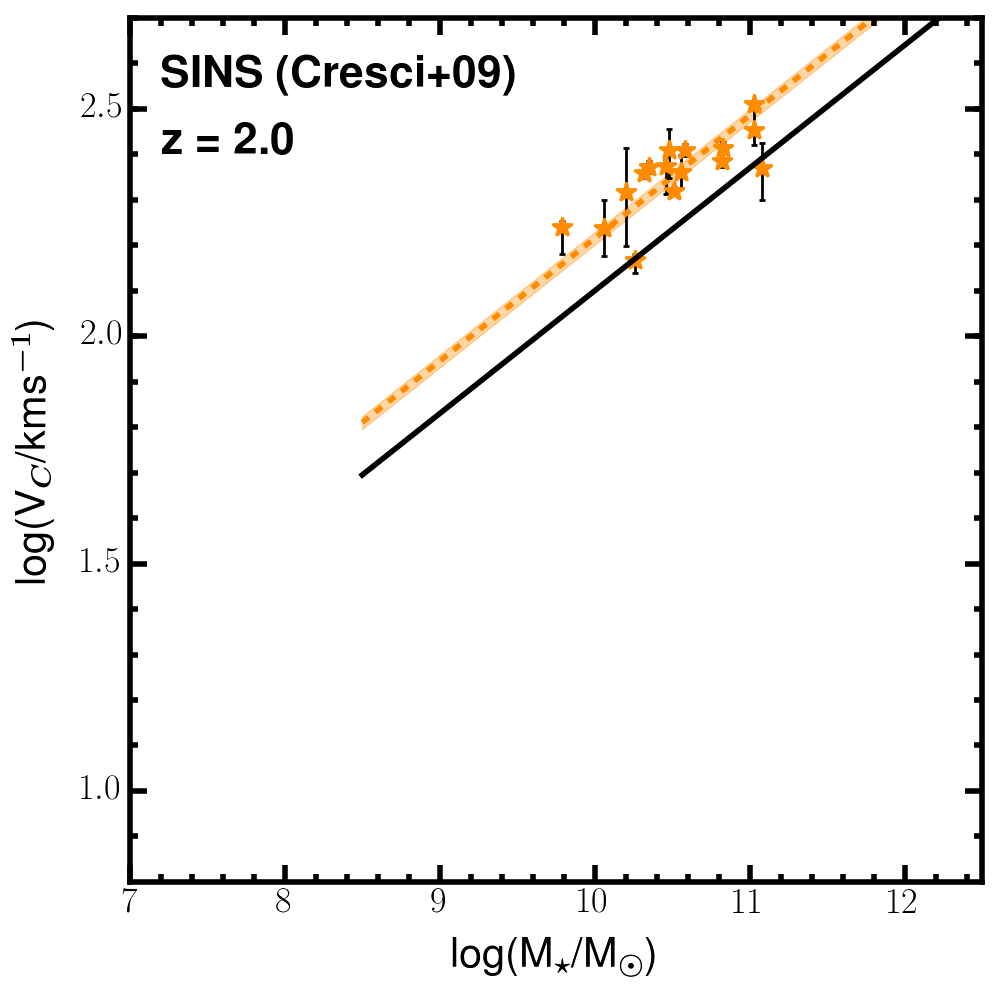}
    \end{subfigure} \hspace{0.15cm}
    \begin{subfigure}[h!]{0.246\textwidth}
        \centering
        \includegraphics[height=1.825in]{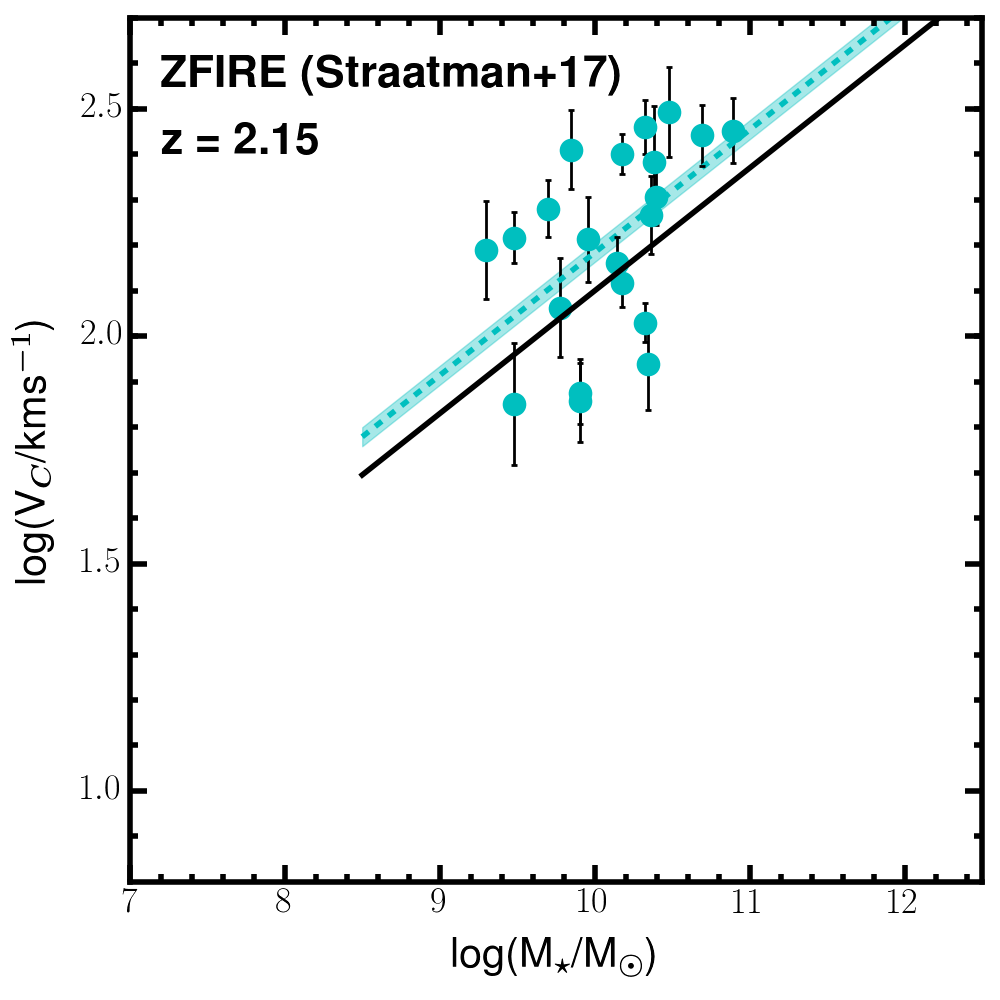}
    \end{subfigure}
    \begin{subfigure}[h!]{0.246\textwidth}
        \centering \hspace{-2.85cm}
        \includegraphics[height=1.825in]{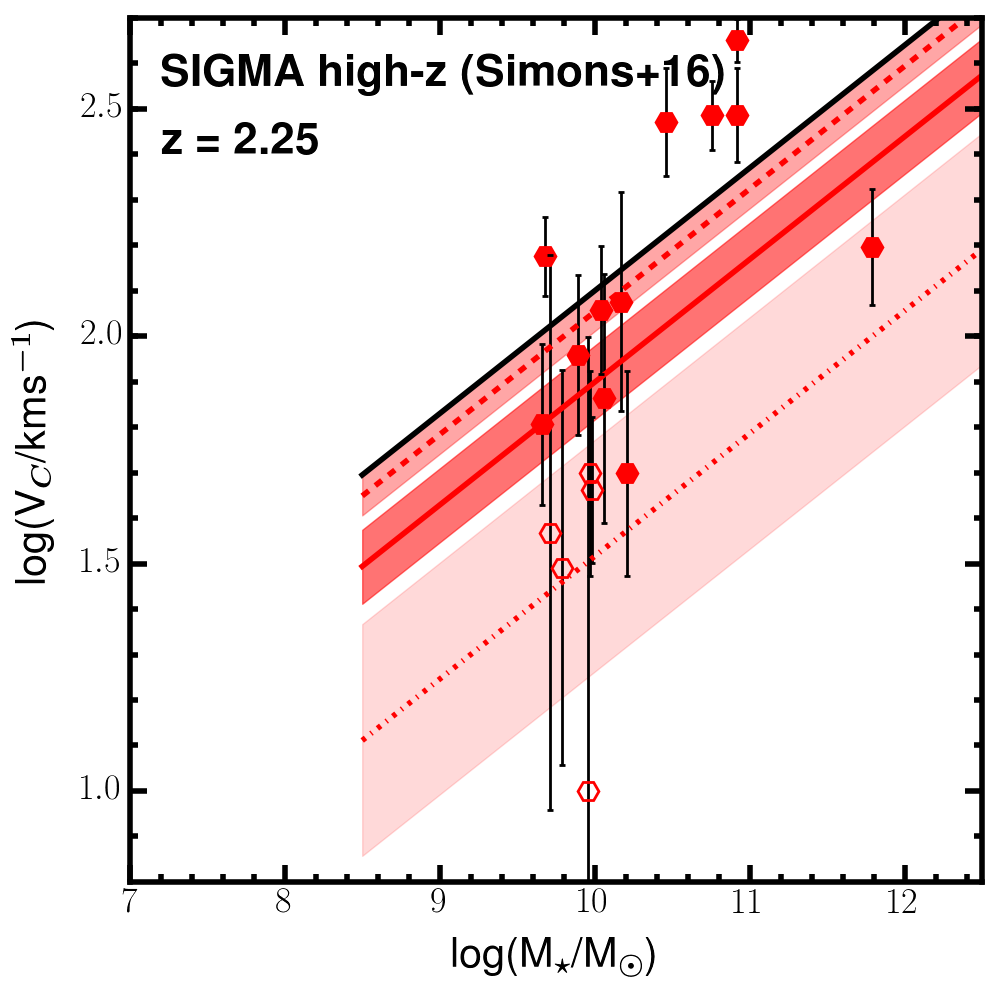}
    \end{subfigure} \hspace{-1.4cm}
    \begin{subfigure}[h!]{0.246\textwidth}
        \centering
        \includegraphics[height=1.825in]{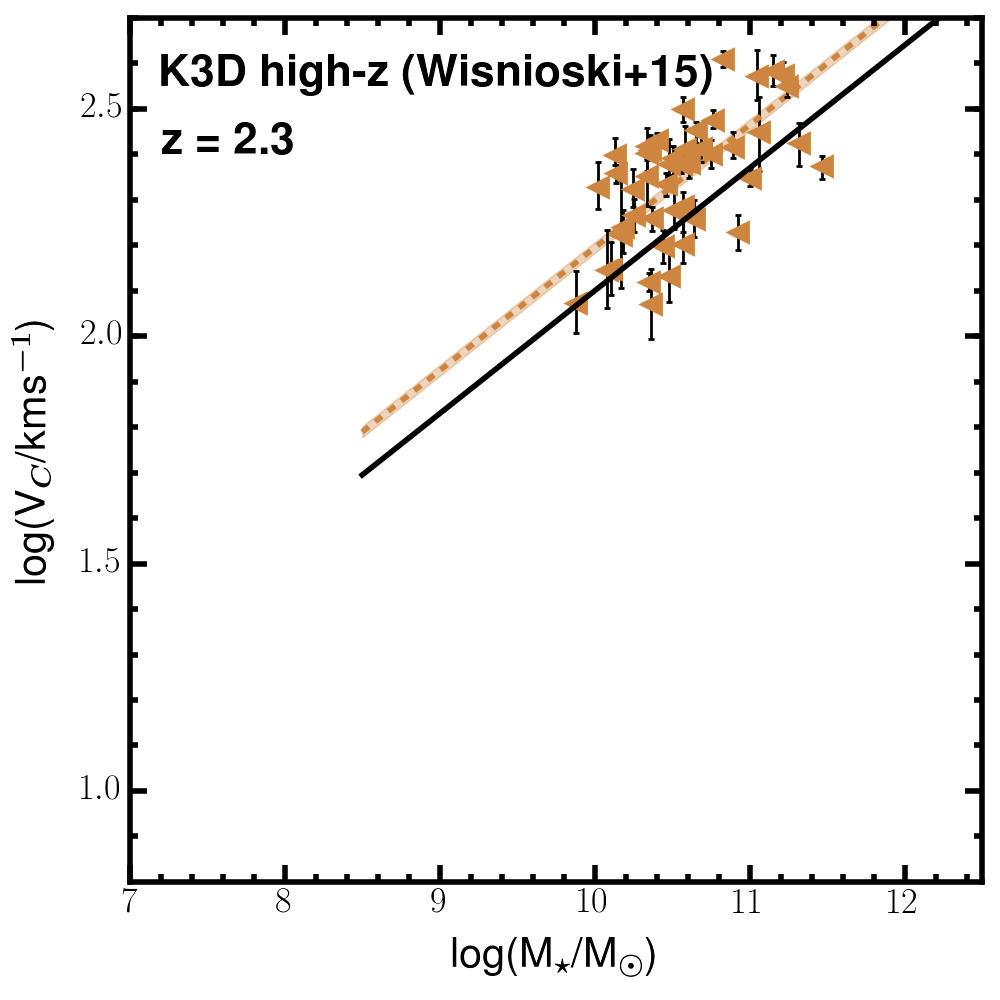}
    \end{subfigure} \hspace{0.15cm}
    \begin{subfigure}[h!]{0.246\textwidth}
        \centering  
        \includegraphics[height=1.825in]{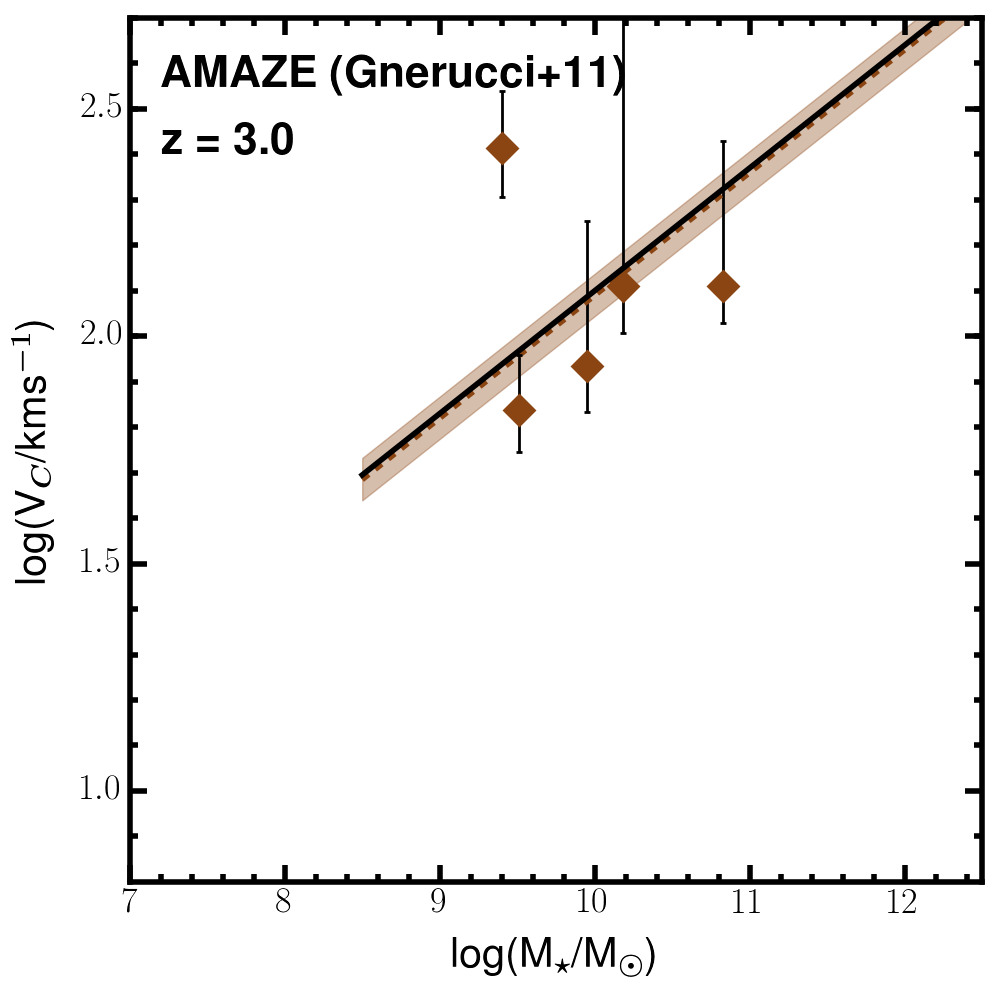}
    \end{subfigure} \hspace{0.15cm}
    \begin{subfigure}[h!]{0.246\textwidth}
        \centering
        \includegraphics[height=1.825in]{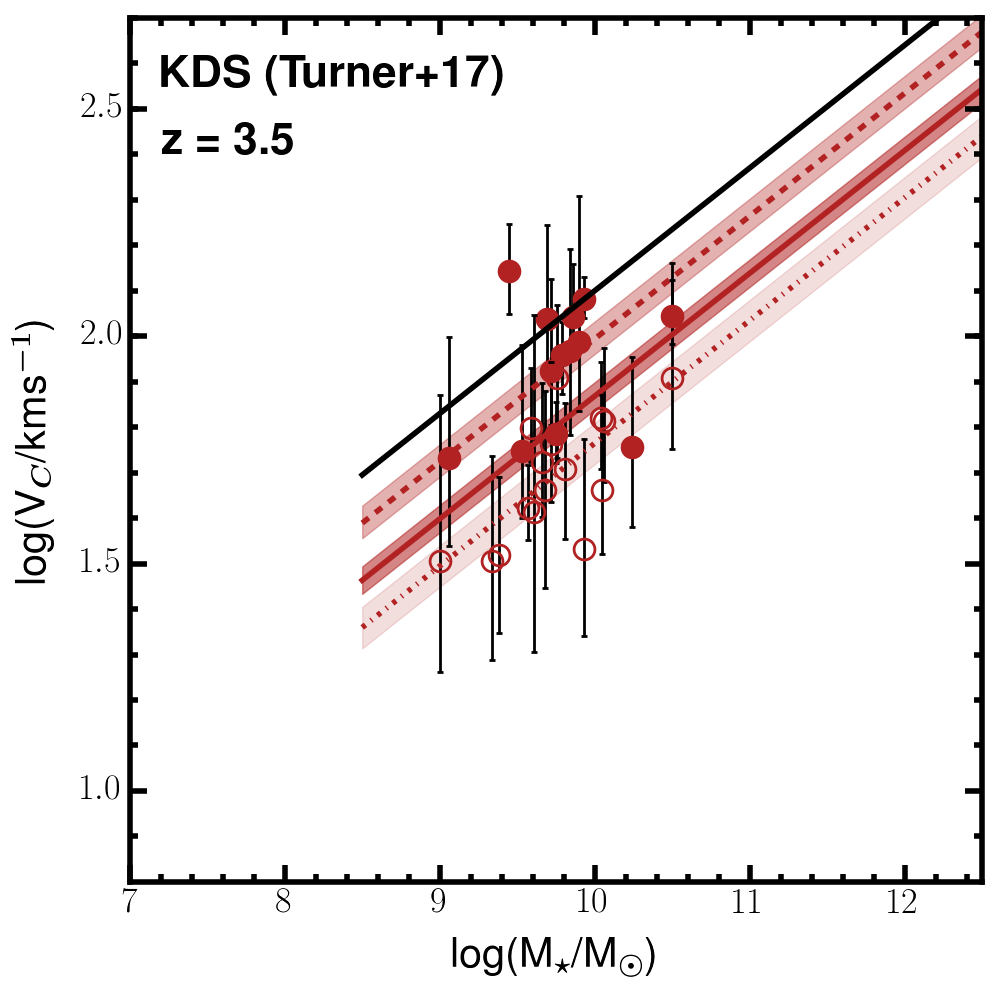}
    \end{subfigure}
    \caption{Best fits of the relation $\textrm{log}({\it V_{\textrm{C}}}) = \upbeta + \upalpha[\textrm{log}({\it M_{\star}}) - 10.1]$ to the comparison samples, using fixed slope $\upalpha=0.270$, with the recovered $\upbeta$ values used throughout Fig.\,\ref{fig:velocity_evolution}.
    Rotation-dominated and dispersion-dominated galaxies have filled and hollow symbols respectively.
    The solid line shows the fit to the full samples, the dashed-line the fit to the rotation-dominated galaxies and the dot-dashed line the fit to the dispersion-dominated galaxies.
    Shaded regions represent the $1-\sigma$ uncertainty on the fits.}
    \label{fig:velocity_fits}
\end{figure*}

\begin{figure*}
    \centering \hspace{-1.65cm}
    \begin{subfigure}[h!]{0.246\textwidth}
        \centering
        \includegraphics[height=1.825in]{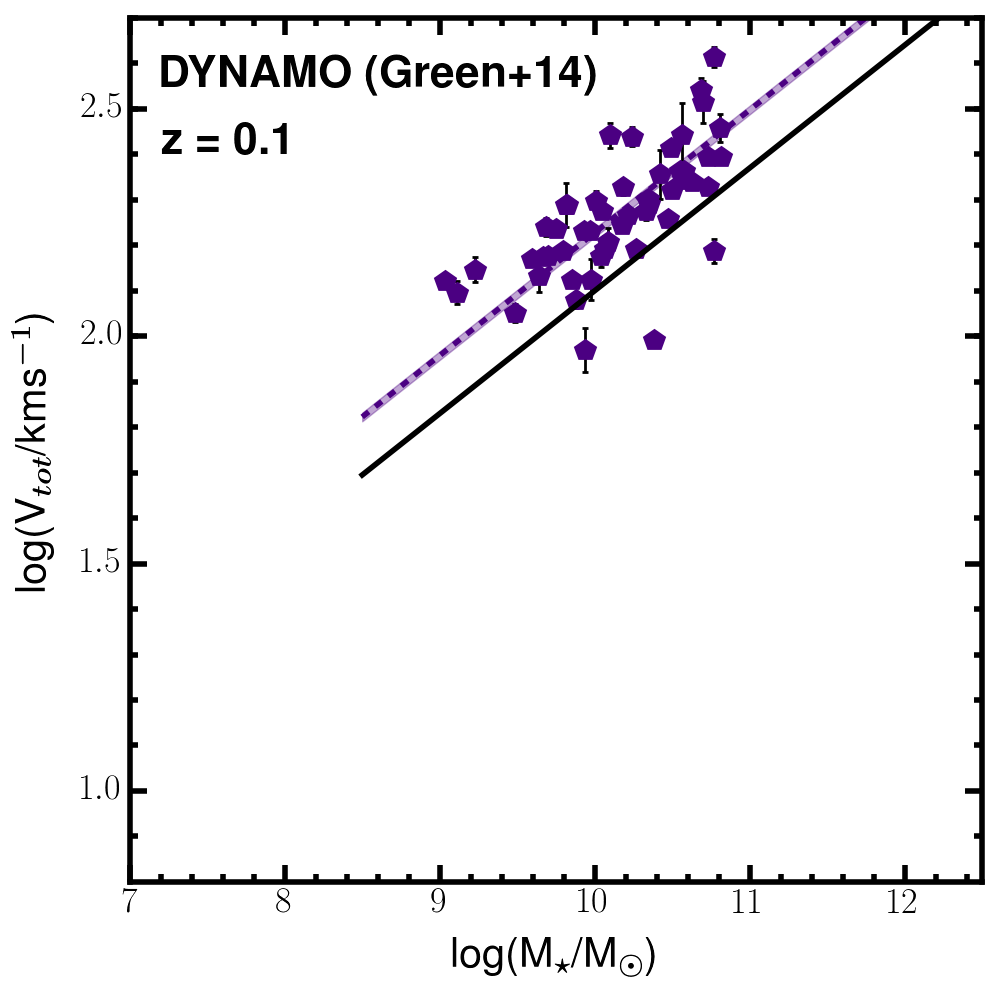}
    \end{subfigure} \hspace{0.15cm}
    \begin{subfigure}[h!]{0.246\textwidth}
        \centering
        \includegraphics[height=1.825in]{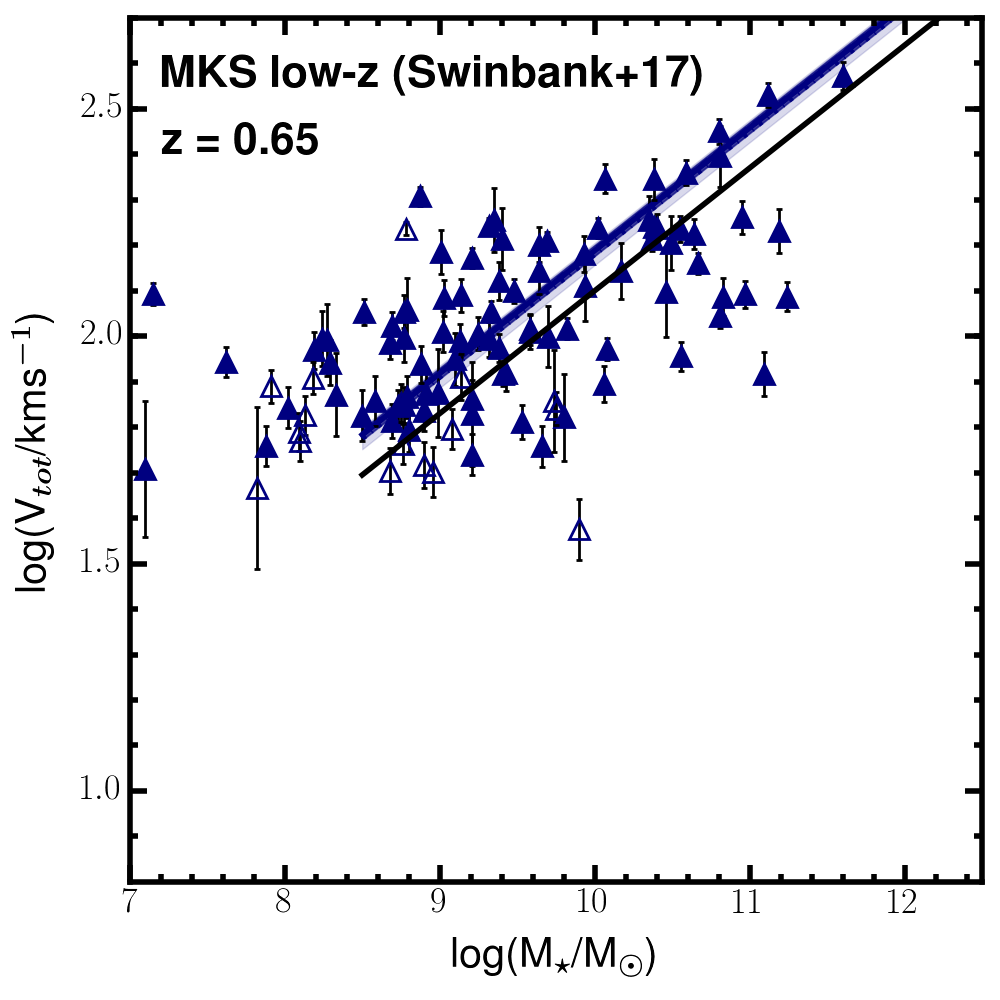}
    \end{subfigure} \hspace{0.15cm}
   \begin{subfigure}[h!]{0.246\textwidth}
        \centering
        \includegraphics[height=1.825in]{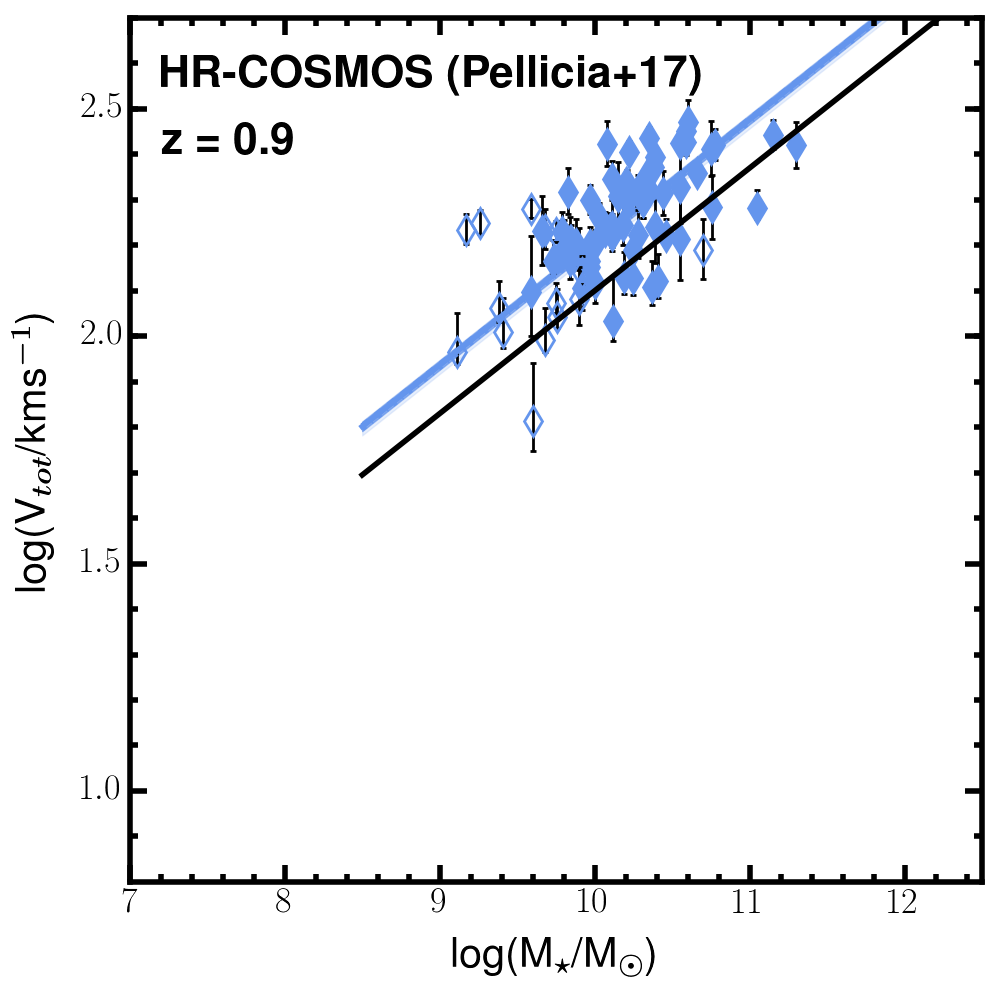}
    \end{subfigure} \hspace{0.15cm}
    \begin{subfigure}[h!]{0.246\textwidth}
        \centering
        \includegraphics[height=1.825in]{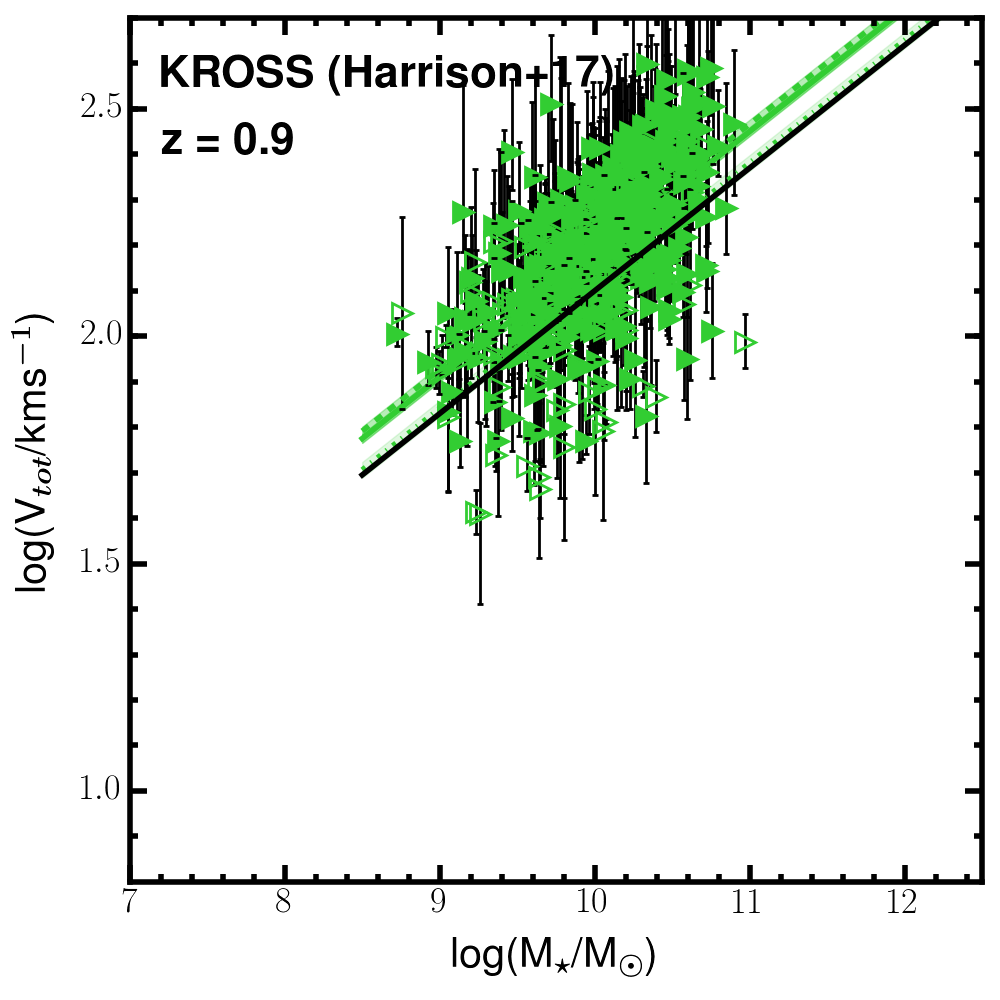}
    \end{subfigure} \hspace{0.15cm}
    \begin{subfigure}[h!]{0.246\textwidth}
        \centering \hspace{-2.85cm}
        \includegraphics[height=1.825in]{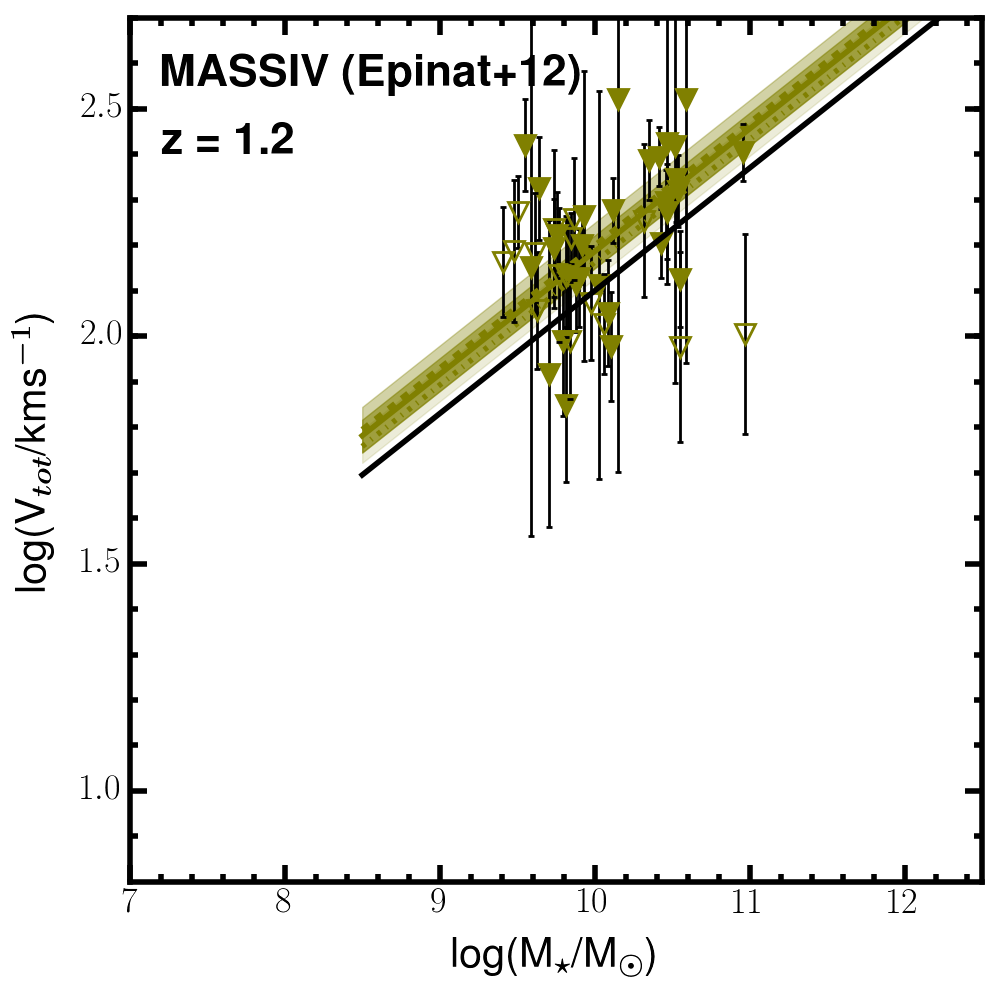}
    \end{subfigure} \hspace{-1.4cm}
    \begin{subfigure}[h!]{0.246\textwidth}
        \centering 
        \includegraphics[height=1.825in]{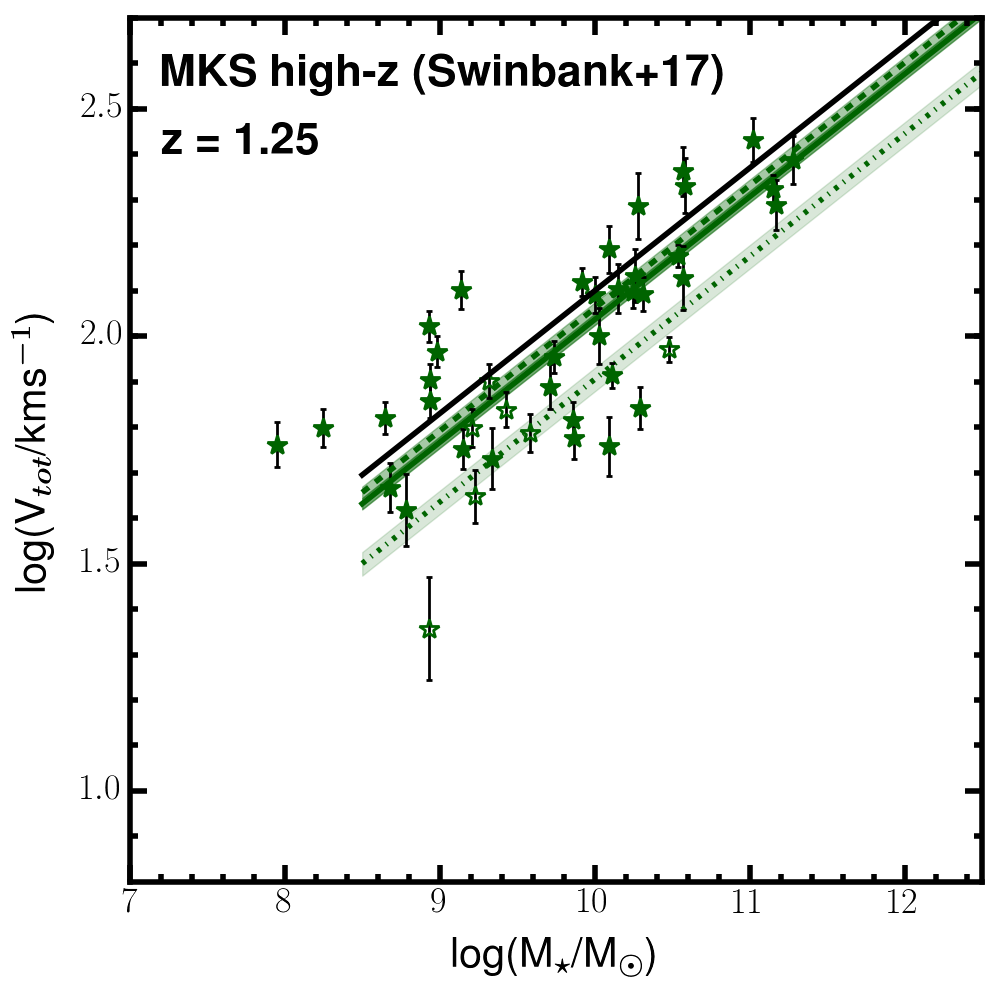}
    \end{subfigure} \hspace{0.15cm}
    \begin{subfigure}[h!]{0.246\textwidth}
        \centering
        \includegraphics[height=1.825in]{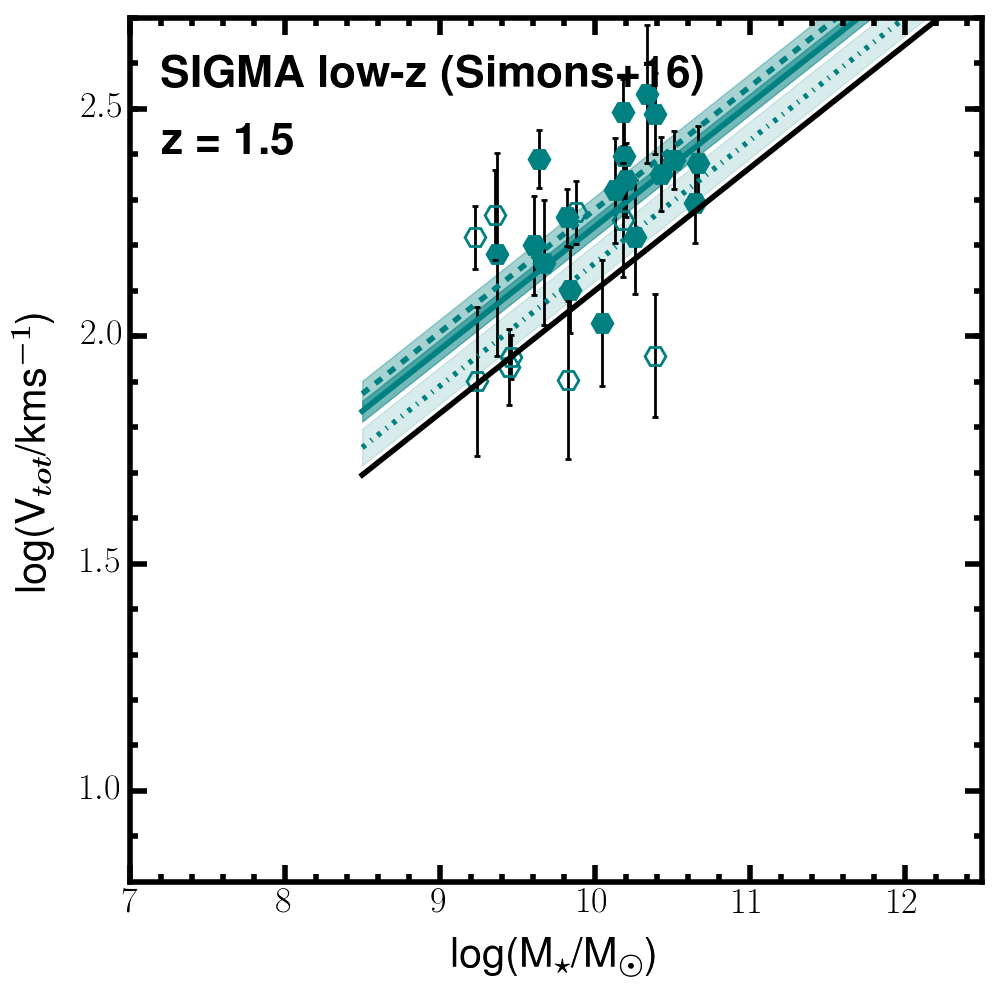}
    \end{subfigure} \hspace{0.15cm}
    \begin{subfigure}[h!]{0.246\textwidth}
        \centering
        \includegraphics[height=1.825in]{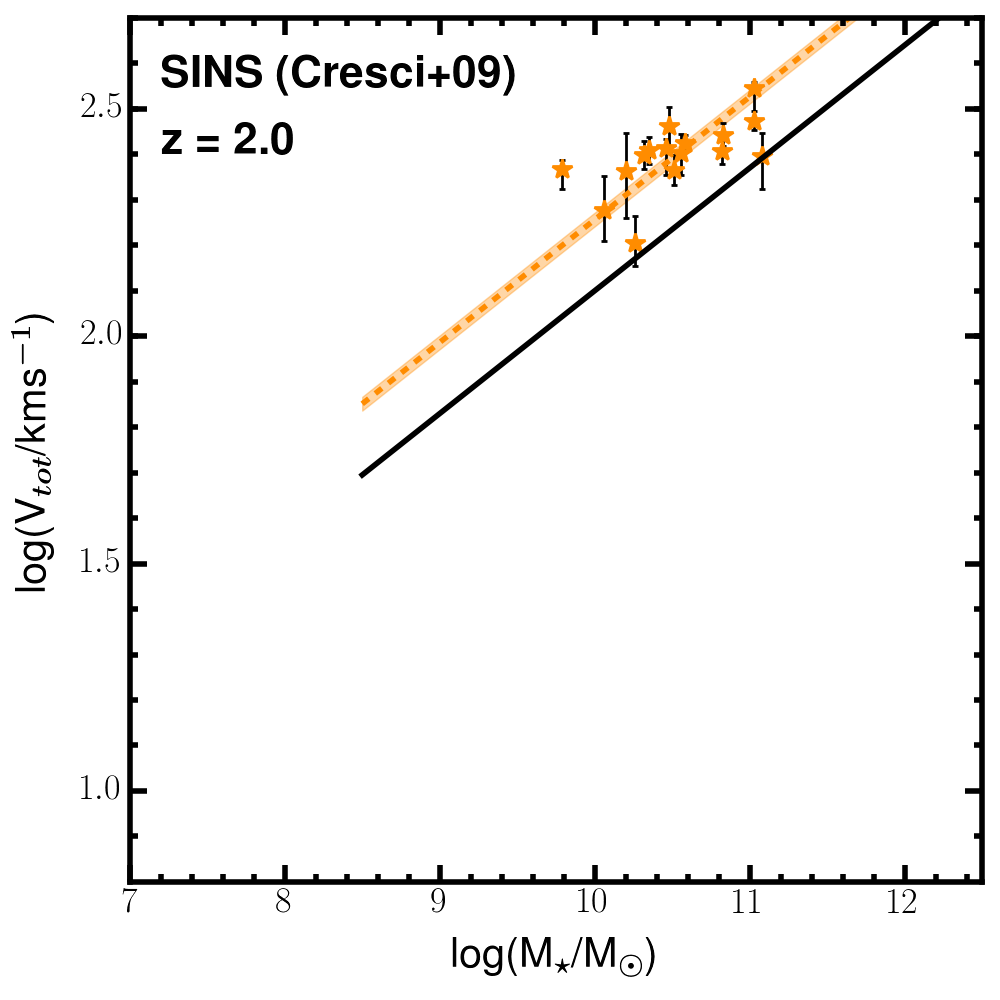}
    \end{subfigure}
    \begin{subfigure}[h!]{0.246\textwidth}
        \centering \hspace{-2.85cm}
        \includegraphics[height=1.825in]{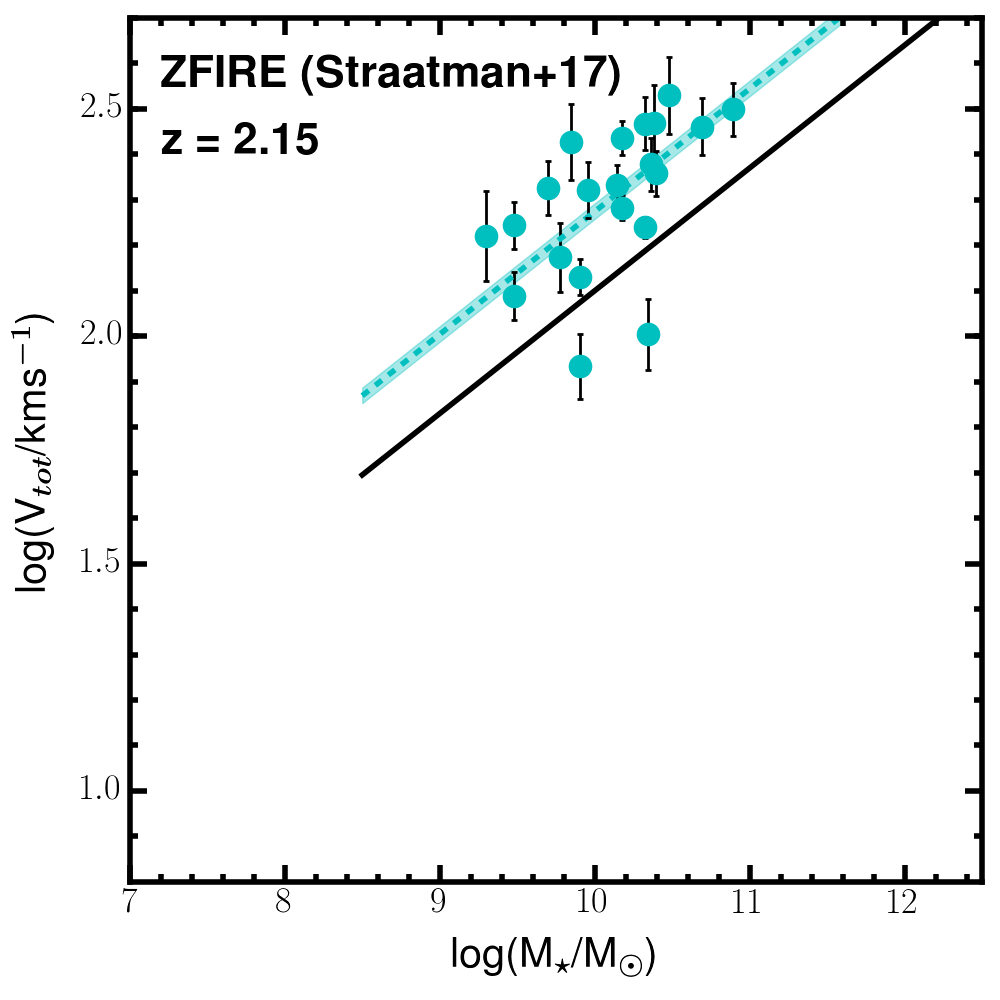}
    \end{subfigure} \hspace{-1.4cm}
    \begin{subfigure}[h!]{0.246\textwidth}
        \centering
        \includegraphics[height=1.825in]{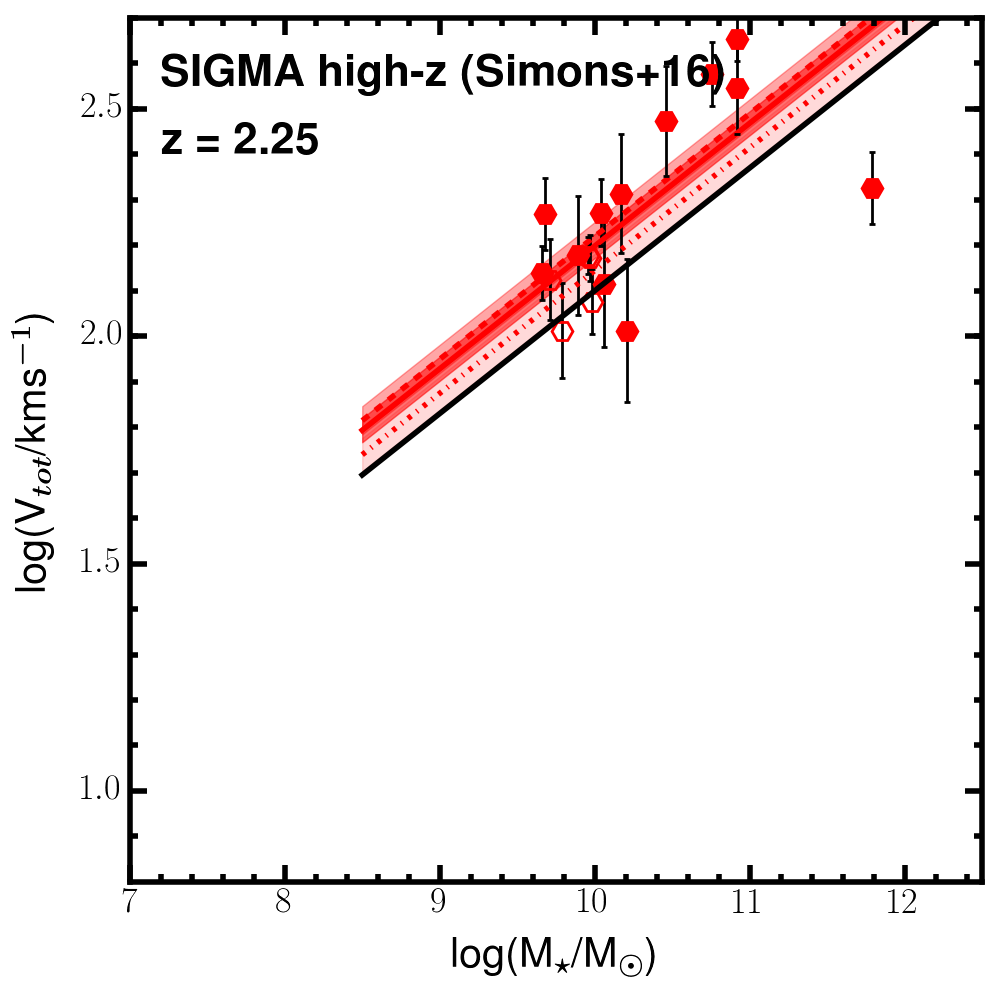}
    \end{subfigure} \hspace{0.15cm}
    \begin{subfigure}[h!]{0.246\textwidth}
        \centering
        \includegraphics[height=1.825in]{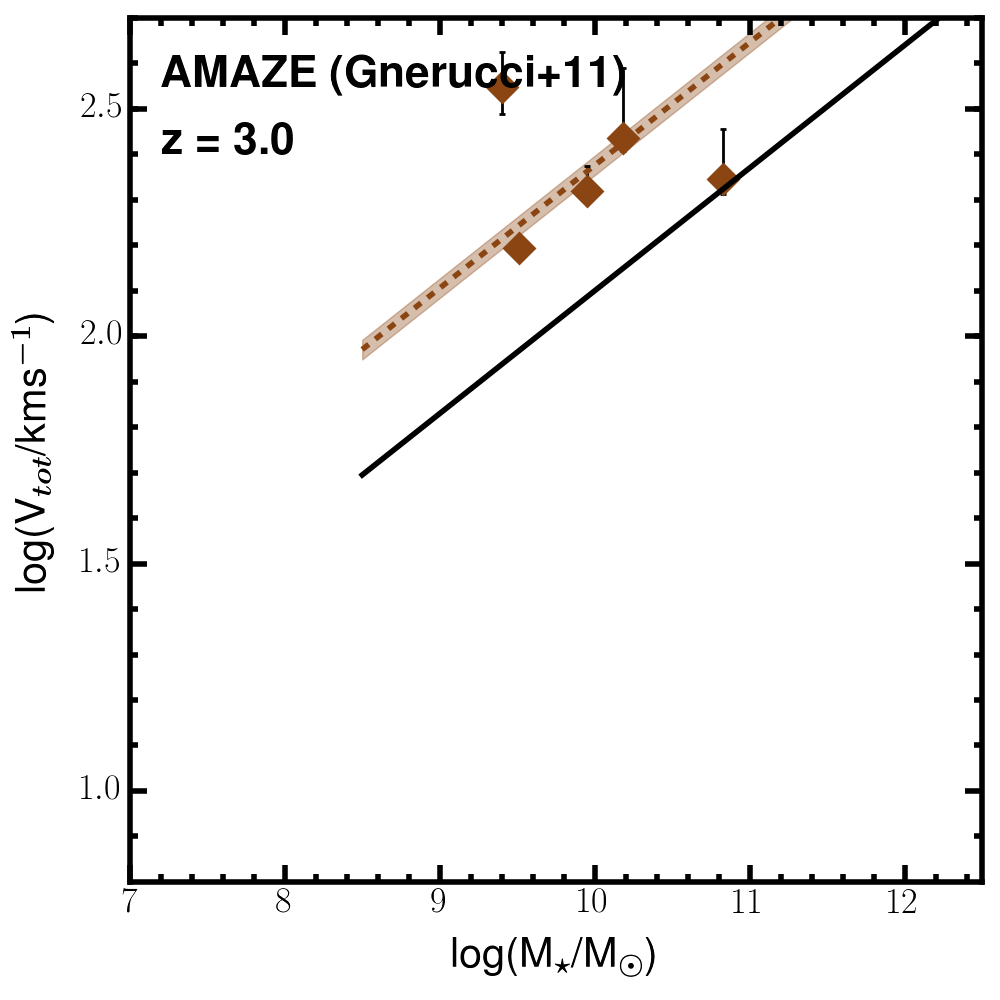}
    \end{subfigure} \hspace{0.15cm}
    \begin{subfigure}[h!]{0.246\textwidth}
        \centering 
        \includegraphics[height=1.825in]{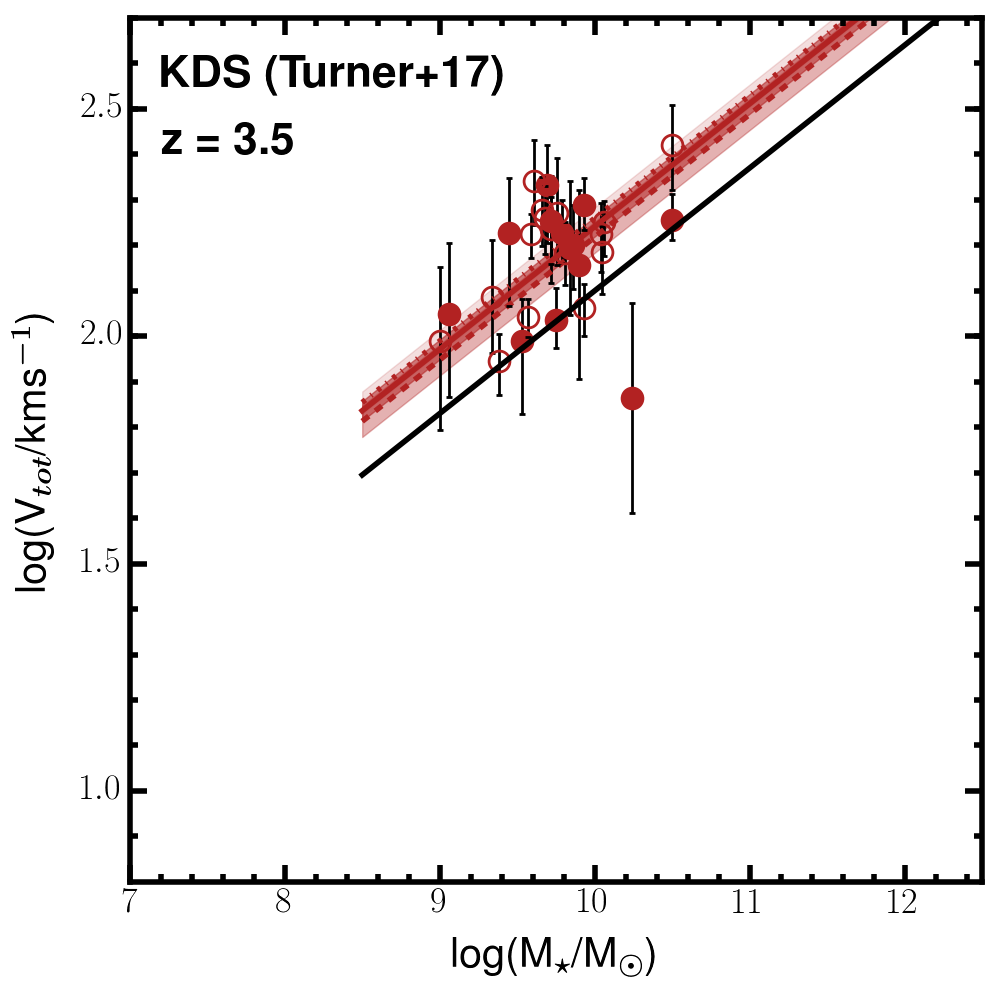}
    \end{subfigure}
    \caption{Best fits of the relation $\textrm{log}({\it V_{\textrm{tot}}}) = \upbeta + \upalpha[\textrm{log}({\it M_{\star}}) - 10.1]$ to the comparison samples, using fixed slope $\upalpha=0.270$, with the recovered $\upbeta$ values used throughout Fig.\,\ref{fig:vtot_evolution}.
    The symbol convention is equivalent to Fig.\,\ref{fig:velocity_fits}.} 
    \label{fig:v_tot_fits}
\end{figure*}

We include for reference the fits to the comparison sample data in the $V_{\textrm{C}}$ vs. $M_{\star}$ plane (Fig.\,\ref{fig:velocity_fits}) and the $V_{\textrm{tot}}$ vs. $M_{\star}$ plane (Fig.\,\ref{fig:v_tot_fits}).
In each subplot the reference sample and redshift are indicated, with solid points indicating rotation-dominated galaxies, hollow points dispersion-dominated (where applicable), the solid coloured line the fit to `All' galaxies, the coloured-dashed line the fit to the rotation-dominated galaxies and the coloured-dash-dot line the fit to the dispersion-dominated galaxies (where applicable).
The shaded regions around these lines indicate the $1-\sigma$ uncertainties on the fits.
The solid black line shows the local reference relation from fitting the \cite{Reyes2011} sample, which we show enlarged in Fig.\,\ref{fig:app_reyes_fit} (and see \ref{fig:rom_tf_relation}) which is used for comparison when constructing Figs \ref{fig:velocity_evolution} and \ref{fig:vtot_evolution}.
\bsp
\label{lastpage}

\clearpage
\thispagestyle{empty}

\clearpage
    \thispagestyle{empty}

\end{document}